\documentclass[10pt]{article}
\usepackage{amsfonts}
\usepackage{amssymb}
\usepackage{amsthm}

\newcommand{\sss}{\setcounter{equation}{0}}
\newtheorem{theorem}{THEOREM}[section]

\newtheorem{lemma}[theorem]{LEMMA}

\newtheorem{corollary}[theorem]{COROLLARY}

\newtheorem{remark}[theorem]{REMARK}

\newtheorem{prop}[theorem]{PROPOSITION}

\newtheorem{definition}[theorem]{DEFINITION}

\newtheorem{example}[theorem]{EXAMPLE}

\newtheorem{assumption}[theorem]{ASSUMPTION}

\newcommand{\ere}{ {\mathbb R}}
\newcommand{\ZETA}{{\mathbb Z}}
\newcommand{\ese}{{\mathbb S}}
\newcommand{\CE}{{\mathbb C}}

\newcommand{\pot}{ \mathcal A_{\Phi}(B)}
\def\nb {\mathcal A_{\Phi}(0)}
\newcommand {\erc}{\mathcal R}
\def\p2{\mathcal A_{\Phi,2\pi}(B)}
\def\0p2{\mathcal A_{\Phi,2\pi}(0)}
\def\sp2{\mathcal A_{\Phi,2\pi,\hbox{\rm SR}}(B)}
\def\hr{H_{1,\hbox{\rm rec}}(\Lambda;\ere)}
\def\beq{\begin{equation}}
\def\ene{\end{equation}}
\def \ds {\displaystyle}
\newcommand{\bull}{\hfill $\Box$}
\def\qed{\ifhmode\unskip\nobreak\fi\ifmmode\ifinner
\else\hskip5pt\fi\fi\hbox{\hskip5pt\vrule width4pt height6pt
depth1.5pt\hskip1pt}}
\def\v{\mathbf v}
\def\hu{\hat{\mathbf u}}
\def\hv{\hat{\mathbf v}}
\def\hn{\hat{\mathbf n}}
\def\hw{\hat{\mathbf w}}
\def\X{\natural\, k\, T}
\def\curl{\, \hbox{ \rm curl}\,}
\def\mo{\mathbf p}
\def\ac{A_{C}}

\def \tf{\tilde{\phi}}
\def \ts{\tilde{\psi}}

\def\rec{\Lambda_{\rm \hbox{\rm rec}}}
\begin{document}
\baselineskip=20 pt
\parskip 6 pt

\title{High-Velocity Estimates for the Scattering Operator and Aharonov-Bohm Effect in Three Dimensions
 \thanks{ Research partially supported by
 CONACYT under Project P42553­F.}}
 \author{ Miguel Ballesteros and Ricardo Weder \thanks{On leave of absence from Departamento de M\'etodos
 Matem\'aticos  y Num\'ericos. Instituto de Investigaciones en Matem\'aticas Aplicadas y en Sistemas.
 Universidad Nacional Aut\'onoma de M\'exico. Apartado Postal 20-726, M\'exico DF 01000. Ricardo Weder is
 a Fellow of the Sistema Nacional de Investigadores.}
\\Department of Mathematics and Statistics. University of Helsinki\\
 P.O. Box 68 (Gustaf Hallstromin katu 2b)
FI-00014. Finland
\\ballesteros.miguel.math@gmail.com, weder@servidor.unam.mx}

\date{}
\maketitle
\begin{center}
\begin{minipage}{5.75in}
\centerline{{\bf Abstract}}
\bigskip
We obtain high-velocity estimates with error bounds for the scattering
operator of the Schr\"odinger equation in three dimensions with
electromagnetic potentials in the exterior of bounded obstacles that are handlebodies. A  particular
 case is a finite number of tori. We prove our results with  time-dependent methods.
We consider high-velocity estimates where the direction of the
velocity of the incoming electrons is kept fixed as its absolute value  goes to infinity.
In the case of one torus our results give a rigorous proof that
quantum mechanics predicts the interference patterns observed in the
fundamental experiments of Tonomura et al. that gave a conclusive
evidence of the existence of the  Aharonov-Bohm effect using a toroidal magnet. We give a
method for the reconstruction of the flux of the magnetic field
over a cross-section of the torus modulo $2\pi$. Equivalently, we determine
modulo $2\pi$ the difference in phase for two electrons  that travel
to infinity, when one goes inside the hole and the other outside it. For this purpose we only
need the high-velocity limit of the scattering operator for one direction of the velocity of the incoming
electrons. When there are several tori -or more generally handlebodies-  the information that we obtain in the
fluxes, and on the difference of phases, depends on the relative
position of the tori and on the direction  of the velocities when we take the high-velocity limit of the
incoming electrons. For some locations of the tori we can determine all the fluxes modulo 2$\pi$
 by taking the high-velocity
limit in only one direction.
We also give a method for the unique reconstruction of the electric
potential and the magnetic field outside the handlebodies from the
high-velocity limit of the scattering operator.

\end{minipage}
\end{center}
\newpage
\section{Introduction}
The Aharonov-Bohm effect is a fundamental quantum mechanical phenomenon wherein charged particles,
like electrons, are physically influenced, in the form of a phase shift,  by the existence of magnetic fields
in regions that are inaccessible to the particles. This genuinely quantum mechanical phenomenon was predicted
by Aharonov and Bohm  \cite{ab}. See also Ehrenberg and Siday \cite{es}. This phenomenon has been extensively
studied both, from the theoretical, and the experimental points of view. For a review of the literature see
\cite{op} and \cite{pt}. There has  been a large controversy, involving over three hundred papers, concerning
the existence of the Aharonov-Bohm effect. For a detailed discussion of this controversy see \cite{pt}.
The issue was finally settled by the fundamental experiments of Tononura et al. \cite{to1,to2}, who used toroidal
magnets to enclose a magnetic flux inside them. In  remarkable experiments they were able to superimpose behind
the magnet an electron beam that traveled inside the hole of the magnet with another electron beam that
traveled outside the magnet, and
they measured the phase shift produced by the magnetic flux enclosed in the magnet, giving a conclusive
evidence of the existence of the Aharonov-Bohm effect.

In this paper we give a rigorous mathematical analysis of this
scattering problem with time-dependent methods. In particular, we
give a rigorous mathematical proof that quantum mechanics predicts
the phase shifts observed in the Tonomura et  al. experiments
\cite{to1,to2}.

We consider bounded obstacles, $K$, whose connected components are handlebodies. In
particular, they can be the
union of a  finite number of bodies diffeomorphic to tori or to balls. Some  of them can be patched trough
the boundary.

We study the high-velocity limit of the scattering operator in the
complement, $\Lambda$, of the obstacle, $K$, for the Schr\"odinger
equation with magnetic field and electric potential in $\Lambda$ and
with magnetic fluxes enclosed in the obstacle $K$. We obtain
high-velocity estimates with error bounds for the scattering
operator using the time-dependent method of \cite{ew}. We consider
high-velocity limits where the direction of the velocity of the
incoming electrons is kept fixed as its absolute value goes to
infinity.

The leading term of our estimate gives us a reconstruction formula that allows us to reconstruct the
circulation of the magnetic potential modulo $2\pi$ along lines in the direction of the velocity
(the X-ray transform).
From these line integrals we uniquely reconstruct the magnetic field in some region of $\Lambda$.
The error term for the leading order goes to zero as a constant divided by the absolute value of the velocity.

The next term in our high-velocity estimate allows us to reconstruct  the integral of the electric potential
along
lines in the direction of the velocity (the X-ray transform). We uniquely reconstruct the electric potential
in a region of $\Lambda$ from
these lines integrals. The error term for this high-velocity estimate goes to zero as a constant divided by
 a power of the absolute value of the velocity, that depends on the decay rate at infinity of the magnetic
 field and of the electric potential. If we have  enough decay  this power is one, as for the leading order.

The leading-order  high-velocity estimate is given in Theorem \ref{th-5.6} and the next term in our
high-velocity estimate is given in Theorem \ref{th-5.8}. The unique reconstruction of the magnetic field and
the electric potential in a region of $\Lambda$ is given in Theorem \ref{th-6.3}. The reconstruction method is
summarized in Remark \ref{rem-6.4}.

Then, we consider the Aharonov-Bohm effect. We assume that the magnetic field in $\Lambda$ is identically
zero. On the contrary, the electric potential is not assumed to be zero. In other words, we  analyze the
Aharonov-Bohm effect in the presence of an electric potential. We use for reconstruction only the
leading-order high-velocity estimate. As for high-velocities the
electric potential gives  a lower-order contribution, it plays no role in the Aharonov-Bohm effect. However,
to allow for a non-trivial electric potential could  be of interest from the experimental point of view.

In Theorem \ref{th-7.1} we reconstruct the circulation of the magnetic potential, modulo $2 \pi$, over a set of
closed paths in $\Lambda$ and in Remark \ref{rm-7.3} we  reconstruct the projection of the
 de Rham cohomology class of the magnetic potential onto a subspace of $H_{\hbox{\rm de R}}^1(\Lambda)$ in
 the sense
 that we  reconstruct, modulo $2\pi$, the expansion coefficients of the projection into the subspace of the
 de Rham cohomology class of the magnetic potential in any basis of the subspace.
The set of circulations  and the projection of the de Rham cohomology class of the magnetic potential that
we can reconstruct depend on the relative position of the handlebodies and on the direction of the velocity
of the
incoming electrons. In Theorem \ref{th-7.12}, Corollary \ref{cor-7.13} and Remark \ref{rem-7.13} we give our
method for the reconstruction of the fluxes inside the obstacle $K$,  modulo $2\pi$. Since the scattering
operator is invariant under short-range gauge transformations that change the fluxes by multiples of
$2\pi$, the fluxes can only be reconstructed modulo $2\pi$. Again, the fluxes that we reconstruct  depend
on the relative position of the handlebodies and on the direction of the velocity of the
incoming electrons. In Example \ref{ex-7.14} we give  obstacles that consist of a finite number of tori and
manifolds diffeomorphic to balls, where from the high-velocity limit of the scattering operator  in only one
direction we reconstruct modulo $2\pi$ all the circulations in $\Lambda$ of the magnetic potential, its
de Rham cohomology
class modulo $2\pi$, and the flux modulo $2\pi$ of the magnetic field over the cross section of all the tori.

Finally, we discuss the fundamental experiments of Tonomura et al. \cite{to1, to2} in Section 8. We show that
our results give a rigorous proof that quantum mechanics predicts the interference patterns between electron
beams that go inside and outside the torus, that where observed in these remarkable experiments.

The paper is organized as follows. In Section 2 we give a precise definition  of the obstacle, $K$, and
we study in a detailed way the homology and the cohomology of $K$ and $\Lambda$. This allows us to construct
homology and cohomlogy basis that have a clear physical significance.  Using these results  we construct in
Section 3 classes of magnetic potentials characterized by the magnetic field in $\Lambda$ and by the fluxes
of the magnetic field in the cross sections of the components of $K$ that have holes. We construct classes
of magnetic potentials where the fluxes are fixed, and classes where the fluxes are only fixed modulo $2\pi$.
We study the gauge transformations between these magnetic potentials. In Section 4 we define the Hamiltonian
of our system. In Section 5 we study our  direct scattering problem. We prove the existence of the wave
operators and we define the scattering operator. We analyze how the wave and scattering operators change under
 the change of the magnetic potential when the fluxes are only fixed modulo $2\pi$. We also prove our
 high-velocity estimates. In Section 6 we give our method for the reconstruction of the magnetic field and
 the electric  potential in a region of $\Lambda$. In Section 7 we obtain our results in the Aharonov-Bohm
 effect and in
 Sections 8 we discuss the Tonomura et al. experiments \cite{to1,to2}. In Appendixes A and B we prove results
 in homology that we need.

For the Aharonov-Bohm effect in scattering in two dimensions see \cite{ni} and \cite{we1}. For inverse
scattering by magnetic fields in all space see \cite{ar1,ar2,ar3}, \cite{ju1,ju2,ju3}. For properties of the
scattering matrix for scattering by Aharonov-Bohm potentials in all space see \cite{rou}, \cite{ry1} and
\cite{ya1,ya2}. For the Ahanov-Bohm effect in inverse boundary-value problems see \cite{e1,e2,e3,e4},
\cite{kk} and \cite{kl}.

Finally, some words in our notations and definitions.
We use notions of homology and cohomology as defined, for example, in \cite{br}, \cite{gh}, \cite{h},
 \cite{dr} and \cite{w}. In particular, we consider  homology and cohomology groups on open sets of
$\ere^n, n=2,3$
 with coefficients in $\ZETA$ and in $\ere$.  As these singular homology and
cohomology groups are isomorphic to the $C^\infty$ homology and cohomology groups,
\cite{br} page 291,   we will identify them.
We also use differential forms, or just forms, in open sets of $\ere^3$ with regular boundary -or in their
closure- with the Euclidean metric, as defined, for example, in \cite{dr}, \cite{sch}, \cite{w}. For
such a set, $O$, we denote by $\Omega^k(O)$   the set of
all $k-$ forms in $ O$.

We use the standard identification between concepts of vector calculus and differential forms in three
dimensions in the interior of $O$, that we denote by $\stackrel{o}O$, \cite{sch}. Let $\{x^i\}_{i=1}^3$
be the Euclidean coordinates of $\ere^3$.

We identify vectors and $1-$ differential forms as
$$
  (A_1,A_2,A_3)  \Longleftrightarrow   \sum_{i=1}^3 A_j dx^j .
$$
We identify vectors and $2-$differential forms as
$$
(B_1,B_2,B_3) \Longleftrightarrow B_3 dx^1 \wedge dx^2- B_2 dx^1\wedge dx^3+ B_1 dx^2\wedge dx^3.
$$
We identify scalars and $3-$differential forms as
$$
f \Longleftrightarrow f dx^1\wedge dx^2\wedge dx^3.
$$

The exterior derivative, $d$, in $1-$ forms is equivalent to the curl of the associated vector,  and in
$2-$ forms is equivalent to the divergence of the associated vector. In particular, a $1-$ form, $A$, is closed
if $dA=0$, or equivalently, if the associated vector has curl zero, and  a $2-$ form, $B$, is  closed
if $dB=0$, or equivalently, if the associated vector has divergence zero. For $0-$ forms the exterior derivative
coincides with the gradient $\nabla$.

We will always assume that the coefficients of our forms are at least locally integrable in any coordinate
chart. Hence, they define
distributions or currents \cite{dr}. We say that a form belongs to some space if its coefficients in any
coordinate chart belong to that space. For example,
we say that a form is continuous if it has continuous coefficients or that is $L^p$ if its coefficients
are in $L^p$. In the case of a $2-$ form, $B$, we will say that $ B\in L^p\Omega^2(O)$,
or, equivalently,  that the associated vector $B \in L^p(\stackrel{o}O)$.
For forms defined in $O$ that are not differentiable in the classical sense the derivatives are taken in
distribution sense in $O$, if $O$ is open, or in $\stackrel{o}O$ if it is closed.

For any $ x\in \ere^3, x \neq 0$, we denote, $\hat{x}:= x/|x|$. By $B^{\ere^n}_r(x_0), n=2,3$ we denote
the open ball of center $x_0$ and radius $r$. By $\ese^2$ we denote the unit sphere in $\ere^3$.
For any set $O$ we denote by $F(x \in O)$ the operator of
multiplication by the characteristic function of $O$. The symbol $\cong$ means isomorphism, the symbol
$\simeq$ means homotopic equivalence and the symbol $\approx$ means homeomorphism.

We define the Fourier transform as a unitary operator on $L^2(\ere^3)$ as follows,
$$
\hat {\phi}(p):= F \phi(p):= \frac{1}{(2 \pi)^{3/2}} \int_{\ere^3} e^{-i p\cdot x} \phi (x)\, dx.
$$
We define functions of the operator $\mo:=-i \nabla$ by Fourier transform,

$$
f(\mo) \phi:= F^\ast f(p) F \phi, \, D(f(\mo)):= \{ \phi \in L^2(\ere^3): f(p) \,\hat{\phi}(p) \in L^2(\ere^3)
\},
$$
for every measurable function $f$.

\section{The Obstacle}\sss
\noindent{\bf 2.1 Handlebodies}

\noindent
 Let us designate  by $\ese^1$  the unit circle. We denote by $T:= \ese^1 \times
 \overline{ B_1^{\ere^2}(0)}$ the solid torus of dimension $3$. We orient $T$ assuming that  the inverse of
 the following function is a chart that  belongs to the orientation of $T$,

\beq
\mathcal U: (0,1) \times  B_1^{\ere^2}(0) \rightarrow T,
\,\, \mathcal U(t,x,y)= (e^{2\pi i t},x,y),
\label{1.1}
\ene

The  boundary sum of $T$ with itself is defined as follows. See  \cite{gs}, page 19.
Let $D_1\subseteq \partial T$ be a disc contained in a chart, $(U_1,\phi_1)$, belonging to  the orientation of
$T$ and
let $D_2 \subseteq \partial T$ be a disc contained in a chart, $(U_2,\phi_2)$, belonging to the opposite
orientation. We
define the boundary sum  $T\natural T$ as the disjoint union of $T$
with itself, identifying $D_1$ in the first torus with $D_2$ in the second torus by means of the charts,
in such a way that $T\natural T$ is an oriented differentiable manifold,
the inclusion $l_1:
T \hookrightarrow T\natural T$ in the first torus is an homeomorphism onto its image whose restriction to
 $ \stackrel{o}T$ is a diffeomorphism that preserves
orientation and the inclusion   $l_2: T \hookrightarrow T\natural T$ in the second torus is an homeomorphism onto its image whose
restriction to $ \stackrel{o}T$ is a diffeomorphism that inverts
orientation. We define the
 boundary sum of $k$ tori by induction. Suppose that
we already defined the boundary sum  $\natural (k-1) \,T:=T \natural T\cdots \natural T, \ k-1\, \hbox{\rm times}$
of $k-1$ tori. Let $l_j, j=1,2,\cdots,k-1$ be the inclusion of $T$ on the $j$th torus. As before,
Let $D_1\subseteq \partial T$ be a disc contained in a chart $(U_1,\phi_1)$ belonging to  the orientation
of $T$ if $k-1$ is odd, or belonging to the opposite orientation if $k-1$ is even. Moreover, we assume
that $l_{k-1}(U_1)$ does not intersect any of the  union charts in $\natural\, (k-1)T$. This is always possible
choosing the union charts small enough.
Let $D_2 \subseteq \partial T$ be a disc contained in a chart $(U_2,\phi_2)$ belonging to the opposite
orientation of $T$. Then, the boundary sum
$\natural k \,T:=T \natural T\cdots \natural T, \, k\, \hbox{\rm times}$ is obtained from   $
\natural T \cdots \natural T,\, k-1\, \hbox{\rm times}$ identifying $l_{k-1}(D_1)$ with $D_2$ by means of the
charts
$(l_{k-1}(U_1), \phi_1\circ l^{-1}_{k-1})$ and $(U_2,\phi_2)$   in such a way that
$\natural\, k T$ is an oriented differentiable manifold,
the inclusion $
\natural \, (k-1) T \hookrightarrow \natural\, k T$ in the first $k-1$ tori is an homeomorphism onto its image
whose restriction to
 the interior  is a diffeomorphism that preserves
orientation and the inclusion   $l_k: T \hookrightarrow \natural\, k T$ in the last torus is an homeomorphism onto its image whose
restriction to $ \stackrel{0} T$ is a diffeomorphism that inverts
orientation.
The structure of $\natural\, kT$  as oriented differentiable manifold  does not depend on the discs used to
join the
 tori \cite{gs}, page 19. We will say that any oriented differentiable manifold diffeomorphic  to
 $\natural k T$ is a handlebody with $k$ handles, where the diffeomorphism is oriented. We
 will denote by $\natural 0 T$ any oriented manifold that is diffeomorphic to the closed ball in $\ere^3$
  of center zero and radius one. Note that the
 inclusions $ l_j: T \hookrightarrow \natural k T$  onto the $j$th torus are homeomorphisms onto their images
 whose restriction to the interior are diffeomorphisms that preserve orientation if $j$ is odd and change
 orientation if $j$ is even.

\noindent {\bf 2.2 Homology  of  Handlebodies }

 We define the functions $\gamma_\pm : [0,1] \rightarrow T: \gamma_\pm(t)
=(e^{\pm 2\pi it},0,0)$ and

\beq
Z_j(t):= \left\{\begin{array}{c}l_j\circ\gamma_+(t)\,\,\, \hbox{\rm if}\, j \,  \hbox{\rm is odd},
\\\\ l_j\circ\gamma_-(t)
\,\,\, \hbox{\rm if}\,  j \, \hbox{\rm is even}.\end{array} \right.
\label{1.2}
\ene

For any $\xi \in \ese^1$ we define

\beq
B_\xi:= \left( \{\xi\}\times \overline{B_1(0)} \right)\subseteq T.
\label{1.3}
\ene
We orient $B_\xi$ by requiring that inverse of the inclusion $B_1(0) \hookrightarrow B_\xi$ belongs to the orientation
of $B_\xi$, i.e., the inverse of the inclusion is a chart.

The image of $Z_j$ in $\natural k T$ is a submanifold that we orient by means of the
curve $Z_j$. We assume that $l_j(B_\xi)$ does not intersect any of the union charts, what is always possible
if the union charts are small enough. We orient the submanifold $l_j(B_\xi)$ by the orientation of $B_\xi$. Let
$v_1 \in T_{l_j(\xi,0,0)} (Z_j([0,1])) \subseteq T_{l_j(\xi,0,0)}(\natural k T)$ be a tangent vector
in the orientation of $Z_j([0,1])$,
and let $v_2,v_3  \in T_{l_j(\xi,0,0)}(l_j
(B_\xi)) \subseteq T_{l_j(\xi,0,0)}(\natural k T)$  with $(v_2,v_3)$ in the positive orientation of $l_j(B_\xi)$.
Then, $(v_1,v_2,v_3)$ is positively oriented in the tangent space $T_{l_j(\xi,0,0)}(\natural k T)$. This means that
$Z_j([0,1])$ and $l_j(B_\xi)$ intersect in a positive way.

Let us denote by $H_1(\natural k T; \ere)$ the first group of singular homology of $\natural k T$ with
coefficients in $\ere$. See \cite{gh}, page 47. In  Appendix A  we give a proof, for the reader's convenience,
that $\left\{ [Z_j]_{H_1(\natural k T; \ere)}\right\}_{j=1}^k$ is a basis  of $H_1(\natural k T; \ere)$.

\noindent {\bf 2.3 Definition of the Obstacle}

\begin{assumption}
\label{ass-1}
We assume that the obstacle $K$ is a compact submanifold of $\ere^3$
of dimension three oriented with the orientation of $\ere^3$.
Moreover, $K= \cup_{j=1}^L K_j$ where $K_j,  1\leq j \leq L$ are the
connected components of $K$. We assume that the $K_j$ are
handlebodies.
\end{assumption}

By our assumption there exist numbers $m_j \in \mathbb N \cup {0}$ and oriented diffeomorphisms $F_j: \natural \, m_j
\, T \rightarrow  K_j, 1 \leq j\leq L$. We denote by ${\mathcal J}_j$ the inclusion $K_j \hookrightarrow K$. The
diffeomorphisms $F_j$ induce a diffeomorphism
$$
G: \bigvee_{j=1}^L\, \natural m_j T \rightarrow K,
$$
where the symbol $\bigvee$ means disjoint union. We denote,
$$
J:= \{ j \in \{1,2,\cdots, L\}: m_j >0  \}, \quad m:= \sum_{j=1}^L m_j,
$$
\beq \{\gamma_k\}_{k=1}^m:= \left\{{\mathcal J}_j\circ F_j\circ Z_i| j \in J, i
\in\{1,2,\cdots m_j\}  \right\}.
\label{1.4}
\ene
Choose a $\xi \in
\ese^1$  such that $l_i(B_\xi)$ does not intersect any chart of
union in $\natural\, m_j\, T, \forall j \in J, \forall i \in
\{1,2,\cdots,m_j\} $. This is always possible by choosing the charts
of union in a proper way. If $\gamma_k= {\mathcal J}_j\circ F_j\circ Z_i$ we
define  $B_k:= {\mathcal J}_j\circ F_j \left(l_i(B_\xi) \right)$. $B_k$ is a
manifold that we orient by means of the orientation of $B_\xi$. As
$F_j$ is a oriented diffeomorphism and $Z_i$ intersects $l_i(B_\xi)$
in a positive way, it follows that $\gamma_k$ intersects $B_k$ in a
positive way.

We define, $\mathcal W_\xi :[0,1]\rightarrow T: \mathcal W_\xi(t):=
(\xi, \cos t,\sin t)$ and
\beq
\tilde{\gamma}_k:= {\mathcal J}_j\circ F_j\circ l_i\circ \mathcal W_\xi.
\label{1.5}
\ene
Take  $\varepsilon >0$ such that $\{x|\,\hbox{\rm distance}(x,\partial K)< \varepsilon \}$ is diffeomorphic to
$\partial K \times (-\varepsilon, \varepsilon)$. This is possible by the  tubular neighborhood theorem.
See theorem 11.4, page 93 of \cite{br}. We define,
\beq
\hat{\gamma}_k(t):= \tilde{\gamma}_k(t)+\frac{\varepsilon}{2} N(\tilde{\gamma}_k(t)),
\label{1.6}
\ene
where $N(\tilde{\gamma}_k(t))$ is the exterior normal to $K$ at the point $\tilde{\gamma}_k(t)$.
Note that $\partial B_k= \tilde{\gamma}_k([0,1])$, the orientation on $ \tilde{\gamma}_k([0,1])$ induced by
$B_k$ is the orientation induced by the curve $\tilde{\gamma}_k$.

\noindent {\bf 2.4 The Homology of the Obstacle}
\begin{prop}\label{prop-2.0}
 $\{[\gamma_k]_{H_1(K;\ere)}\}_{k=1}^m$ is a basis
of $H_1(K;\ere)$.
\end{prop}

\noindent{\it Proof:}  As $G: \bigvee_{j=1}^L\, \natural m_j T \rightarrow K,$ is a diffeomorphism and since
$$
H_1\left(\bigvee_{j=1}^L\, \natural m_j T ;\ere\right)\cong
\oplus_{j=1}^L H_1\left( \natural m_j T ;\ere\right),
$$
by Proposition 9.5, page 47 of \cite{gh}, it follows from Proposition \ref{prop-a.3} of Appendix A
that $\{[\gamma_k]_{H_1(K;\ere)}\}_{k=1}^m$ is a basis
of $H_1(K;\ere)$.

\noindent{\bf 2.5 The Cohomology of the Obstacle}

\noindent  As $K$ is an ANR (absolute neighborhood
retract, page 225 and Theorem 26.17.4 of \cite{gh}) we have that

\beq
\check{H}^{ 1}(K;\ere)\cong H^1(K;\ere),
\label{1.7}
\ene
by Proposition 27.1, page 230 of \cite{gh} (see also  page 347, Theorem 7.15 of \cite{br}).

By Alexander's duality theorem (see Theorem 27.5, page 233 of \cite{gh})

\beq
\check{H}^{ 1}(K;\ere)\cong H_2(\ere^3,\ere^3\setminus K;\ere).
\label{1.8}
\ene
By Theorem 14.1, page 75 of \cite{gh} we have the following exact sequence.
$$
H_2(\ere^3;\ere)\rightarrow H_2(\ere^3,\ere^3\setminus K;\ere)\rightarrow H_1(\ere^3\setminus K;\ere)\rightarrow
H_1(\ere^3;K).
$$
As $\ere^3$ is homotopically equivalent to a point,  it follows from Theorem 11.3, page 59 and Example 9.4,
page 47 of \cite{gh} that $H_2(\ere^3;\ere)=0$ and $ H_1(\ere^3;\ere)=0$. Then,  we have that the exact
sequence

$$
0\rightarrow H_2(\ere^3,\ere^3\setminus K;\ere)\rightarrow H_1(\ere^3\setminus K;\ere)\rightarrow 0,
$$
and then,
\beq H_2(\ere^3,\ere^3\setminus K;\ere)\cong
H_1(\ere^3\setminus K;\ere).
\label{1.9}
\ene

By
(\ref{1.7},\ref{1.8}, \ref{1.9}), \beq H^1(K;\ere)\cong
H_1(\ere^3\setminus K;\ere). \label{1.10} \ene By the theorem of
universal coefficients, page 198 of \cite{h}, $H^1(K; \ere) \cong
\hbox{\rm Hom}_{\ere} \left(H_1(K;\ere),\ere\right)$. Then,  it follows
that,
\beq \hbox{\rm dim}\, H_1(K;\ere)= \hbox{\rm dim}\,
H_1(\ere^3\setminus K;\ere)=m.
\label{1.11} \ene

We denote,

$$
\Lambda := \ere^3 \setminus K.
$$

We will prove in Corollary \ref{cor-2.3} that $\{[\hat{\gamma}_k]_{H_1(\Lambda;\ere)}\}_{k=1}^m$ is a basis of
$H_1(\Lambda;\ere)$.

\noindent{\bf  2.6 de Rham Cohomology of $\Lambda$ }

\noindent Let us define,

\beq G^{(j)}(x):= \curl \frac{1}{4\pi}\, \int_{\gamma_j}
\frac{1}{|x-y|} d\vec{\gamma _j}:= \curl \frac{1}{4\pi}\,
\int \frac{1}{|x-\gamma_j(t)|}\, \dot{{\gamma_j}}(t)\, dt.
\label{1.12}
\ene
Then, $\curl G^{(j)}(x)=0, \, x \in \ere^3 \setminus \gamma_j$ and

\beq
\int_{\hat{\gamma}_k} \,G^{(j)} = \delta_{k,j}, j,k= 1,2,\cdots,m.
\label{1.13}
\ene
Equation (\ref{1.12}) is the law of Biot-Savart that gives the magnetic field created by a  current
in $\gamma_j$ and (\ref{1.13}) is Ampere's law. For a proof see Satz 1.4, page 33, of \cite{ma}.

\begin{prop}\label{prop-2.2}
$\left\{ \left[G^{(j)}\right]_{H^1_{\hbox{\rm de R}}}(\Lambda)\right\}_{j=1}^m$ is a basis of
$H^1_{\hbox{\rm de  R}}(\Lambda)$.
\end{prop}

\noindent{\it Proof:} We first prove that they are linearly independent. Suppose that
$\sum \alpha_j G^{(j)}=0$ then, $\sum \alpha_j G^{(j)} = d\lambda$ for some $0-$ form $\lambda$.
Hence,
$$
\int_{\hat{\gamma}_k} \,\sum \alpha_j G^{(j)}= \alpha_k=0.
$$

By de Rham's Theorem (Theorem 4.17, page 154 of \cite{w}) the dual
space to $H^1_{\hbox{\rm de  R}}(\Lambda)$ is isomorphic  to $H_1(\Lambda;\ere)$ and
viceversa. The isomorphisms are given by

$$
\langle [\alpha]_{H_1(\Lambda;\ere)},
[A]_{H^1_{deR}(\Lambda)}\rangle:= \int_{\alpha} A.
$$
Then, by (\ref{1.11})
$$
\hbox{\rm dim} H^1_{\hbox{\rm de  R}}(\Lambda)= \hbox{\rm dim}H_1(\Lambda;\ere)=m,
$$
and this proves the Proposition.

\begin{corollary}\label{cor-2.3}
$\left\{ \left[\hat{\gamma}_{r}\right]_{H_1(\Lambda;\ere)}\right\}_{r=1}^m$ is a basis of
$H_1(\Lambda;\ere)$.
\end{corollary}

\noindent{\it Proof:} By (\ref{1.13})
$\left\{ \left[\hat{\gamma}_{r}\right]_{H_1(\Lambda;\ere)}\right\}_{r=1}^m$ is the dual basis -in the sense
of de Rham's Theorem- to the \linebreak basis  $\left\{ \left[G^{(r)}\right]_{H^1_{\hbox{\rm de  R}}}
(\Lambda)\right\}_{r=1}^m$
 of $H^1_{\hbox{\rm de  R}}(\Lambda)$.

\bull

\begin{prop} \label{prop-2.4}
Let $A$ be a closed $1-$ form with continuous coefficients defined in $\Lambda$ and such that
$$
\int_{\hat{\gamma}_{r}}\, A =0, r=1,2,\cdots,m.
$$
 Then, there is a continuously differentiable  $0- \ form $,   $\lambda$, such that $A=d\lambda$. Moreover, we can take
 $\lambda(x):=\int_{C(x_0,x)}A$ where $x_0 $ is any fixed point in $\Lambda$ and $C(x_0,x)$ is any
 curve from $x_0$ to $x$.
\end{prop}

\noindent {\it Proof:} By Theorem 12, page 68, of \cite{dr} there is a regularization $R(\epsilon)$
and an operator $\Gamma (\epsilon)$ such that if $\alpha$ is a continuous $k- $ form on $\Lambda$, $R\alpha$
is a $C^\infty \ k- $ form on $\Lambda$ and $\Gamma \alpha$ is a continuous $(k-1)- $ form on $\Lambda$. Moreover,
$\lim_{\epsilon \rightarrow 0} R\alpha=\alpha$ uniformly on compact sets in $\Lambda$. Furthermore,

\beq
R\alpha-\alpha= b \Gamma \alpha+\Gamma b \alpha,
\label{1.14}
\ene
where $ b\alpha := (-1)^{\hbox{\rm grade}\,(\alpha)-1}\, d$.
Multiplying (\ref{1.14}) on the left by $b$ and applying it to $b\alpha$ we prove that $Rb=bR$.
As $A$ is closed, it follows from (\ref{1.14}) that $RA-A= b\Gamma A$. In particular, this implies that
$b\Gamma A$ is continuous. Let $C$ be a closed curve. Then, by Stokes theorem,

$$
\int_C \, b\Gamma A =\lim_{\epsilon \rightarrow 0} \int_C\, Rb\Gamma A = \lim_{\epsilon \rightarrow 0} \int_C\,
bR\Gamma A = 0,
$$
and then,
$$
\int_C\, RA = \int_C A,
$$
and in particular,
$$
\int_{\hat{\gamma}_{r}}\, RA = \int_{\hat{\gamma}_{r}}\, A=0, r=1,2,\cdots,m.
$$
As $RA$ is $C^\infty$ and closed, and since $\left\{ [\hat{\gamma}_{r}]_{H_1(\Lambda; \ere)}\right\}_{r=1}^m$ is a basis of
$H_1(\Lambda ;\ere)$ it follows from de Rham's Theorem (Theorem 4.17, page 154, \cite{w}) that there is an
infinitely differentiable $ 0- $form $  \alpha $ such that $ RA=b\alpha$. But then, using Stokes theorem again,

$$
\int_C \,RA = \int_C \, b\alpha=0,
$$
and we obtain that,
$$
\int_C\, A=0,
$$
for any closed curve $C$ and we can define $\lambda:= \int_{C(x_0,x)} A$. Clearly, $A=d\lambda$.

\bull

Recall that $\{K_j  \}_{j=1}^L$ is the set of connected components of $K$. For each $j\in\{1,2,\cdots,L\}$
we choose a $x_j$ in the interior of $K_j$. We define the vector,

\beq
D_j:=\, - \hbox{\rm grad} \frac{1}{4\pi}\frac{1}{|x-x_j|},\, x \in \ere^3\setminus \{x_j\}.
\label{1.15}
\ene
and according to our convention, we denote by the same symbol the associated $2-$ form.
Note that $\hbox{\rm div}\, D_j(x)=d D_j = -\Delta \frac{1}{4\pi} \frac{1}{|x-x_j|}= 0, x\neq x_j, j=1,2,\cdots,L$ and that,
\beq
\left| D_j(x)\right| \leq C (1+|x|)^{-2}, \, x \in \Lambda.
\label{1.15b}
\ene

For any $ r >0$ such  that $ K \subset  B_r^{\ere^3}(0)$ we denote,
$$
\Lambda_r := \Lambda \cap B_r^{\ere^3}(0),\, \hbox{\rm and}\,  \,\,\, \Lambda_\infty:= \Lambda.
$$

\begin{prop}\label{prop-1.6}
$\left\{[D_j]_{H^2_{\hbox{ \rm de  R}}(\Lambda_r)}\right\}_{j=1}^L$ is a basis of $H^2_{\hbox{\rm de  R}}
(\Lambda_r)$
for $ r \leq \infty$.\end{prop}

\noindent{\it Proof:}  Let us first consider the case $r=\infty$.
As in the proof of (\ref{1.11}) we prove that
$$
\hbox{\rm dim}\,H_0(K;\ere)=\, \hbox{\rm dim}\, H_2(\Lambda;\ere).
$$
But by  Proposition 9.6, page 48 of \cite{gh}
$$
H_0(K;\ere)\cong \oplus_{j=1}^L \ere.
$$
Moreover, by de Rham's Theorem (Theorem 4.17, page 154 of \cite{w})
\beq
H_{\hbox{\rm de  R}}^2(\Lambda)\cong H_2(\Lambda;\ere).
\label{1.16}
\ene
Then,
\beq
\hbox{\rm dim}\, H^2_{\hbox{\rm de  R}}(\Lambda)=\, \hbox{\rm dim}\, H_2(\Lambda;\ere)=L.
\label{1.17}
\ene
Let us now consider $ r < \infty$.
We define, $f : \Lambda \rightarrow \Lambda_r$

$$
f(x):=\left\{\begin{array}{c} r_1 \frac{x}{|x|},\,\hbox{\rm  if}\, |x|\geq \ r_1, \\\\
x,\, \hbox{\rm if}\, |x| \leq r_1,
\end{array}\right.
$$

and $ H(x,t): \left( \Lambda \times [0,1]\right) \rightarrow \Lambda  $
$$
H(x,t):= \left\{ \begin{array}{c}    x+ t(r_1 \frac{x}{|x|}-x), \, {\rm if}\, |x| \geq r_1, \\ \\
x, \, {\rm if} \, |x| \leq r_1,
\end{array}\right.
$$
where $ r_1 < r$ and  $ K \subset  B_{r_1}^{\ere^3}(0)$.
Let $l$ be the inclusion $ l: \Lambda_r \hookrightarrow \Lambda$. Then as $ l\circ f (x)= l\circ H(x,1)= H(x,1)$
and $H(x,0)=I(x)$, we have that $l\circ f$ is homotopic to the identity.
Let us denote by $\tilde{H}(x,t)$ the restriction of $H(x,t)$ to $\Lambda_r$.
Then, $f\circ l(x)= \tilde{H}(l(x),1)= \tilde{H}(x,1)$, and as $\tilde{H}(x,0)=I(x)$ we also have that
$f\circ l$ is homotopic to the identity. Hence, by Theorem 11.3, page 59 \cite{gh} the inclusion $l$ induces an
isomorphism in homology. In particular, $H_2(\Lambda_r; \ere)\cong H_2(\Lambda;\ere)$ and then,

\beq
\hbox{\rm dim}\, H_2(\Lambda_r;\ere)=\, \hbox{\rm dim}\, H_2(\Lambda;\ere)=L.
\label{1.20}
\ene

It follows from Stoke's theorem and as $-\Delta \frac{1}{4\pi}\frac{1}{|x-x_j|}=\hbox{\rm div}\, D_j(x)=\delta(x-x_j)$ that,

\beq
\int_{\partial K_i}\, D_j = \int_{\partial B^{\mathbb{R}^{3}}_{\rho}(x_i)} D_j= \delta_{i,j},
\label{1.18}
\ene
for $ \rho $  small enough and $ i,j = 1,2, \cdots, L $.
This easily implies that the set $\left\{[D_j]_{H^2_{\hbox{\rm de R}}(\Lambda_r)}\right\}_{j=1}^L$ is linearly
independent.

\begin{lemma}\label{lemm-2.7}
Suppose that $ \{[S_j]_{H_2(\Lambda_r;\ere)
}\}_{j=1}^L,$ is a basis of $H_2
(\Lambda_r;\ere)$ for $ r \leq \infty$. Let $D$ be a closed $2-$ form with continuous coefficients
in $\overline{\Lambda_r}$. Then,

\beq
\int_{\partial K_j}\, D=0, \forall j \in \{1,2,\cdots,L\} \Longleftrightarrow \int_{S_j}\, D=0,  \forall j \in \{1,2,\cdots,L\}.
\label{1.19}
\ene
\end{lemma}
\noindent{\it Proof:} Denote $ K_{j,\varepsilon}:= \{x\in \ere^3: \,\hbox{\rm dist}(x,K_j) < \varepsilon \}$
where $\varepsilon$ is so small that the tubular neighborhood theorem applies and let $R$ be
the regularization operator. Suppose that the left side of (\ref{1.19}) holds. Then, as
$D$ is closed we prove using the Stokes theorem that,

$$
\int_{\partial K_{j,\varepsilon}}\, R D =0, \, \forall j \in \{1,2,\cdots ,L\}.
$$
As $RD$ is $C^\infty$ and closed, since $bR=Rb$,  there are coefficients
$\lambda_j, j=1,2,\cdots, L$ and a $1-$ form $ \alpha$ such that,

$$
R D= \sum_{j=1}^L \lambda_j D_j+ d\alpha.
$$
Then, it follows from (\ref{1.18}) (with $ K_{j,\varepsilon} $ instead of $ K_j $ ) and Stoke's theorem that

$$
0= \int_{\partial K_{j, \varepsilon}}\, RD =\lambda_j,
$$
and we obtain that
$$
RD= d\alpha.
$$
Furthermore, using the regularization operator and Stoke's theorem we prove that,
$$
\int_{S_j} D = \int_{S_j}RD= \int_{S_j}d\alpha =0, j \in\{1,2,\cdots,L\}.
$$

Assume now that $\int_{S_j} D=0, j \in\{1,2,\cdots,L\}$. We prove as above that,
$\int_{S_j} RD=0, j \in\{1,2,\cdots,L\}$, and by de Rham's Theorem (Theorem 4.17, page 154 of \cite{w})
there is a $1-$ form $ \alpha$ such that
$$
R D = d\alpha.
$$
Hence,
$$
\int_{\partial K_{j,\varepsilon}} D =\int_{\partial K_{j,\varepsilon}}RD =\int_{\partial K_{j,\varepsilon}} d\alpha=0,
$$
and then,
$$
\int_{\partial K_{j}} D = \lim_{\varepsilon \rightarrow 0}\int_{\partial K_{j,\varepsilon}} D=0, j \in \{1,2,\cdots,L\}.
$$

\section{Magnetic Field and Magnetic Potentials}
In this section we introduce the class of magnetic fields that we consider and we construct a class of
associated magnetic potentials with nice behaviour at infinity that will allows us to solve our scattering
problems.

\begin{definition}
We say that a form $B $ in $\overline{\Lambda}$ is continuous in a neighborhood of $\partial K$ if there is a
$\varepsilon >0$ such that the coefficients of $B$ are  continuous
in $\overline{\Lambda}\cap K_\varepsilon$ where
$K_\varepsilon:=\{x\in \ere^3: \,\hbox{\rm dist}(x, K) < \varepsilon\}$.
\end{definition}

Below we assume that the magnetic field, $B$, is a $2-$ form that is continuous in a neighborhood of $\partial K$ and satisfies

\beq
\int_{\partial K_{j}} B=0,j \in \{1,2,\cdots,L\}.
\label{3.1}
\ene
This condition means that the total contribution of magnetic monopoles inside each component $ K_j $ of  the obstacle is $ 0 $. In a formal way we can
use Stokes theorem to conclude that
$$
\int_{\partial K_j} B =0 \Longleftrightarrow \int_{K_j} \, \hbox{\rm div}\, B=0,j \in \{1,2,\cdots,L\}.
$$
As $\hbox{\rm div}\, B$ is the density of magnetic charge, $\int_{\partial K_j} B$ is the total magnetic charge
inside $K_j$, and our condition (\ref{3.1}) means that the total magnetic charge inside $K_j$ is zero, this condition in fulfilled if there is no magnetic monopole inside $K_j, j \in \{1,2,\cdots,L\}$.

\begin{theorem} \label{th-3.2}
Let $B$ be a $2-$ form in $L^p_{\hbox{\rm loc}}\Omega^2(\overline{\Lambda} ),\, p \geq 2$ that is continuous in a
neighborhood of $\partial K$ and satisfies (\ref{3.1}). Suppose that the restriction of B to $ \Lambda $ is closed ($ d B|_{\Lambda} = 0 $) as a distribution (or current \cite{dr})  . Then, $B$ has an extension to a closed $2-$ form
$\overline{B} \in L^p_{\hbox{\rm loc}}\Omega^2(\ere^n)$ such that, $\overline{B}|_{\overline{\Lambda}}=B$.
\end{theorem}

\noindent{\it Proof:}
Let us denote $ M:= \overline{\Lambda_r}, \, r < \infty$.
$ M$ is a compact manifold. We denote by $B_{ M}$ the restriction of $B$ to  $M$.
As $d B|_{\Lambda}=0$, it follows from Green's formula (Proposition 2.12, page 60, \cite{sch})
that
\beq
 \langle\langle B_M, \delta \eta\rangle\rangle=0,\,  \forall \eta \in C^\infty_0\Omega^3
(\stackrel{\circ}M).
\label{3.2}
\ene
We denote (Definition 2.4.1, page 80 \cite{sch})

$$
C^k(M):= \left\{\delta \eta|\eta  \in H^{1}\Omega^{k+1}_N(M)\right\},
$$
and (Definition 2.2.1, page 67 \cite{sch})
$$
H^1\Omega^k_N(M):= \left\{ \eta\in H^1\Omega^k(M)|{\mathbf n }\eta=0 \right\}.
$$
Let us recall (page 27 \cite{sch}) that given $ \eta  \in \Omega^3(M)$ and tangent vectors $ v_{i} \in T_{x}(M) $,  $x \in \partial M, \, i \in \{ 1, 2, 3 \} $,
$$
{\mathbf t} \eta(v_{1}, v_{2}, v_{3}) = \eta(v^{\parallel}_{1}, v^{\parallel}_{2}, v^{\parallel}_{3}),
$$
where $ v_{i}^{\parallel} $ is the projection of $ v_{i} $ into $ T_{x}(\partial M) $.
As $ \eta $ is a multi-linear function and $ \left\{  v_{1}^{\parallel} ,v_{2}^{\parallel},
v_{3}^{\parallel}  \right\} $
are linearly dependent,

$$
{\mathbf t} \eta = 0.
$$
By the definition in page 27 \cite{sch},
$$
{ \mathbf n} \eta := \eta - {\mathbf t} \eta = \eta.
$$
It follows that
$$
{\mathbf n} \eta = \eta, \,  \eta \in H^{1}\Omega^{3}(M).
$$
Let $ \eta \in H^{1}\Omega^{3}_N(M)  $, then there exists $ f \in W^{1,2}(M) $ such that
$$
\eta|_{\stackrel{o}M} = f|_{\stackrel{o}M}dx^{1}\wedge dx^{2} \wedge dx^{3}.
$$
As ${\mathbf n} \eta = \eta = 0 $, it follows that $ f|_{\partial M} = 0  $ in trace sense. Hence
(Theorem 4.7.1,
page 330, \cite{tr}), $f$ can be
approximated
in the $W^{1,2}(M)$ norm by functions in $C^\infty_0\left(\stackrel{\circ}M\right)$, and then $ \eta  $ can be
approximated in the $ H^{1}\Omega^{3}(M) $ norm by forms in $ \Omega^{3}(\stackrel{0}M) $ with compact support.
Whence,  it follows from
(\ref{3.2}) that
\beq
\langle\langle B_M, \delta \eta\rangle\rangle=0, \forall \eta \in C^2(M).
\label{3.3}
\ene
By Corollary 2.4.9, page 87  \cite{sch}
\beq
B_M= d\alpha+ \delta \beta+d\epsilon+\gamma \in  \mathcal E^2(M)\oplus  C^2(M)
\oplus L^2 \mathcal H  ^2_{\hbox{\rm ext}}(M) \oplus \mathcal H^2_N(M),
\label{3.4}
\ene
where (Definition 2.4.1, page 80 \cite{sch})
$$
\mathcal E^k(M):= \left\{ d\alpha | \alpha \in H^1\Omega^{k-1}_D(M)\right\}\,
$$
and (Definition 2.2.1, page 67 \cite{sch})
$$
H^1\Omega^k_D(M):= \left\{ \eta\in H^1\Omega^k(M)|{\mathbf t }\eta=0 \right\}.
$$
Furthermore (page 86 \cite{sch}),
$$
\mathcal H^k_{\rm ext}(M):= \left\{ \eta \in \mathcal H^k(M)| \eta= d \epsilon \right\},
$$
and (Definition 2.2.1, page 67 \cite{sch})

$$
 \mathcal H^k(M):= \left\{ \eta\in H^1\Omega^k(M)|d\eta= 0, \, \delta \eta =0 \right\}.
$$
are the harmonic fields, and

$$
\mathcal H^k_N (M):=   \mathcal H^k(M) \cap H^1\Omega^k_N(M).
$$

Note that Theorem 2.2.7, page 72  \cite{sch} implies that $\mathcal H^2_N(M)$ consists of $C^\infty$ forms.
  Furthermore by Lemma 2.4.11 page 90 \cite{sch}  we can choose $\alpha \in
W^{1,p}\Omega^1_D(M)$, and by Theorem 2.4.8, page 86 and Theorems 2.2.6 and 2.2.7, page 72 \cite{sch}
$ \epsilon \in W^{1,p}\Omega^{1}_N(M)$.
Moreover, the decomposition (\ref{3.4})
is orthogonal in $L^2(M)$, and then by (\ref{3.3})  $\delta \beta=0$.

Let $R$ be the regularization operator in $\Lambda_r= \stackrel{\circ}M$. Then, as in the proof
of Lemma \ref{lemm-2.7} we prove that
$$
\int_{\partial K_{j,\varepsilon}} R B=0.
$$

Hence,
$$
0=\int_{\partial K_{j,\varepsilon}} R B= \int_{\partial K_{j,\varepsilon}} d(R\alpha +R\epsilon)+
\int_{\partial K_{j,\varepsilon}}
R \gamma = \int_{\partial K_{j,\varepsilon}}
R \gamma.
$$
Then,
$
\int_{\partial K_{j,\varepsilon}}R \gamma=0, \, j \in \{1,2,\cdots, L\}
$
and when the parameter of the regularization tends to zero we obtain
$
\int_{\partial K_{j,\varepsilon}} \gamma=0, \, j \in \{1,2,\cdots, L\}
$.

As  $\gamma$ is harmonic it is closed and it follows from Stokes theorem that

$$
\int_{\partial K_{j}}\gamma=0, j \in \{1,2,\cdots, L\}.
$$
Then, by Lemma \ref{lemm-2.7} $\int_{ S_{j}}\gamma=0, j \in \{1,2,\cdots, L\}$. By de Rham's Theorem
$ \gamma =d \lambda, \lambda \in \Omega^1(\stackrel{\circ}M)$. Denote $M_\varepsilon:=\{x \in M :\,
\hbox{\rm dist}\,(x,\partial M)\geq\varepsilon\}$. Let $\gamma_\varepsilon$ be the restriction of $\gamma$ to
$M_\varepsilon$. Then $\gamma_\varepsilon$ is exact and by
Lemma 3.2.1, page 119 \cite{sch}, and its proof, $\gamma_\varepsilon= d\omega_\varepsilon$ with $\omega_\varepsilon \in H^1\Omega^1(
M_\varepsilon)$ and
$$
\|\omega_\varepsilon\|_{ H^1\Omega^1(
M_\varepsilon) }\leq C \|\gamma_\varepsilon\|_{L^2\Omega^2(M_\varepsilon)}\leq C \|\gamma
\|_{L^2\Omega^2(M)},
$$
where the constant $C$ can be taken independent of $\varepsilon$ for $ 0 < \varepsilon < \varepsilon_0$ for
$\varepsilon_0$ small enough. Let us denote by $\Lambda^k(M), \Lambda^k(M_\varepsilon)$, respectively,
the exterior $k-$ form bundle of $M, M_\varepsilon$ (see Definition 1.3.8 in page 39 of \cite{sch}). For
 any vector bundle, $\mathbb F$, over a manifold $N$ we denote by $\Gamma (\mathbb F)$ the space of all
 smooth sections of $\mathbb F$ (see Definition 1.1.9, page 17 of \cite{sch}) .
Note that the norm, $C_1$, of the trace operator (Theorem 1.3.7, page
38 \cite{sch})
from $ H^1(\Omega^k(M_\varepsilon))$ into $L^2\Gamma\left(\Lambda^k( M_\varepsilon)|_{\partial M_\varepsilon}
\right)$
can be taken independent of
$\varepsilon$ for  $ 0 < \varepsilon < \varepsilon_0$. By Green's formula and as $\delta \gamma_\varepsilon=0$,
\beq
\langle\langle \gamma_\varepsilon,\gamma_\varepsilon\rangle\rangle=
\langle\langle d\omega_\varepsilon,\gamma_\varepsilon\rangle\rangle=
\int_{\partial M_\varepsilon} {\mathbf t} \omega_\varepsilon \wedge * {\mathbf n}\gamma_\varepsilon.
\label{3.5}
\ene
But as
$$
\| {\mathbf t} \omega_\varepsilon \|_{L^2\Gamma\left(\Lambda^1( M_\varepsilon)|_{\partial M_\varepsilon}
\right)}\leq C_1
\|\omega_\varepsilon\|_{ H^1\Omega^1(
M_\varepsilon) }\leq C_1 C \|\gamma_\varepsilon\|_{L^2\Omega^2(M_\varepsilon)}\leq C_1 C \|\gamma
\|_{L^2 \Omega^2(M)},
$$
and
$$
\lim_{\epsilon \rightarrow 0}\| {\mathbf n}\gamma_\varepsilon  \|_
{L^2\Gamma\left(\Lambda^2( M_\varepsilon)|_{\partial M_\varepsilon}
\right)}=0,
$$
it follows from (\ref{3.5}) and Schwarz inequality that

$$
\|\gamma\|^2_{L^2\Omega^2(M)}= \lim_{\varepsilon \rightarrow 0}\|\gamma_\varepsilon\|^2_{L^2\Omega^2(M)}=0.
$$
Then $ \gamma=0$ and we have  that

\beq
B_M=d A_M
\label{3.6}
\ene
where $A_M:= \alpha+\epsilon \in W^{1,p}\Omega^1(\stackrel{o}M)$. It follows from Theorem 4.2.2, page 311 \cite{tr} that
there is
$ \overline{A_M}\in W^{1,p}\Omega^1( \overline{B^{\ere^3}_r(0)})$ such that $ \overline{A_M}|_M= A_M$.
We define
\beq
\overline{B}(x)=\left\{\begin{array}{c} d\overline{A}_M(x),\, \hbox{\rm if}\, x \in B^{\ere^3}_r(0),\\\\
B(x),\,\hbox{\rm if}\  x \in \ere^3\setminus B^{\ere^3}_r(0).
\end{array}\right.
\label{3.7}
\ene
Hence, $\overline{B}$ is the required extension.

\bull

Recall that the functions $\{ \hat{\gamma}_j\}_{j=1}^m$ where defined in (\ref{1.6}). We introduce now a function
that gives the magnetic flux across  surfaces that have  $\{\hat{\gamma}_j\}_{j=1}^m$ as their boundaries.

\begin{definition} \label{def-3.4}
The flux, $\Phi$ is a function $\Phi: \{\hat{\gamma}_j\}_{j=1}^m \rightarrow \ere$.
\end{definition}
We now define a class of magnetic potentials with a given flux.

\begin{definition} \label{def-3.5}
Let $B\in L^p \Omega^2(\overline{\Lambda}), p>3,$ be a closed $2-$ form that is continuous in a neighborhood of $\partial K$
 where $K$ is as in  Assumption \ref{ass-1}. Assume, furthermore, that  (\ref{3.1}) holds. We denote by
 $\mathcal A_\Phi(B)$ the set of all continuous $1-$ forms in $ \overline{\Lambda}$ that satisfy.
\begin{enumerate}
\item
\beq
|A(x)| \leq C \frac{1}{1+|x|}, \,\, a(r):=\hbox{\rm max}_{x \in \Lambda, |x| \geq r}\,
\{ |A(x)\cdot \hat{x}| \} \in L^1(0,\infty).
\label{3.7b}
\ene
\item
\beq
\int_{\hat{\gamma_j}}\, A= \Phi (\hat{\gamma_j}), \, j \in \{1,2,\cdots, m \}.
\label{3.7c}
\ene
\item
\beq
d A|_{\Lambda}=B|_{\Lambda}.
\label{3.7d}
\ene
\end{enumerate}
\end{definition}

The definition of  the flux $\Phi$ depends on the particular choice of the curves $\{\hat{\gamma}_j\}_{j=1}^m$.
 However, the class $\mathcal{A}_\Phi(B)$ is independent of this particular choice as we prove below.

Recall that by Corollary \ref{cor-2.3} $\beta:=\{[\hat{\gamma}_j]_{H_1(\Lambda;\ere)}\}_{j=1}^m$ is a basis of $H_1(\Lambda;\ere)$. Let
$\beta':=\{[C_j]_{H_1(\Lambda;\ere)}\}_{j=1}^m$ be another  basis of $H_1(\Lambda;\ere)$. We define
$\Phi_{\beta'} : \{C_j\}_{j=1}^m \rightarrow \ere$ as follows. As $\beta$ is a basis of $H_1(\Lambda;\ere)$
there are real numbers $b^i_j$ and chains $ \sigma_j$ such that

\beq
C_j = \sum_{i=1}^m b^i_j \hat{\gamma}_i+\partial \sigma_j.
\label{3.8}
\ene
We define,
\beq
\Phi_{\beta'}(C_j):= \sum_{i=1}^m b^i_j \Phi(\hat{\gamma}_i)+\int_{\sigma_j} B.
\label{3.9}
\ene

We denote by $\mathcal{A}_{\Phi_{\beta '}}(B)$ the set of continuous $1-$ forms
$A$ in $\overline{\Lambda}$ that satisfy
$1$ and $3$ of Definition \ref{def-3.5} and moreover,
$$
\int_{C_j} A =\Phi_{\beta'}(C_j), \, j=1,2,\cdots,m.
$$

\begin{prop}\label{prop-3.5}

 $\mathcal A_{\Phi_{\beta '}}(B)= \mathcal A_\Phi(B)$.
\end{prop}

\noindent {\it Proof:} Let $A \in \mathcal A_\Phi(B)$. Then, by (\ref{3.8})
$$
\int_{C_j} A = \sum_{i=1}^m b^i_j \int_{\hat{\gamma}_i} A+ \int_{\sigma_j} d A = \Phi_{\beta'}(C_j),
j=1,2,\cdots,m,
$$
and it follows that $ A \in \mathcal A_{\Phi_{\beta'}}(B)$.

Suppose now that $ A \in \mathcal A_{\Phi_{\beta'}}(B)$. As $\beta$ and $\beta'$ are basis, the numbers
$b^i_j, i,j= 1,2,\cdots,m$ determine an invertible matrix. We denote by $ \tilde{b}^j_i$ the entries of the
inverse matrix. Hence,
$$
\hat{\gamma}_i = \sum_{j,s=1}^m \tilde{b}^j_i \, b^s_j \hat{\gamma}_s =
\sum_{j=1}^m \tilde{b}^j_i(C_j-\partial \sigma_j),
$$
and then by (\ref{3.9}),
$$
\int_{\hat{\gamma}_i} A = \sum_{j=1}^m \tilde{b}^j_i \left( \Phi_{\beta'}(C_j)-\int_{\sigma_j} B \right)=
\Phi(\hat{\gamma}_i).
$$
This implies that $ A \in  \mathcal A_\Phi(B)$.

\bull

By Stoke's theorem the circulation $\int_{\hat{\gamma}_{j}} A$ of a potential  $ A \in \pot$ represents the flux
 of the magnetic field $B$ in any surface whose boundary is $ \hat{\gamma}_j, j= 1,2,\cdots,m$. As the magnetic
field is {\it a priori} known outside the obstacle, it is natural to specify the magnetic potentials fixing
fluxes of the magnetic field in surfaces inside the obstacle. This is accomplished fixing the circulations
$\int_{\tilde{\gamma}_j} A$ instead of the circulations $\int_{\hat{\gamma}_j} A$, as we prove below.
Recall that $\tilde{\gamma}_j$ is defined in (\ref{1.5}). With $\varepsilon$ as in (\ref{1.6}) we define,
$$
S_j:= \left\{ \tilde{\gamma}_j(t)+ s \frac{\varepsilon}{2} N(\tilde{\gamma}_j(t))| t,s \in[0,1] \right\}.
$$
We give $S_j$ the structure of an oriented surface with boundary $\hat{\gamma}_j-\tilde{\gamma}_j$.
By Stoke's theorem and regularizing we prove that,
$$
\int_{\tilde{\gamma}_j} A = \int_{\hat{\gamma}_j} A - \int_{S_j} B.
$$
We define the fluxes $ \tilde{\Phi}: \{\tilde{\gamma}_j\}_{j=1}^m \rightarrow \ere$  accordingly,
$$
\tilde{\Phi}(\tilde{\gamma}_j)= \Phi(\hat{\gamma}_j)-\int_{S_j}B.
$$
We denote by $ \tilde{\mathcal A}_{\tilde{\Phi}}(B)$ the set of continuous $1-$ forms, $A$, in
$\overline{\Lambda}$ that satisfy
$1$ and $3$ of Definition \ref{def-3.5} and moreover,
$$
\int_{\tilde{\gamma}_j} A =\tilde{\Phi}(\tilde{\gamma}_j), j=1,2,\cdots,m.
$$

\begin{prop} \label{prop-3.6}
$\mathcal A_{\tilde{\Phi}}(B)= \pot$.
\end{prop}

\noindent{\it Proof:}
Let $ A \in \pot$. By Stoke's theorem and regularizing,
$$
\int_{\tilde{\gamma}_j} A = \int_{\hat{\gamma}_j} A- \int_{S_j}B= \tilde{\Phi}_j.
$$
Then, $ A \in \mathcal A_{\tilde{\Phi}}(B)$. We prove in the same way that $  A \in
\mathcal A_{\tilde{\Phi}}(B)  \Rightarrow  A \in \pot  $.

\bull

Note that for $1-$ forms $A= \sum_{i=1}^3 A_i d x^i$,  $\delta A= -\sum_{i=1}^3 \frac{\partial}{\partial x_i} A_i= -\hbox{\rm div}\, A$ \cite{sch}. We use the definition
of divergence of a vector field, $A$,  as it is usual in vector calculus. The definition given in \cite{sch}
differs from our's in a $-$ sign.

\begin{theorem}{\bf(Coulomb Potential)} \label{th-3.7}
Let $B\in L^p \Omega^2(\overline{\Lambda}), p>3,$ be a closed $2-$ form that is continuous in a neighborhood of
$\partial K$,
 where $K$ is as in  Assumption \ref{ass-1}. Assume, furthermore, that  (\ref{3.1}) holds and that
for some $r$ with $ K \subset B^{\ere^3}_r(0)$,
\beq
|B(x)| \leq C (1+|x|)^{-\mu}, |x| \geq r, \mu > 2.
\label{3.10}
\ene
 Then, for any
 flux, $\Phi$, there is a potential $A_{C} \in \pot$ such that $ A_{C}= A_{(C,1)}+ A_{(C,2)}$ where
 $A_{(C,1)}$ is
 continuous on $\overline{\Lambda}$, $A_{(C,2)}$ is $C^\infty$ on $\overline{\Lambda}$, and
 $ \delta A_{(C,j)}=-{\rm div}\, A_{(C,j)}=0, \, j=1,2$.
 Furthermore,
\beq
|A_{(C,1)}(x)|\leq C (1+|x|)^{- {\rm min}(2-\varepsilon,\mu-1) },\,\forall \varepsilon >0,
\label{3.11}
\ene

\beq
|A_{(C,2)}(x)|\leq C (1+|x|)^{-2}.
\label{3.12}
\ene
\end{theorem}

\noindent {\it Proof:} Let $\overline{B}$ be the extension to $\ere^3$ of $B$ given by Theorem \ref{th-3.2}.
by Proposition 2.6 of \cite{ju3} and its proof we can take as $A_{(C,1)}$ the Coulomb gauge of $\overline{B}$.

\beq
A_{(C,1)}:=-\frac{1}{4\pi} \int_{\ere^3} \, \frac{x-y}{|x-y|^3} \times \overline{B}(y)\, dy,
\label{3.13}
\ene
where we use the notation of vector calculus. We define $A_{(C,2)}$ as follows,

\beq
A_{(C,2)}:= \sum_{j=1}^m \left(  \Phi(\hat{\gamma}_j)- \int_{\hat{\gamma}_j} A_{(C,1)}\right) G^{(j)},
\label{3.14}
\ene
where $G^{(j)}, j=1,2,\cdots, m$ are defined in (\ref{1.12}) and we used (\ref{1.13}).
Clearly, $G^{(j)} \in C^\infty (\overline{\Lambda})$ and $|G^{(j)}(x)| \leq C(1+|x|)^{-2}$.

\bull

Note that in $\ere^3$  $\ac$ is the Coulomb potential that corresponds to the magnetic field

$$
\overline{B}+ \sum_{j=1}^m   \left( \Phi(\gamma_j) -\int_{\hat{\gamma_j}} A_{(C,1)}\right) \delta(x-\gamma_j)
d\vec{\gamma_j},
$$
with
$$
\left\langle \delta(x-\gamma_j) d\vec{\gamma_j},\, \phi\right\rangle:= \int_{\gamma_j}\,
\phi\, d\vec{\gamma_j}.
$$

The  div-curl problem in exterior domains in the case of $C^1$ vector fields with H\"older continuous first
derivatives was considered in \cite{vw}.

\begin{lemma} \label{lemm-3.8}{\bf (Gauge Transformations)}
Suppose that  $A,\tilde{A} \in \pot$. Then, there is a $C^1 \, 0-$ form $\lambda$ in $\overline{\Lambda}$
such that,
$\tilde{A} - A= d \lambda$.  Moreover, we can take $\lambda(x):=\int_{C(x_0,x)}(\tilde{A}-A)$ where $x_0 $
is any fixed
point in $\Lambda$ and $C(x_0,x)$ is any
 curve from $x_0$ to $x$. Furthermore, $\lambda_\infty(x):=\lim_{r\rightarrow \infty} \lambda(rx)$ exists
 and it is continuous in $  \mathbb{R}^3 \setminus \{ 0 \}  $ and homogeneous of order zero, i.e.
  $\lambda_\infty(r x)=\lambda_\infty(x), r >0,
  x \in  \ere^3 \setminus\{0\}$. Moreover,
\beq
\begin{array}{c}
|\lambda_\infty(x)-\lambda(x)|\leq \int_{|x|}^\infty \, b(|x|), \hbox{\rm for some}\,\, b(r)\in L^1(0,\infty),
\, \\\\ \hbox{\rm and} \,\,  |\lambda_\infty (x+y)-\lambda_\infty(x)|\leq C |y|, \forall x: |x|=1, \,
 {\rm and}\, \forall y: |y| < 1/2.
\end{array}
\label{3.15}
\ene
\end{lemma}

\noindent {\it Proof:}
The existence of $\lambda$ follows from Proposition \ref{prop-2.4}. The existence of $\lambda_\infty$ and
the first equation in (\ref{3.15}) follow from
condition 1 in Definition \ref{def-3.5}. The homogeneity follows from the definition.
 Denote $G:= \tilde{A} - A$. Take $m > 1$ such  that $ K \subset B^{\ere^3}_{m/2}(0)$.
Suppose that $|x|=1$ and that $|y|< 1/2$.

 Denote, $x':= m x, y':= m \frac{x+y}{|x+y|}-x'$.
Then, $\lambda_\infty(x)= \lambda_\infty(x'), \lambda_\infty(x+y)=\lambda_\infty(x'+y')$.
Hence,
$$
\lambda_\infty (x+y)-\lambda_\infty(x)=\lambda_\infty (x'+y')-\lambda_\infty(x')= \int_{x'}^{x'+y'} G+\int_{x'+y'}^\infty G- \int_{x'}^\infty G=
\lim_{r \rightarrow \infty}\int_{rmx}^{rm (x+y)/|x+y|}\, G,
$$
where we used Stoke's theorem and $d G=0$. Then,

$$
|\lambda_\infty (x'+y')-\lambda_\infty(x')| \leq \lim_{r \rightarrow \infty}\int_{rmx}^{rm (x+y)/|x+y|}\, |G|
\leq C |y|.
$$
This proves (\ref{3.15}).

We now consider potentials that satisfy the flux condition modulo $2\pi$.

\begin{definition} \label{def-3.6}
Let $B$ be as in Definition \ref{def-3.5}. We denote by $\p2$ the set of all continuous $1-$ forms in
$ \overline{\Lambda}$ that satisfy $1$ and $3$ of Definition \ref{def-3.5} and moreover,
$$
\int_{\hat{\gamma_j}}\, A= \Phi (\hat{\gamma_j})+ 2\pi n_j(A), \, n_j(A) \in \ZETA,  j \in \{1,2,\cdots, m \}.
$$
\end{definition}
Given $A \in \p2$ we define,
\beq
A_\Phi:= A- \sum_{j=1}^m 2\pi n_j(A)\, G^{(j)}.
\label{3.16}
\ene

 By (\ref{1.13}) $A_\Phi \in \pot$.

Suppose that $A,\tilde{A} \in \p2$. Then, $A_\Phi, \tilde{A}_\Phi \in \pot$, and  by Lemma \ref{lemm-3.8}
\beq
\tilde{A}_\phi- A_\Phi= d \lambda
\label{3.17}
\ene
and it follows that,
\beq
\tilde{A}- A= d \lambda+ A_\ZETA,
\label{3.18}
\ene
where
\beq
A_\ZETA:= \sum_{j=1}^m 2\pi (n_j(\tilde{A})- n_j(A)) \,G^{(j)}.
\label{3.18b}
\ene

Let $C$
be any closed curve in
$\Lambda$. Then, by Proposition \ref{prop-b.1} in Appendix B,
$$
C:= \sum_{j=1}^m n_j(C) \hat{\gamma}_{j}+ \partial \sigma, n_j(C)\in \ZETA.
$$
Hence,
\beq
\int_C (\tilde{A}-A)= 2 \pi N, \, \hbox{\rm for some}\, N \in \ZETA.
\label{3.19}
\ene
Whence, we can define the non-integrable factors \cite{Wu},
\beq
U_{\tilde{A},A}(x):= e^{i\int_{C(x_0,x)} (\tilde{A}-A)}= e^{i( \lambda(x)+\int_{C(x_0,x)} A_\ZETA)},
\label{3.20}
\ene
where $x_0$ is any fixed point in $\Lambda$ and $C(x_0,x)$ is any curve in $\Lambda$ from $x_0$ to $x$.
Clearly, $U_{\tilde{A},A}\in C^1(\overline{\Lambda})$. Moreover, if $ \tilde{A}, A \in \pot$ we have that
 $A_\ZETA=0$, and then,
 \beq
U_{\tilde{A},A}(x)= e^{i \lambda(x)}, \tilde{A}, A \in \pot.
\label{3.20b}
\ene

\begin{lemma} \label{lemm-3.10}
Suppose that $ \tilde{A}, A \in \p2$. Then, for $ x \neq 0$,
\beq
\lim_{r \rightarrow \infty} U_{\tilde{A},A}(rx)= e^{i(\lambda_\infty(x)+ C_{\tilde{A}, A})},
\label{3.21}
\ene
  with $\lambda_\infty(x):= \lim_{r \rightarrow \infty}\lambda(rx)$ given by Lemma \ref{lemm-3.8} with
  $\lambda$ as in (\ref{3.17}),
  and where $C_{\tilde{A},A}$ is a real number that is independent of $x$. Furthermore,

\beq
\left| U_{\tilde{A},A}(x)- e^{i(\lambda_\infty(x)+ C_{\tilde{A}, A})}\right|\leq
 \int_{|x|}^\infty \, c(|x|), \hbox{\rm for some}\,\, c(r)\in L^1(0,\infty).
\label{3.22}
\ene
Moreover, if $ \tilde{A}, A \in \pot$ we have that $C_{\tilde{A},A}=0$.
\end{lemma}

\noindent{\it Proof:} Let $r_0$ be such that $K \subset B_{r_0}^{\ere^3}(0)$.
Take in (\ref{3.20}) any curve from $x_0$ to $ r_0 \hat{x}$ and then  the straight line from $r_0 \hat{x}$
to $r \hat{x}$ with $ r_0 \leq r < \infty$. By (\ref{1.12}, \ref{3.7b}, \ref{3.15}, \ref{3.18})
$$
\lim_{r \rightarrow \infty} U_{\tilde{A},A}(rx)= e^{i\lambda_\infty(x)}
\lim_{r \rightarrow \infty} e^{i \int_{C(x_0, r\hat{x})} A_\ZETA}
$$
and
$$
\left| U_{\tilde{A},A}(x)- e^{i\lambda_\infty(x)}
\lim_{r \rightarrow \infty} e^{i \int_{C(x_0, r\hat{x})} A_\ZETA} \right|\leq
 \int_{|x|}^\infty \, c(|x|), \hbox{\rm for some}\,\, c(r)\in L^1(0,\infty).
$$

For any $ y \neq 0, y\neq  \pm x$ let $C(r\hat{x}, r \hat{y})$ be the straight line from $ r\hat{x}$ to
$r \hat{y}$. Then,
$$
\lim_{r \rightarrow \infty} e^{i (\int_{C(x_0, r\hat{x})} A_\ZETA- \int_{C(x_0, r\hat{y})} A_\ZETA)}
= \lim_{r \rightarrow \infty} e^{-i \int_{C(r \hat{x}, r\hat{y})} A_\ZETA}=1,
$$
and it follows that,
$$
\lim_{r \rightarrow \infty} e^{i \int_{C(x_0, r\hat{x})} A_\ZETA}=
\lim_{r \rightarrow \infty} e^{i \int_{C(x_0, r\hat{y})} A_\ZETA}= e^{i C_{\tilde{A}, A}}
$$
for some $ C_{\tilde{A},A} \in \ere$ that is independent of $x$. If $ \tilde{A}, A \in \pot$, $n_j(\tilde{A})=
n_j(A)=0, j= ,1,2,\cdots, m$ and hence, $A_\ZETA=0$, what implies that $C_{\tilde{A},A}=0$.

\section{The Hamiltonian}\sss

\noindent Let us denote  $\mo:=-i\nabla$.
The Schr\"odinger equation for an electron in $\Lambda$  with electric potential $\mathbf V$ and magnetic
field $\mathbf B$   is given by
\beq
i\hbar \frac{\partial}{\partial t} \phi = \frac{1}{2 M} ({\mathbf P}- \frac{q}{ c}\mathbf A)^2+q \,\mathbf V,
\label{4.1}
\ene
where $\hbar$ is Planck's constant, ${\mathbf P}:=\hbar \mo$ is the momentum operator, $c$ is the
speed of light, $M$ and $q$ are,
respectively, the mass and the charge of the electron and $ \mathbf A$ a magnetic potential with
$\hbox{\rm curl} \mathbf A= \mathbf B$. To simplify the notation
we multiply both sides of (\ref{4.1}) by $\frac{1}{\hbar}$ and we write Schr\"odinger's equation as follows

\beq
i \frac{\partial} {\partial t} \phi= \frac{1}{2m} (\mo- A)^2 \phi+V\phi,
\label{4.2}
\ene
with $m:= M/ \hbar, A= \frac{q}{\hbar c}\mathbf A$ and $V:= \frac{q}{\hbar} \mathbf V$.
Note that since we write Schr\"odinger's equation in this form our Hamiltonians below are the physical
Hamiltonians divided by $\hbar$. We fix the flux modulo $2\pi$ by taking $A \in \mathcal A _{\Phi, 2 \pi}$ where $B:=\frac{q}{\hbar c}
{\mathbf B}$. Note that this corresponds
to fixing the circulations of $\mathbf A$ modulo $\frac{\hbar c}{q} 2 \pi$, or equivalently, to fixing
the fluxes
of the magnetic field $\mathbf B$ modulo  $\frac{\hbar c}{q} 2 \pi$.

 For any open set, $O$, we denote by $\mathcal H_s(O), \,
s=1,2,\cdots$  the Sobolev spaces \cite{ad} and by $\mathcal H_{s,0}(O)$ the closure of $C^\infty_0(O)$
in the norm of $\mathcal H_{s}(O)$. We define the quadratic form,
\beq
h_0(\phi,\psi):= \frac{1}{2 m}(\mo \phi,\mo \psi), \, D(h_0):= \mathcal H_{1,0}(\Lambda).
\label{4.3}
\ene
The associated positive operator in $L^2(\Lambda)$  \cite{ka}, \cite{rs2} is $\frac{-1}{2m} \Delta_D$
where $\Delta_D$ is  the  Laplacian with
Dirichlet boundary condition on $\partial \Lambda$. We define
$H(0,0):= \frac{-1}{2m} \Delta_D$.
By elliptic regularity \cite{ag}, $D(H(0,0))= \mathcal H_2(\Lambda) \cap \mathcal H_{1,0}(\Lambda)$.

For any $ A \in \p2$ we define,
\beq
h_A(\phi,\psi):= \frac{1}{2m}\left( (\mo-A)\phi, (\mo-A) \psi\right)=h_0(\phi,\psi)
+ \frac{1}{2m}(-(\mo \phi,A\psi)-(A\phi,\mo\psi)) + \frac{1}{2m} (A \phi, A \psi ),\, D(h_A)= \mathcal H_{1,0}(\Lambda).
\label{4.4}
\ene
As the quadratic form $ -\frac{1}{2m} ((\mo \phi,A\psi)+(A\phi,\mo\psi))+ \frac{1}{2m} (A \phi, A \psi )$ is $h_0-$ bounded with relative bound zero,
$h_A$ is closed and positive. We denote by $H(A,0)$ the associated positive self-adjoint operator \cite{ka},
\cite{rs2}. $H(A,0)$ is the Hamiltonian with magnetic potential $A$.  Note that as
the operator $ \frac{1}{2m} (-2 \ac\cdot \mo+\ac^2 )$ is $H(0, 0)\,\hbox{\rm compact}$ we have that
$H(0,0)-\frac{1}{m} \ac\cdot \mo+\frac{1}{2m}\ac^2$ is self-adjoint on the domain of $H(0,0)$ and then,
\beq
H(\ac,0)= H(0,0)-\frac{1}{m} \ac\cdot \mo+\frac{1}{2m}\ac^2, \, D(H(\ac,0))= \mathcal H_2(\Lambda)\cap \mathcal H_{1,0}(\Lambda).
\label{4.5}
\ene

The electric potential $V$ is a measurable real-valued function defined on $\Lambda$. We assume that
$|V|$ is $h_0-$ bounded with relative bound zero.
Under this condition \cite{ka}, \cite{rs2} the quadratic form,

\beq
h_{A,V}(\phi, \psi):= h_A( \phi,\psi)+(V \phi,\psi), \, D(h_{A,V})= \mathcal H_{1,0}(\Lambda),
\label{4.6}
\ene
is closed and bounded from below. The associated operator, $H(A,V)$, is self-adjoint and bounded from below.
$H(A,V)$ is the Hamiltonian with magnetic potential $A$ and electric potential $V$. If furthermore,
$V$ is $-\Delta_D$ compact, the operator $H(0,0)-\frac{1}{m} \ac\cdot \mo+\frac{1}{2m}\ac^2 + V$ is
self-adjoint on the domain of
$H(0,0)$ and
then,

\beq
H(\ac, V)= H(0,0)-\frac{1}{m} \ac\cdot \mo+\frac{1}{2m}\ac^2 + V, \, D(H(\ac,V))= \mathcal H_2(\Lambda)\cap \mathcal H_{1,0}(\Lambda).
\label{4.7}
\ene

We will  denote by $ U_{\tilde{A},A}$ the operator of multiplication by $ U_{\tilde{A},A}(x)$. See (\ref{3.20}).
 Note that $U_{\tilde{A},A}$ is unitary in $L^2(\Lambda)$ and that $U_{\tilde{A},A}^\ast$ is the operator
 of multiplication by $U_{A,\tilde{A}}(x
 )$ .

\begin{theorem}\label{th-4.1}
Suppose that $\tilde{A}, A \in \p2$. Then $ H(\tilde{A},V)$ and $H(A,V)$ are unitarily equivalent,
\beq
H(\tilde{A},V)= U_{\tilde{A},A}\, H(A,V)\, U_{\tilde{A},A}^\ast, \, \,\, D(H(\tilde{A},V))=
U_{\tilde{A},A}\,D(H(A,V)).
\label{4.8}
\ene
\end{theorem}
\noindent {\it Proof:}  As $U_{\tilde{A},A}$   and
 $U_{\tilde{A},A}^\ast$ are bijections on $\mathcal H_{1,0}(\Lambda)$ we have that

 $$
 h_{\tilde{A},V}(\phi,\psi)= h_{A,V}\left(U_{\tilde{A},A}^\ast\phi,U_{\tilde{A},A}^\ast  \psi \right), \,
 \phi,\psi \in \mathcal H_{1,0}(\Lambda).
 $$
 Suppose that $ \phi \in D(H(\tilde{A},V))$.
Then, for every $ \chi\in \mathcal H_{1,0}(\Lambda)$,
$$
(U_{\tilde{A},A}^\ast\, H(\tilde{A},V)\phi, \chi)= h_{A,V}(U_{\tilde{A},A}^\ast\phi, \chi).
$$
This implies that $ U_{\tilde{A},A}^\ast\, \phi \in D(H(A,V))$ and that

$$
H(A,V)U_{\tilde{A},A}^\ast\, \phi = U_{\tilde{A},A}^\ast H(\tilde{A},V)\phi.
$$
what proves the theorem.

\section{Scattering}\sss
In the following assumptions we summarize the conditions on the magnetic field and the electric potential that
we use. We denote by $\Delta$ the self-adjoint realization of the Laplacian in $L^2(\ere^3)$ with domain
$\mathcal H_2(\ere^3)$. Below we assume that $V$ is $\Delta$- bounded with relative bound zero. By this we
mean that the extension of $V$ to $\ere^3$ by zero is $\Delta-$ bounded with relative bound zero.
Using a extension operator from $\mathcal H_2(\Lambda)$ to $H_2(\ere^3)$ \cite{tr} we prove that this is
equivalent to require that $V$ is  bounded from $\mathcal H_2(\Lambda)$ into $L^2(\Lambda)$ with relative bound
zero.
We denote by $\| \cdot \|$ the operator norm in $L^2(\ere^3)$.
\begin{assumption}\label{ass-5.1}
We assume that the magnetic field, $B$, is a real-valued, bounded $2-$ form in $ \overline{\Lambda}$,
that is continuous in
a neighborhood of $\partial K$, where $K$ satisfies Assumption \ref{ass-1}, and furthermore,
\begin{enumerate}
\item
$ B \, \hbox{\rm is closed}:    d B|_{\Lambda} \equiv \,\hbox{\rm div} B=0$.

\item
There are no magnetic monopoles in $K$:
\beq
\int_{\partial K_{j}} B=0, \, j \in \{1,2,\cdots,L\}.
\label{5.1}
\ene

\item
\beq
\left|B(x)\right| \leq C (1+|x|)^{- \mu}, \, \hbox{\rm for some}\,\, \mu > 2.
\label{5.2}
\ene
\item

  $d* B|_{\Lambda} \equiv \hbox{\rm curl}\, B $ is  bounded and,
\beq
|\hbox{\rm curl}\, B |  \leq C (1+|x|)^{- \mu}.
\label{5.3}
\ene
\item
The electric  potential, $V$, is a real-valued function, it is  $\Delta-$bounded, and
\beq
\left\|   F(|x|\geq r) V(-\Delta+I)^{-1}\right\|  \leq C (1+|x|)^{- \alpha}, \,\hbox{\rm for some}\, \alpha >1 .
\label{5.4}
\ene
\end{enumerate}
\end{assumption}

Note that (\ref{5.4}) implies that $V$ is $h_0-$bounded with relative bound zero. Furthermore, condition
(\ref{5.4}) is equivalent to the following assumption \cite{rs3}

\beq
\left\|V (-\Delta+I)^{-1}  F(|x|\geq r)  \right\|  \leq C (1+|x|)^{- \alpha}, \,\hbox{\rm for some}\, \alpha >1 .
\label{5.5}
\ene
Condition (\ref{5.4}) has a clear intuitive meaning, it is
 a condition on   the decay of $V$ at infinity.
However, in the proofs below we use the equivalent statement (\ref{5.5}).

Let us define,
$$
H_0:= -\frac{1}{2m} \Delta, \, D(H_0)= \mathcal H_2(\ere^3).
$$

Let $J$ be the identification operator from $L^2(\ere^3)$ onto $L^2(\Lambda)$ given by multiplication by the
characteristic function of $\Lambda$. The wave operators are defined as follows,

\beq
W_\pm(A,V):= \hbox{\rm s-}\lim_{t \rightarrow \pm \infty} e^{it H(A,V)}\, J\, e^{-it H_0},
\label{5.6}
\ene
provided that the strong limits exist. We first prove that they exist in the Coulomb gauge.

\begin{prop} \label{prop-5.1}
Suppose that $B$ and $V$ satisfy Assumption \ref{ass-5.1}. Then, the wave
operators $W_\pm(A_C, V)$ exist and are
isometric.
\end{prop}

\noindent{\it Proof:} Let $ \chi \in C^\infty(\ere^3)$ satisfy $\chi(x)=0$ in a neighborhood of $K$ and
$\chi (x)=1$ for $ |x| \geq r_0$ with $r_0$ large enough. Then, since $(1-\chi(x))(H_0+I)^{-1}$ is compact,

$$
W_\pm(A_C,V)= \hbox{\rm s-}\lim_{t \rightarrow \pm \infty} e^{it H(A_C,V)}\, \chi(x)\, e^{-it H_0}.
$$
By Duhamel's formula, for $ \phi \in D(H_0)$,

\beq
W_\pm(A_C,V)\phi = \chi(x) \phi(x) +\int_{0}^{\pm \infty} i\, e^{itH(A_C,V)}\left[ H(A_C,V)\chi(x)- \chi(x)H_0
\right]\,\phi(x) \,dt.
\label{5.7}
\ene
By Theorem \ref{th-3.7}  the proof that the integral in the right-hand side of (\ref{5.7}) is absolutely
convergent is standard.
For example, it follows from Lemma 2.2 of \cite{ew} taking  $\phi= e^{im\v\cdot x}\varphi$, with
  $\v \in \ere^3, |\v| \geq 4 \eta > 0$, and
 $ \hat{\varphi}\in  C^\infty_0(B^{\ere^3}_{m \eta}(0))$, what is a dense set in $L^2(\ere^3)$.

\bull

\begin{lemma}{\bf (Gauge Transformations)} \label{lemm-5.2}
Suppose that Assumption \ref{ass-5.1} is true. Then, for every $ A \in \p2$ the wave operators $W_\pm(A, V)$
exist and are isometric. Moreover, if $ \tilde{A}\in \p2$, then,
\beq
W_\pm(\tilde{A},V)= e^{-i C_{\tilde{A},A}}\, U_{\tilde{A},A}\, W_\pm(A,V)\, e^{-i \lambda_\infty(\pm\mo)}.
\label{5.8}
\ene
\end{lemma}

\noindent {\it Proof:} Since we already know that $W_\pm(\ac,V)$ exist and are isometric it is enough to prove
the gauge transformation formula (\ref{5.8}). We argue as in the proof of Lemma 2.3 of \cite{we1}.
By (\ref{4.8})

$$
W_\pm(\tilde{A},V)= U_{\tilde{A},A}\, \hbox{\rm s-}\lim_{t \rightarrow \pm \infty} e^{it H(A,V)}
U_{A, \tilde{A}} J e^{-it H_0}=  U_{\tilde{A},A}\, \hbox{\rm s-}\lim_{t \rightarrow  \pm\infty} e^{it H(A,V)}
\, J \,e^{-i( \lambda_\infty(x)+ C_{\tilde{A},A})}   e^{-it H_0},
$$
where we used that by Lemma \ref{lemm-3.10} and Rellich selection  theorem $U_{A,\tilde{A}}-
e^{-i( \lambda_\infty(x)+ C_{\tilde{A},A})}$ is a compact operator from $D(H_0)$ into $L^2(\ere^3)$.
We finish the proof of the lemma as in the proof of equation (2.29) of \cite{we1}, using the second
equation in (\ref{3.15}).

The scattering operator is defined as

$$
S(A,V):= W_+^\ast(A,V)\, W_-(A,V).
$$
By (\ref{5.8})

\beq
S(\tilde{A},V)= e^{i \lambda_\infty(\mo)}\, S(A, V)\, e^{-i \lambda_\infty(-\mo)}, \, \tilde{A}, A  \in \p2.
\label{5.9}
\ene

\begin{definition} \label{def-5.3}
We say that $ A \in \p2$ is short-range if
\beq
\left| A(x)\right| \leq C(1+|x|)^{-1-\varepsilon},\, \, \hbox{\rm for some}\, \varepsilon > 0.
\label{5.10}
\ene
We denote the set of all short-range potentials in $\p2$ by $\sp2$.
\end{definition}
Note that if $\tilde{A}, A \in \p2$ and $\tilde{A}-A$ satisfies (\ref{5.10}),
$\lambda_\infty$ is constant, and then,

\beq
S(\tilde{A},V)=  S(A,V), \, \tilde{A}, A \in \p2\, \hbox{\rm and $ \tilde{A}-A$ satisfies (\ref{5.10})}.
\label{5.11}
\ene

This implies that,
\beq
S(A_\Phi,V)= S(A,V), \,\hbox{\rm for any}\, A \in \p2,
\label{5.12}
\ene
where $A_\Phi$ is defined in (\ref{3.16}).
Remark that (\ref{5.11}) holds if $ \tilde{A}, A \in \sp2$.

We quote below the following result of \cite{we1} that we will often use.

\begin{lemma} \label{lemm-5.4}
For any $f \in C^\infty_0(B^ {\ere^3}_{m\eta}(0)), 0 \leq \rho <1$, and for any $j=1,2,\cdots$ there is a
constant
$C_j$ such that
\beq
\left\| F\left(|x- \v t| > \frac{|\v t|}{4}\right) e^{-it H_0} \, f\left(\frac{\mo-m \v}{v^\rho}\right)\,
F\left( |x| \leq |\v t|/8\right)\right\| \leq
C_j (1+|\v t|)^{-j},
\label{5.13}
\ene
for $v:=|\v| > (8\eta)^{1/(1-\rho)}$.
\end{lemma}

\noindent {\it Proof:} Corollary 2.2 of \cite{we1}.

\noindent {\bf 5.1 High-Velocity Estimates I. The Magnetic Potential}

We denote,
\beq
\Lambda_{\hv}:= \{x \in \Lambda: x+\tau \hv \in \Lambda,\, \forall \tau \in \ere\},\, \hbox{\rm for}\, \v \neq 0.
\label{5.14}
\ene

\beq
L_{A,\hv}(t):= \int_0^t\, \hv\cdot A(x+\tau \hv) d\tau, -\infty \leq t \leq \infty.
\label{5.15}
\ene
Remark that under translation in configuration or momentum space generated, respectively, by $\mo$ and x
we obtain

\beq \label{5.15.0}
e^{i \mo \cdot \v t} \, f(x) \, e^{-i \mo \cdot \v t}= f(x+\v t),
\ene
\beq \label{5.15.1}
e^{-im\v\cdot x}\, f(\mo )\, e^{im\v\cdot x}= f(\mo +m \v ),
\ene
and, in particular,
\beq
e^{-im\v\cdot x}\, e^{-it H_0 } \, e^{im\v\cdot x}= e^{-i mv^2 t/2}\, e^{-i\mo\cdot \v t}\,
 e^{-i tH_0}.
\label{5.15b}
\ene
The purpose of the obstacle $K$ is to shield the incoming electrons from the magnetic field inside the
obstacle. In order to separate the scattering effect of the magnetic potential from that of the magnetic field inside the obstacle $K$, we consider
asymptotic configurations that have negligible interaction with $K$ for all times in the high-velocity
limit. For any non-zero $\v \in \ere^3$ we take asymptotic configurations $ \phi$ with compact support
in $\Lambda_{\hv}$. The free evolution boosted by $\hv$ is given by (\ref{5.15b}) and -to a good approximation-
in the limit when $v \rightarrow \infty$ with $\hv$ fixed this can be replaced
(modulo an unimportant phase factor) by the classical translation $e^{-i\mo\cdot\v t}$. Then,
in the high-velocity limit it is a good approximation to assume that the free evolution of our asymptotic
configuration is given by $e^{-i\mo\cdot \v t}\phi_0= \phi_0(x-\v t)$, and as $\phi_0$ has support in
$\Lambda_{\hv}$, it has negligible interaction with $K$ for all times. Note that instead of boosting the
observables we can boost the asymptotic configurations and consider the high-velocity asymptotic configurations
$$
\phi_{\v}:= e^{i m \v\cdot x}\phi_0.
$$

\begin{lemma}\label{lemm-5.5}
 Suppose that $B,V$ satisfy Assumption \ref{ass-5.1}. Let  $\Lambda_0$ be a compact subset of $\Lambda_{\hv} $, with
$\v \in \ere \setminus \{0\}$. Then, for all $\Phi$ and  all $A \in \p2$
there is a constant $C$ such that,
\beq
\left\| \left(e^{-im\v\cdot x}\, W_\pm(A,V)\, e^{im\v\cdot x} -e^{-i L_{A,\hv}(\pm \infty)}\right) \phi
\right\|_{L^2(\ere^3)}\leq
 C \frac{1}{v} \| \phi\|_{\mathcal H_2(\ere^3)},
\label{5.16}
\ene
and if moreover, $ \hbox{\rm div}A \in L^2_{\hbox{\rm loc}}\left(\overline{\Lambda}\right)$,
\beq
\left\| \left(e^{-im\v\cdot x}\, W_\pm^\ast(A,V)\, e^{im\v\cdot x} -e^{i L_{A,\hv}(\pm \infty)}\right) \phi
\right\|_{L^2(\ere^3)}\leq
 C \frac{1}{v} \| \phi\|_{\mathcal H_2(\ere^3)},
\label{5.17}
\ene
for all  $\phi \in \mathcal H_2(\ere^3) $ with $ \hbox{\rm support}\, \phi \subset
\Lambda_0 $.
 \end{lemma}

 \noindent {\it Proof:} We follow the proof of Lemma 2.4 of \cite{we1}. We first give the proof in the case
 of the Coulomb potential $\ac$. We give the proof for $W_+(\ac,V)$. The proof for   $W_-(\ac,V)$ follows
 in the same way.

Let $g \in C^\infty_0(\ere^3)$ satisfy $g(p)=1, |p| \leq 1, g(p)=0, |p| \geq 2$. Denote
\beq \label{5.17.0}
\tf:= g(\mo/ v^\rho) \,\phi, \, \frac{1}{2} \leq \rho < 1.
\ene
Then,
\beq
\left\| \tf- \phi\right\|_{L^2(\ere^3)} \leq \frac{1}{v^{2 \rho}} \|\phi\|_{\mathcal H_2(\ere^3)}.
\label{5.17b}
\ene
Hence, it is enough to prove (\ref{5.16}) for $\tilde{\phi}$.

 By our assumption there is a function
 $\chi \in C^\infty(\ere^3)$ such that $\chi \equiv 0$ in a neighborhood of $K$ and $\chi(x)=1, x \in
 \{ x: x= y+ \tau \hv, y \in \,\hbox{\rm support}\, \phi , \, \tau \in \ere\} \cup \{ x: |x| \geq M\}$ for some
 $M$ large enough. We use the following notation,

\beq \label{5.18.0}
 H_1:= \frac{1}{v} e^{-im\v\cdot x}H_0 \,e^{im\v\cdot x},\,\,  H_2:= \frac{1}{v} e^{-im\v\cdot x}H(\ac,V)
 \, e^{im\v\cdot x}.
\ene
Note that,
\beq
 \left(e^{-im\v\cdot x}\, W_+(\ac,V)\, e^{im\v\cdot x} -\chi(x)e^{-i L_{\ac,\hv}\left( \infty \right)} \right)\tf =
 \hbox{\rm s-}\, \lim_{t \rightarrow \infty}\left[ e^{it H_2}\chi (x)e^{-it H_1}- \chi(x)e^{-iL_{\ac,\hv}
 \left(t\right)}
 \right]
 \tf .
\label{5.18}
\ene
Denote,
\beq
P(t,\tau):= e^{i\tau H_2} i \left[H_{2}e^{-i L_{\ac, \hv}(t-\tau)}\chi(x)-e^{-i L_{\ac,\hv}(t-\tau)} \chi(x)
\left(H_1-\hv\cdot\ac(x+(t-\tau)\hv)\right)\right] e^{-i\tau H_1}\tf.
\label{5.19}
\ene
Then, by Duhamel's formula,
\beq
\left[ e^{it H_2}\chi (x)e^{-it H_1}- \chi(x)e^{-iL_{\ac,\hv}
 \left(t\right)}
 \right]
 \tf = \int_0^t\, d \tau\, P(t,\tau).
 \label{5.20}
 \ene
We designate,
\beq
b(x,t):= \ac(x+t\hv)+\int_0^t (\hv\times B)(x+\tau\hv)d\tau.
\label{5.21}
\ene
For ${\mathbf f}: \ere^3\times \ere \rightarrow \ere^3$ with $ \mathbf f_t(x):={\mathbf f}(x,t) \in L^1_{\hbox{\rm loc}}
(\ere^3,\ere^3)$ we define,
\beq
\Xi_{\mathbf f}(x,t):= \frac{1}{2m}\chi (x)\left [
-\mo\cdot \mathbf f(x,t)-\mathbf f(x,t)\cdot \mo
+(\mathbf f(x,t))^2\right].
\label{5.22}
\ene
We have that \cite{we1},
\beq
P(t,\tau)= T_1+T_2+T_3
\label{5.23},
\ene
with
\beq
T_1:= \frac{1}{v} e^{i \tau H_2}i e^{-iL_{\ac ,\hv}(x,t-\tau)}\, \left( \Xi_b(x,t-\tau)+ \chi V(x)\right)e^{-i\tau H_1}\tf ,
\label{5.24}
\ene
\beq
\begin{array}{c}
T_2:= \frac{1}{2mv}e^{i \tau H_2}i e^{-i L_{\ac , \hv}(x,t-\tau)}\left\{-(\Delta \chi)+2(\mo \chi)\cdot \mo
-2 b(x, t-\tau) \cdot (\mo \chi)\right\} e^{-i\tau H_1}\tf,
\label{5.25}
\end{array}
\ene

\beq
T_3:=e^{i \tau H_2}i e^{-i L_{\ac,\hv}(x,t-\tau)} \left[ (\mo \chi)\cdot \hv  \right]
e^{-i\tau H_1}\tf.
\label{5.26}
\ene
Note that (\cite{ar1}, equation (2.18))
\beq
\left\| \int_0^{t-\tau} d\nu (\hv \times B)(x+\nu \hv)\, F(|x-\tau \hv|\leq |\tau|/4)
\right\|_{L^\infty(\ere^3)} \leq
C \frac{1}{(1+|\tau|)^{\mu-1}},
\label{5.27}
\ene

\beq
\begin{array}{l}
\left\| \int_0^{t-\tau} d\nu (\nabla\cdot(\hv \times B))(x+\nu \hv)\, F(|x-\tau \hv|\leq |\tau|/4)
\right\|_{L^\infty(\ere^3)}
=\left\| \int_0^{t-\tau} d\nu (\hv \cdot\curl B)(x+\nu \hv)\right. \\\\
\left. F(|x-\tau \hv|\leq |\tau|/4)\right
\|_{L^\infty(\ere^3)}
\leq   C \frac{1}{\ds (1+|\tau|)^{\mu-1}}.
\end{array}
\label{5.28}
\ene
Using Theorem \ref{th-3.7}, Lemma \ref{lemm-5.4}, (\ref{5.2}, \ref{5.3}, \ref{5.5}, \ref{5.27}, \ref{5.28})
we prove as in the proof of Lemma 2.4 of \cite{we1} that,

\beq
\left\| T_1(\tau) \right\|_{L^2(\ere^3)} \leq \frac{C}{v} \frac{1}{ (1+|\tau|)^{\min (2-\varepsilon,
 \mu-1, \alpha)}} \|\phi\|_{\mathcal H_2(\ere^3)}  ,
\label{5.29}
\ene
\beq
\left\| T_2(\tau)\right\|_{L^2(\ere^3)} \leq \frac{C_j}{v} \frac{1}{(1+|\tau|)^j} \|\phi\|_{\mathcal H_2(\ere^3)},   j=1,2,\cdots,
\label{5.30}
\ene
\beq
\int_{-\infty}^\infty \, d \tau \,  \left\| T_3(\tau)\right\|_{L^2(\ere^3)} \leq \frac{C}{v}
\|\phi\|_{\mathcal H_2(\ere^3)}.
\label{5.31}
\ene
For the reader's convenience we estimate one of the terms. Denote by

\beq
\eta (x,t):= \int_0^t (\hv\times B)(x+\tau\hv)d\tau.
\label{5.32}
\ene

Then, by Lemma \ref{lemm-5.4} and (\ref{5.27}),
$$
\begin{array}{c}
\left\|\frac{1}{mv} e^{-iL_{\ac ,\hv}(x,t-\tau)} \eta(x,t-\tau) e^{-i\tau H_1} \cdot \mo  \tf\right
\|_{L^2(\ere^3)}
\leq \frac{C}{v}\left[  \|\eta(x,t-\tau)| F(|x-\tau \v|> |\tau|/4) e^{-i H_0 \tau/v} \right. \\ \\ \left.
g(\frac{\mo- m\v}{v^\rho})
F(|x|\leq |\tau|/ 8 )\| \|\phi\|_{\mathcal H_2(\ere^3)}   +
\|\eta(x,t-\tau) F(|x-\tau \v|\leq |\tau|/4)\|_{L^\infty(\ere^3)} \|\phi\|_{\mathcal H_2(\ere^3)}    +
\| F(|x|\geq |\tau|/8)\mo \cdot \tf \|_{L^2(\ere^3)}\right]\\\\
\leq   \frac{C}{(1+|\tau|)^{\mu-1}} \|\phi\|_{\mathcal H_2(\ere^3)} .
\end{array}
$$

By (\ref{5.20}, \ref{5.23}, \ref{5.29}, \ref{5.30}, \ref{5.31})
\beq
\left\|\left[ e^{it H_2}\chi (x)e^{-it H_1}- \chi (x)e^{-iL_{\ac,\hv}
 \left(t\right)}
 \right] \tf \right\|_{L^2(\ere^3)} \leq \frac{C}{v} \|\phi\|_{\mathcal H_2(\ere^3)}.
 \label{5.32b}
 \ene
By (\ref{5.18}) this proves (\ref{5.16}) for $\ac$. Given $A \in \p2$  we define $A_\Phi$ as in (\ref{3.16}).
As $A_\Phi \in \pot$,  we prove that (\ref{5.16}) holds for $A_\Phi$ as in the proof of Lemma 2.4 of \cite{we1}
using the formulae for change of gauge (\ref{5.8}). Then, we prove that it is true for $A$ using the gauge
transformation formulae between $A$ and $A_\Phi$, note that in this case
$\lambda\equiv\lambda_\infty\equiv 0$ ,
 observing that

\beq
e^{-i C_{A,A_\Phi}}= e^{-i (\int_{C(x_0,x)} A_\ZETA + \int_0^{\pm \infty}
\hv\cdot A_\ZETA(x+\tau \hv) d\tau)} = (U_{A, A_{\Phi}})^{*}e^{-i\int_0^{\pm \infty}
\hv\cdot A_\ZETA(x+\tau \hv) d\tau},
\label{5.32.0}
\ene
and using (\ref{3.18}) with $\lambda \equiv 0$.

We now prove (\ref{5.17}). Note that (\cite{ar1}, equation 2.12)

\beq
(\mo- A(x)) e^{-i L_{A,\hv}(t)}= e^{-i L_{A,\hv}( t)} \left(\mo -A(x+t\hv)-
\int_0^t (\hv\times B)(x+\tau\hv)d\tau \right).
\label{5.33}
\ene
Then, since $\hbox{\rm div} A \in L^2_{\hbox{\rm loc}}\left(\ \overline{\Lambda}  \right)$ if follows from
Sobolev's imbedding theorem \cite{ad} that,
\beq \label{5.33.0}
\| e^{i L_{A,\hv}(\pm \infty)} \phi\|_{\mathcal H_2(\ere^3)} \leq C \|\phi\|_{\mathcal H_2(\ere^3)}.
\ene
For simplicity we denote below  $W_\pm(A,V)$ by $W_\pm$ and we define,
\beq
W_{\pm,\hv}:= e^{-im\v\cdot x}\, W_\pm   \, e^{im\v\cdot x}.
\label{5.34}
\ene
As the wave operators are isometric, $W_{\pm,\hv}^\ast W_{\pm,\hv}=I$ and then,

$$
\begin{array}{c}
\left\| \left( W_{\pm,\hv}^\ast -e^{i L_{A,\hv}(\pm \infty)}\right) \phi
\right\|_{L^2(\ere^3)}= \left\|  W_{\pm,\hv}^\ast \phi -
W_{\pm,\hv}^\ast W_{\pm,\hv} e^{i L_{A,\hv}(\pm \infty)} \phi
\right\|_{L^2(\ere^3)} \\ \\ \leq
\left\| \left( W_{\pm,\hv} -e^{-i L_{A,\hv}(\pm \infty)}\right) e^{i L_{A,\hv}(\pm \infty)}  \phi
\right\|_{L^2(\ere^3)}\leq
 C \frac{1}{v} \| \phi\|_{\mathcal H_2(\ere^3)}.
\end{array}
$$

\bull

We now state the main result of this subsection.

\begin{theorem} \label{th-5.6}{\bf ( Reconstruction Formula I)}
 Suppose that $B,V$ satisfy Assumption \ref{ass-5.1}. Let  $\Lambda_0$ be a compact subset of $\Lambda_{\hv},
 $ with $\v \in \ere \setminus \{0\}$. Then, for all $\Phi$ and all $A \in \p2$  there is a constant $C$ such that,
\beq
\left\| \left( e^{-im\v\cdot x}\, S(A,V)\, e^{im\v\cdot x} -
e^{i \int_{-\infty}^\infty \, \hv \cdot A(x+\tau \hv )\,d\tau}
 \right) \phi
\right\|_{L^2(\ere^3)}\leq
 C \frac{1}{v} \| \phi\|_{\mathcal H_2(\ere^3)},
\label{5.35}
\ene

\beq
\left\| \left(e^{-im\v\cdot x}\, S(A,V)^\ast\, e^{im\v\cdot x} -  e^{-i \int_{-\infty}^\infty \,
\hv \cdot A(x+\tau \hv )\,d\tau} \right )  \phi
\right\|_{L^2(\ere^3)}\leq
 C \frac{1}{v} \| \phi\|_{\mathcal H_2(\ere^3)},
\label{5.36}
\ene
for all  $\phi \in \mathcal H_2(\ere^3) $ with $ \hbox{\rm support}\, \phi \subset
\Lambda_0 $. \\

 \end{theorem}

\noindent {\it Proof:} We use the same notation as in the end of the proof of Lemma \ref{lemm-5.5}.

First we prove (\ref{5.35}) and (\ref{5.36}) for $A_{C}$,

$$
\begin{array}{c}
\left\| \left(e^{-im\v\cdot x}\, S(A_{C}, V)\, e^{im\v\cdot x} -  e^{i \int_{-\infty}^\infty \,
\hv \cdot A_{C}(x+\tau \hv )\,d\tau} \right )  \phi
\right\|_{L^2(\ere^3)}= \left\|  W_{+,\hv}^\ast W_{-,\hv} \phi - W_{+,\hv}^\ast W_{+,\hv}
 e^{i (L_{\ac,\hv}(\infty)- L_{\ac, \hv} (-\infty))}\right.\\\\
\left.
   \phi \right\|_{L^2(\ere^3)}
 \leq \left\|   \left[W_{-,\hv}  - e^{-iL_{\ac,\hv}(-\infty)}\right] \phi
- \left[ W_{+,\hv}-e^{-i L_{\ac,\hv}(\infty)}\right] e^{i (L_{\ac, \hv}(\infty)- L_{\ac,\hv}( -\infty))}
  \phi \right\|_{L^2(\ere^3)} \leq \frac{C}{v}  \left\| \phi \right \|_{\mathcal H_2(\ere^3)}.
\end{array}
  $$
The proof for $ S(A_{C},V)^\ast$ follows in the same way.

Now we prove (\ref{5.35}) for $ A \in \p2 $, the proof of (\ref{5.36}) follows in the same way. \\
By (\ref{5.12}), $  S(A, V) = S(A_{\Phi},V) $. From (\ref{5.32.0}) it follows that
$$
e^{i \int_{- \infty}^{\infty} \hat{\v}\cdot A_{\mathbb{Z}}(x+\tau  \hat{\v} ) d\tau} =
e^{-iC_{A, A_{\Phi}}}e^{iC_{A, A_{\Phi}}} = 1,
$$
and thus,
\beq \label{wep}
e^{i \int_{- \infty}^{\infty} \hat{\v}\cdot A(x+\tau  \hat{\v} ) d\tau} =
e^{i \int_{- \infty}^{\infty} \hat{\v}\cdot A_{\Phi}(x+\tau  \hat{\v} ) d\tau}.
\ene
Then it is enough to prove (\ref{5.35}) for $ A= A_{C} + \nabla \lambda $. \\
By  (\ref{5.9}), (\ref{5.15.1}) and as $ \lambda $ in homogenous of order zero,

$$
\begin{array}{c}
 \| (e^{-im \v \cdot x}S(A,V)e^{im \v \cdot x} - e^{i \int_{- \infty}^{\infty} \hat{\v}\cdot A(x+\tau  \hat{\v} ) d\tau}) \phi \|_{L^{2}(\mathbb{R}^{3})}
= \\ \\
 \| (e^{i \lambda_{\infty}( \frac{\mo}{mv}+ \hat{\v} )}
e^{-im \v \cdot x}S(A_{C},V)e^{im \v \cdot x} e^{-i \lambda_{\infty}(-\frac{\mo}{mv}- \hat{\v})} - e^{i \int_{- \infty}^{\infty} \hat{\v}\cdot A(x+\tau  \hat{\v} ) d\tau} )\phi \|_{ L^{2}(\mathbb{R}^{3})} \leq \\ \\
\|( e^{i \lambda_{\infty}( \frac{\mo}{mv}+ \hat{\v} )}
e^{-im \v \cdot x}S(A_{C},V)e^{im \v \cdot x} ( e^{-i \lambda_{\infty}(-\frac{\mo}{mv}- \hat{\v})} -
e^{-i \lambda_{\infty}(-\hat{\v})}) \phi  \|_{L^{2}(\mathbb{R}^{3})} + \\ \\
\| ( e^{i \lambda_{\infty}( \frac{\mo}{mv}+ \hat{\v} )}
(e^{-im \v \cdot x}S(A_{C},V)e^{im \v \cdot x} - e^{i \int_{- \infty}^{\infty} \hat{\v}\cdot A_{C}(x+\tau  \hat{\v} ) d\tau})e^{-i \lambda_{\infty}(-\hat{\v})} ) \phi \|_{L^{2}(\mathbb{R}^{3})} + \\ \\
\|  (e^{i \lambda_{\infty}( \frac{\mo}{mv}+ \hat{\v} )}-e^{i \lambda_{\infty}(\hat{\v} )})
e^{i \int_{- \infty}^{\infty} \hat{\v}\cdot A_{C}(x+\tau  \hat{\v} ) d\tau}e^{-i \lambda_{\infty}(-\hat{\v})} \phi\|_{L^{2}(\mathbb{R}^{3})}  \leq C \frac{1}{v}\| \phi \|_{\mathcal H_{2}(\mathbb{R}^{3})}.
\end{array}
$$
The last inequality follows from (\ref{3.15}), (\ref{5.33.0}) and (\ref{5.35}) for $ A_{C} $.

\noindent {\bf 5.2 High-Velocity Estimates II. The Electric Potential}

Recall that $ \tilde{\phi} $ is defined in (\ref{5.17.0}) and that  $ H_{1} $ is given by (\ref{5.18.0}).
\begin{lemma}\label{v}
Let $ h : \mathbb{R}^{3} \to \mathbb{R} $ be a  bounded function with compact  support  contained in
$ \mathbb{R}^{3} \setminus \Lambda_{\hat{\v}} $, and let $ \phi$ be a function
in $\mathcal H_{6}(\mathbb{R}^{3})  $ with
 compact support contained in $ \Lambda_{\hat{\v}} $.
Then, for any $ l \in \mathbb{N} $ there exists  constant $ C_l $ such that following inequalities hold:

\begin{itemize}
\item[i)] $ \| h e^{-i \tau H_{1}} \tilde{\phi}  \|_{L^{2}(\mathbb{R}^{3})}
\leq C_l \frac{1}{(1+|\tau|)^{l}} \frac{1}{v^{3- \epsilon}} \| \phi \|_{\mathcal H_{6}(\mathbb{R}^{3})}
\,
\forall
\epsilon > 0$.
\end{itemize}

\begin{itemize}
\item[ii)] $ \| h \mo e^{-i \tau H_{1}} \tilde{\phi}  \|_{L^{2}(\mathbb{R}^{3})}
\leq C_l \frac{1}{(1+|\tau|)^{l}} \frac{1}{v^{2- \epsilon}} \| \phi \|_{\mathcal H_{5}(\mathbb{R}^{3})} \,
\forall
\epsilon > 0$.
\end{itemize}
\end{lemma}

\noindent {\it Proof:}
We prove i), ii) follows in a similar way. \\
\noindent Clearly,
\beq \label{5.36b}
\|  \tilde{\phi} - \phi \|_{L^{2}(\mathbb{R}^{3})} \leq \frac{1}{v^{6 \rho}} \|  \phi
\|_{\mathcal H_{6}(\mathbb{R}^{3})}, \,\hbox{\rm where}\, \rho \geq 1/2.
\ene
It follows from (\ref{5.15b}) and the properties of the support of $ h $ and $ \phi $ that

$$
\|  h e^{-i \tau H_{1}} \phi  \|_{L^{2}(\mathbb{R}^{3})} = \| h e^{-i\tau \mo \cdot \hat{\v}}
(e^{-i \tau \frac{\mo^{2}}{2mv}} - I - (-i \tau \frac{\mo^{2}}{2mv})-
\frac{1}{2} (-i \tau \frac{\mo^{2}}{2mv})^{2}  ) \phi \|.
$$
Observing that
$$
|(e^{-i \tau \frac{\mo^{2}}{2mv}} - I - (-i \tau \frac{\mo^{2}}{2mv})-
\frac{1}{2} (-i \tau \frac{\mo^{2}}{2mv})^{2} )| \leq C |\tau|^{3} \frac{\mo^{6}}{(2mv)^{3}},
$$
we obtain,

\beq \label{5.36c}
\|   h e^{-i \tau H_{1}} \phi \|_{L^{2}(\mathbb{R}^{3})} \leq C \frac{(1+|\tau|)^{3} }{(2mv)^{3}}\|  \phi \|_{\mathcal H_{6}(\mathbb{R}^{3})}.
\ene

We prove as in (\ref{5.30}) that there exist a constant $C_l$ such that,

\beq \label{5.36d}
\|  h e^{-i \tau H_{1}} \tilde{\phi} \|_{L^{2}(\mathbb{R}^{3})} \leq C_l \frac {1}{(1+ |\tau|)^{l}} \| \phi  \|_{L^{2}(\mathbb{R}^{3})}
\ene
Finally we obtain i) from (\ref{5.36b})  and  interpolating (\ref{5.36c},
\ref{5.36d}).

\bull

\noindent
We denote,

\beq
a(\hv,x):= \int_{-\infty}^\infty A(x+\tau \hv)\cdot \hv \, d\tau,
\label{5.37}
\ene
and for  $\phi_0 \in \mathcal H_6(\ere^3)$ with compact support  in $\Lambda_{\hv}$,
$$
\phi_{\hv} := e^{im\v\cdot x}\phi_0.
$$
Recall that $ \Lambda_{\hv}$ is defined in (\ref{5.14}), that $\eta$ is defined in   (\ref{5.32})
and that $\p2$ is defined in Definition \ref{def-5.3}.

\begin{theorem}\label{th-5.8} {\bf(Reconstruction Formula II)}
 Suppose that $B,V$ satisfy Assumption \ref{ass-5.1}. Let  $\Lambda_0$ be a compact subset of $\Lambda_{\hv},
 $ with $\v \in \ere \setminus \{0\}$. Then, for all $\Phi$ and all $A \in \sp2$
\beq
\begin{array}{l}
v \left( \left[S(A,V)- e^{ia(\hv,x)}\right] \phi_{\hv}, \psi_{\hv}\right)=
\left(-i e^{ia(\hv,x)}\int_{-\infty}^\infty
V(x+\tau\hv)\,d\tau \, \phi_0,\, \psi_0\right)\\\\
+
\left( -i e^{i a(\hv, x) }\int_{-\infty}^0\, \Xi_\eta (x+\tau \hv,-\infty)\,d\tau \,\phi_0,\psi_0\right)+
\left(-i \int_{0}^\infty\, \Xi (x+\tau \hv,\infty)\,d\tau \, e^{ia(\hv, x)} \phi_0,\psi_0\right)+ R (\v, \phi_0, \psi_0),
\end{array}
\label{5.38}
\ene
where,
\beq
\left|R(\v, \phi_0,\psi_0)\right| \leq C \|\phi\|_{\mathcal H_6(\ere^3)}\, \|\psi\|_{\mathcal H_6(\ere^3)}
\left\{ \begin{array}{c}\frac{1}{v^{ \min(\mu-2, \alpha -1)}},\, \, \hbox{\rm if}\, \min (\mu-3, \alpha -2)
 <0, \\ \\
\frac{|\ln v|}{v}, \,\,\hbox{\rm if}\, \min (\mu-3, \alpha -2)=0, \\ \\
\frac{1}{v},\,\,\hbox{\rm if}\, \min (\mu-3, \alpha -2) > 0, \\ \\
\end{array}\right.
\label{5.39}
\ene
for some constant $C$ and all $\phi_0,\psi_0 \in \mathcal H_6(\ere^3)$ with compact support in $\Lambda_0$.
\end{theorem}

\noindent{\it Proof:} We first prove the theorem in  the Coulomb gauge $\ac$.  Note that,
\beq
v \left( \left[S(A,V)- e^{ia}\right] \phi_{\hv}, \psi_{\hv}\right)=
v \left( e^{-iL_{\ac,\hv}(-\infty)} \phi_0,  \mathcal R_+ \psi_0\right)+ v\left(  \mathcal R_- \phi_0,
e^{-i L_{\ac,\hv}(\infty) } \psi_0 \right)+
v\left(  \mathcal R_- \phi_0, \mathcal R_+ \psi_0 \right),
\label{5.40}
\ene
where,

$$
\mathcal R_\pm:= e^{-im\v\cdot x}W_\pm(\ac, V) e^{im\v\cdot x}  - e^{-i L_{\ac, \hv}(\pm \infty)}.
$$
By Lemma \ref{lemm-5.5}
\beq
v \left| \left(  \mathcal R_- \phi_0, \mathcal R_+ \psi_0 \right)\right|
\leq C \frac{1}{v} \|\phi\|_{\mathcal H_6(\ere^3)}\,
\|\psi\|_{\mathcal H_6(\ere^3)}.
\label{5.41}
\ene
We prove below that,
\beq
v \left( e^{-iL_{\ac,\hv}(-\infty)} \phi_0,  R_+ \psi_0\right)=
 \left(-i  \int_{0}^\infty\, (\Xi_\eta (x+\tau \hv,\infty)+ \chi V(x+\tau \hv))\,d \tau \,
e^{ia}\phi_0,\psi_0\right)
+ R_+ (\v, \phi_0, \psi_0),
\label{5.42}
\ene

\beq
v \left(\mathcal R_- \phi_0,  e^{-iL_{\ac,\hv}(\infty)}   \psi_0\right)=
 \left(-i e^{ia} \int_{-\infty}^0\, (\Xi_\eta (x+\tau \hv,-\infty)+ \chi V(x+\tau \hv))\,d \tau \,\phi_0,
\psi_0\right)
+ R_- (\v, \phi_0, \psi_0),
\label{5.43}
\ene

where $ R_\pm$ satisfy (\ref{5.39}). Note that (\ref{5.43}) follows from (\ref{5.42}) by time inversion and
charge conjugation in the magnetic potential, i.e., by taking complex conjugates and changing $\ac$ to
$-\ac$. It can also be proved as in the proof of (\ref{5.42}) that we give below in seven steps.

We use the notation of the proof of Lemma \ref{lemm-5.5}. For simplicity we denote by $ O(r)$ a term that
satisfies

$$
\left|O(r)\right| \leq C   \|\phi\|_{\mathcal H_6(\ere^3)}\, \|\psi\|_{\mathcal H_6(\ere^3)} \,r.
$$

\noindent  Step 1

\beq
\begin{array}{l}
v \left( e^{-iL_{\ac,\hv}(-\infty)} \phi_0,  R_+ \psi_0\right)=
 \left( e^{-iL_{\ac,\hv}(-\infty)} \phi_0, \right. \\\\
 \left. \lim_{t \rightarrow \infty}\int_0^t\, d\tau e^{i\tau H_2} i
e^{-iL_{\ac,\hv}(t-\tau)} \, [\Xi_b (x,t-\tau )+ \chi V(x)] e^{-i\tau H_1} \ts \right)+
O(1/v).
\end{array}
\label{5.44}
\ene
Equation (\ref{5.44}) follows from (\ref{5.18}), (\ref{5.20}), (\ref{5.23}) and the following formula that is easily obtained from Lemma \ref{v}
\beq
 \left\| T_2 + T_3\right\|_{L^2(\ere^3)} \leq C_l \frac{ \| \phi \|_{\mathcal H_6(\ere^3)}
   }{v^{3 - \epsilon } (1+ |\tau|)^{l}}, \, \forall \epsilon > 0, l=1,2,\cdots,
\label{5.45}
\ene
that improves (\ref{5.30}, \ref{5.31}).

\noindent Step 2
\beq
\begin{array}{c}
\lim_{t \rightarrow \infty}\int_0^t\, d\tau e^{i\tau H_2} i
e^{-iL_{\ac,\hv}(t-\tau)} \, [\Xi_b (x,t-\tau )+ \chi V(x)] e^{-i\tau H_1} \ts = \\\\
\lim_{t \rightarrow \infty}\int_0^t\, d\tau e^{i\tau H_2} i
e^{-iL_{\ac,\hv}(t-\tau)} \, [\Xi_\eta (x,t-\tau )+ \chi V(x)] e^{-i\tau H_1} \ts.
\end{array}
\label{5.46}
\ene
This follows from Lebesque's dominated convergence theorem and as
$$
\lim_{t\rightarrow \infty} \left\| \left(\Xi_b (x,t-\tau )-\Xi_\eta (x,t-\tau )\right)
e^{-i\tau H_1}\ts\right\|_{L^2(\ere^3)}=0,
$$
and, moreover,

$$
\left\| \left(\Xi_b (x,t-\tau )-\Xi_\eta (x,t-\tau )\right) e^{-it H_1} \ts
\right\|_{L^2(\ere^3)}\leq h(\tau), \,\hbox{\rm for some}
\, h(\tau) \in L^1(0,\infty).
$$
This estimate is proven as in the proof of Lemma \ref{lemm-5.5}, using Lemma \ref{lemm-5.4}.

\noindent Step 3
\beq
\begin{array}{c}
v \left( e^{-iL_{\ac,\hv}(-\infty)} \phi_0,  R_+ \psi_0\right)= \int_0^t\, d\tau
 \left( e^{-iL_{\ac,\hv}(-\infty)} \phi_0, \right.
\\ \\
 \left. e^{i\tau H_2} i
e^{-iL_{\ac,\hv}(\infty)} \, [\Xi_\eta (x,\infty )+ \chi V(x)] e^{-i\tau H_1} \ts \right)+
O(1/v)+ 0\left(1/ (1+|t|)^{\min (\mu-2, \alpha-1)}\right).
\end{array}
\label{5.47}
\ene
This follows from Steps 1 and 2, and from the following argument. As in the proof of Lemma \ref{lemm-5.5}
we prove that
\beq
\left\| [\Xi_\eta (x,t-\tau )+ \chi V(x)] e^{-i\tau H_1} \ts\right\|_{L^2(\ere^3)}\leq C
\left(1/ (1+|\tau|)^{\min (\mu-1, \alpha)}\right) \|\psi\|_{\mathcal H_2(\ere^3)}.
\label{5.48}
\ene
Then by Fatou's lemma
\beq
\left\| [\Xi_\eta (x,\infty )+ \chi V(x)] e^{-i\tau H_1} \ts\right\|_{L^2(\ere^3)}\leq C
\left(1/ (1+|\tau|)^{\min (\mu-1, \alpha)}\right) \|\psi\|_{\mathcal H_2(\ere^3)}.
\label{5.49}
\ene

Hence, by Lebesque's dominated convergence theorem,
$$
\begin{array}{c}
\lim_{t \rightarrow \infty}\int_0^t\, d\tau e^{i\tau H_2} i
e^{-iL_{\ac,\hv}(t-\tau)} \, [\Xi_\eta (x,t-\tau )+ \chi V(x)] e^{-i\tau H_1} \ts \\ \\
=\int_0^\infty\, d\tau e^{i\tau H_2} i
e^{-iL_{\ac,\hv}(\infty)} \, [\Xi_\eta (x,\infty )+ \chi V(x)] e^{-i\tau H_1} \ts,
\end{array}
$$
where the limit is on the strong topology of $L^2(\ere^3)$. We complete  the proof of (\ref{5.47})
using (\ref{5.49})

We now estimate the integrand in (\ref{5.47}).

\noindent Step 4

\beq
\begin{array}{c}
\left( e^{-iL_{\ac,\hv}(-\infty)} \phi_0, e^{i\tau H_2}
e^{-iL_{\ac,\hv}(\infty)} \,i [\Xi_\eta (x,\infty )+ \chi V(x)] e^{-i\tau H_1} \ts \right)= \\\\
 \left(  e^{ i(L_{\ac,\hv}(\tau)-L_{\ac,\hv}(-\infty))} \phi_0, e^{i\tau H_1}
e^{-iL_{\ac,\hv}(\infty)} \,i [\Xi_\eta (x,\infty )+ \chi V(x)] e^{-i\tau H_1} \ts \right)+
\frac{1}{v} \, \, O( \frac{1}{(1+|\tau|)^{\min (\mu-2, \alpha-1)}} ).
\end{array}
\label{5.50}
\ene
Denote by $\chi_\Lambda$ the characteristic function of $\Lambda$. Then,
$$
\begin{array}{c}
\left( e^{-iL_{\ac,\hv}(-\infty)} \phi_0, e^{i\tau H_2}
e^{-iL_{\ac,\hv}(\infty)} \,i [\Xi_\eta (x,\infty )+ \chi V(x)] e^{-i\tau H_1} \ts \right)= \\\\
\left( e^{i\tau H_1}\chi_\Lambda e^{-i\tau H_2}  e^{-iL_{\ac,\hv}(-\infty)} \phi_0, e^{i\tau H_1}
e^{-iL_{\ac,\hv}(\infty)} \,i [\Xi_\eta (x,\infty )+ \chi V(x)] e^{-i\tau H_1} \ts \right).
\end{array}
$$
Hence, (\ref{5.50}) will be proved if we can replace  $ e^{i\tau H_1}\chi_\Lambda e^{-i\tau H_2}$
by $\chi e^{i L_{\ac,\hv}(\tau)}$ adding the error term. But, this follows from (\ref{5.49}) and the estimate,
\beq
\left\|\left(e^{i\tau H_1}\chi_\Lambda e^{-i \tau H_2} -\chi e^{i L_{\ac,\hv}(\tau)}\right)
e^{-i L_{\ac,\hv}(-\infty)}\phi_0
\right\| \leq C \frac{1+|\tau|}{v} \|\phi_0\|_{\mathcal H_2(\ere^3)},
\label{5.51}
\ene
that we prove below.

We designate,
$$
\varphi_\tau:= e^{i(L_{\ac,\hv}(\tau)-L_{\ac,\hv}(-\infty))}\phi_0.
$$
We have that,
\beq
\begin{array}{c}
\left(e^{i\tau H_1}\chi_\Lambda e^{-i \tau H_2} -\chi e^{i L_{\ac,\hv}(\tau)}\right)
 e^{-i L_{\ac,\hv}(-\infty)}\phi_0 = \\\\
 e^{i\tau H_1}\chi_\Lambda e^{-i\tau H_2}\left( e^{-iL_{\ac,\hv}(\tau)}-
e^{i\tau H_2}\chi e^{-i \tau H_1}\right) \varphi_\tau + \left( e^{i\tau H_1} \chi e^{-i\tau H_1}-\chi \right)
\varphi_\tau.
\end{array}
\label{5.52}
\ene

By (\ref{5.33})
\beq
\|\varphi_\tau\|_{\mathcal H_2(\ere^3)}\leq C  \|\phi_0\|_{\mathcal H_2(\ere^3)} .
\label{5.53}
\ene
Hence, using
$$
\left| e^{-i\tau (p+m \v)^2 / 2mv}-e^{-i\tau (p\cdot \hv+ v^2/2mv)}\right| \leq C \frac{ |\tau| p^2}{2mv},
$$
we prove that,
\beq
\left\|\left( e^{-i\tau H_1 }- e^{-i\tau (\mo \cdot \hv + v^2/2mv)}\right) \varphi_\tau \right\|_{L^2(\ere^3)} \leq C
\frac{|\tau|}{v} \|\phi_0\|_{\mathcal H_2(\ere^3)},
\label{5.54}
\ene
and since $\chi -1\equiv 0$ on the support of $e^{-i\tau (\mo\cdot \hv+ v^2/2mv)} \varphi_\tau$,
\beq
\begin{array}{c}
\left\| \left(e^{i\tau H_1}\chi e^{-i\tau H_1}-\chi\right) \varphi_\tau\right\|_{L^2(\ere^3)}=
\left\| e^{i\tau H_1}(\chi - 1)(e^{-i \tau H_1}- e^{-i\tau (\mo\cdot \hv+ v^2/2mv)}) \varphi_\tau
\right\|_{L^2(\ere^3)} \\\\
\leq C\frac{|\tau|}{v} \|\phi_0\|_{\mathcal H_2(\ere^3)}.
\end{array}
\label{5.55}
\ene
Then, (\ref{5.51}) follows from (\ref{5.32b},\ref{5.52}, \ref{5.53}, \ref{5.55}).

\noindent Step 5.

\noindent We now replace $e^{\pm i\tau H_1}$ by $ e^{\pm i(\tau \mo \cdot \hv+v^2/ 2mv )
 }$. We will prove that,
\beq
\begin{array}{c}
\left(  e^{ i(L_{\ac,\hv}(\tau)-L_{\ac,\hv}(-\infty))} \phi_0, e^{i\tau H_1}
e^{-iL_{\ac,\hv}(\infty)} \,i [\Xi_\eta (x,\infty )+ \chi V(x)] e^{-i\tau H_1} \ts \right)=
 \left(  e^{ i(L_{\ac,\hv}(\tau)-L_{\ac,\hv}(-\infty))} \phi_0, \right.\\\\
 \left.e^{i\tau \mo \cdot \hv}
 e^{-iL_{\ac,\hv}(\infty)} \,i [\Xi_\eta (x,\infty )+ \chi V(x)] e^{-i\tau \mo \cdot \hv} \ts \right)
 + \frac{1}{v} O\left(\frac{1}{(1+|\tau |)^{\min (\mu-2,\alpha-1)}} \right)= \\\\
\left(   \phi_0,  e^{-i a(\hv,x)} \,i [\Xi_\eta (x+\tau \hv,\infty )+ \chi V(x+\tau \hv)] \ts \right)
 + \frac{1}{v} O\left(\frac{1}{(1+|\tau |)^{\min (\mu-2,\alpha-1)}} \right), \tau \geq 0.
\end{array}
 \label{5.56}
 \ene
 Recall that $\varphi_\tau$ is defined below (\ref{5.51}). By (\ref{5.49}) and (\ref{5.54}),
\beq
\begin{array}{c}
\left( e^{-i\tau H_1} \varphi_\tau,
e^{-iL_{\ac,\hv}(\infty)} \,i [\Xi_\eta (x,\infty )+ \chi V(x)] e^{-i\tau H_1} \ts \right)=
 \left( e^{-(i\tau \mo \cdot \hv+ v^2/2mv)} \varphi_\tau,\right.\\\\ \left.
 e^{-iL_{\ac,\hv}(\infty)} \,i [\Xi_\eta (x,\infty )+ \chi V(x)] e^{-i\tau H_1} \ts \right)
 + \frac{1}{v} O\left(\frac{1}{(1+|\tau|)^{\min (\mu-2,\alpha-1)}} \right).
 \end{array}
 \label{5.57}
 \ene
 The first equality in (\ref{5.56}) follows from (\ref{5.54}) and as,
 \beq
 \begin{array}{c}
\|[\Xi_\eta (x,\infty )+ \chi V(x)]e^{iL_{\ac,\hv}(\infty)}
e^{-i(\tau \mo \cdot \hv+ v^2/2mv)} \varphi_\tau \|_{L^2(\ere^3)}
 \leq  C\frac{1}{(1+|\tau|)^{\min (\mu-1,\alpha)}}
 \|\phi_0\|_{\mathcal H_2(\ere^3)}, \tau >0,
 \end{array}
 \label{5.58}
 \ene
 because $\phi_0$ has compact support, $e^{-i \tau \mo \cdot \hat{\v}}$ is just a translation and the decay properties of $ V(x) $ and $ \Xi_\eta (x,\infty ) $ (in the direction $  \hat{\v} $). The second equality is immediate.

By (\ref{5.47}, \ref{5.50}, \ref{5.56})

$$
\nonumber
\begin{array}{c}
v \left( e^{-iL_{\ac,\hv}(-\infty)} \phi_0,  R_+ \psi_0\right)= \int_0^t\, d\tau
\left(   \phi_0,  e^{-i a(\hv,x)} \,i [\Xi_\eta (x+\tau \hv,\infty )+ \chi V(x+\tau \hv)] \ts \right)
+
\end{array}
$$
\beq
O\left(1/v\right)\  + O\left(1/  (1+|t|)^{\min (\mu-2,\alpha-1)}\right)+\left\{
\begin{array}{cc} \frac{1}{v} O\left(\ln (1+ |t|)\right),& \hbox{\rm if}\, \min (\mu-2,\alpha-1)=1,\\
\frac{1}{v} O \left(\frac{1}{(1+|t|)^{\min (\mu-3,\alpha-2,0)}}\right),& \hbox{\rm otherwise}.
\end{array}\right.
 \label{5.59}
 \ene

\noindent Step 6

\noindent We now prove that,
\beq
\begin{array}{c}
 \int_0^t\, d\tau
\left(   \phi_0,  e^{-i a(\hv,x)} \,i [\Xi_\eta (x+\tau \hv,\infty )+ \chi V(x+\tau \hv)] \ts \right)\\\\
-  \int_0^\infty\, d\tau
\left(   \phi_0,  e^{-i a(\hv,x)} \,i [\Xi_\eta (x+\tau \hv,\infty )+ \chi V(x+\tau \hv)] \psi_0 \right)
= O(1/v)+ O \left(\frac{1}{ (1+|t|)^{\min (\mu-2,\alpha-1)}}\right), t >0.
\end{array}
\label{5.60}
\ene
As $\phi_0$ has compact support,
\beq
\left\|[\Xi_\eta (x+\tau \hv,\infty )+ \chi V(x+\tau \hv)]
 e^{ia(\hv,x)}\phi_0 \right\|_{L^2(\ere^3)} \leq  C\frac{1}{(1+|\tau|)^{\min (\mu-1,\alpha)}}
 \|\phi_0\|_{\mathcal H_2(\ere^3)}, \tau >0.
 \label{5.61}
 \ene
 Equations (\ref{5.17b}) and (\ref{5.61}) prove (\ref{5.60}).

By (\ref{5.59}, \ref{5.60})

$$
\nonumber
\begin{array}{c}
v \left( e^{-iL_{\ac,\hv}(-\infty)} \phi_0,  R_+ \psi_0\right)= \int_0^\infty\, d\tau
\left(   \phi_0,  e^{-i a(\hv,x)} \,i [\Xi_\eta (x+\tau \hv,\infty )+ \chi V(x+\tau \hv)] \psi_{0} \right)
+
\end{array}
$$
\beq
O\left(1/v\right)\  + O\left(1/  (1+|t|)^{\min (\mu-2,\alpha-1)}\right)+\left\{
\begin{array}{cc} \frac{1}{v}O\left(\ln (1+ |t|\right)),& \hbox{\rm if}\, \min (\mu-2,\alpha-1)=1,\\
\frac{1}{v}O \left(\frac{1}{ (1+|t|)^{\min (\mu-3,\alpha-2,0)}}\right),& \hbox{\rm otherwise}.
\end{array}\right.
 \label{5.62}
 \ene
Finally, taking $t=v$ we obtain (\ref{5.42}) in the Coulomb gauge, and then, (\ref{5.38}) is proven for $\ac$.

 Suppose that $A\in \sp2$. By (\ref{5.11}) $ S(A, V) = S(A_{C}, V)$. As $ \lambda_{\infty}  $ is constant,
$ e^{i\int_{- \infty}^{\infty}A(x + \tau \hv) \cdot \hv d \tau}  = e^{i\int_{- \infty}^{\infty}A_{C}
(x + \tau \hv) \cdot \hv d \tau}$, and it follows that (\ref{5.38}) holds for $A \in \p2$.

\section{Reconstruction of the Magnetic Field and the Electric Potential Outside the Obstacle }
\sss
 In this section we  obtain  a method for the  unique reconstruction of the magnetic field and the
 electric potential outside the obstacle, $K$, from the high-velocity limit of the scattering operator.
 The method is given in  the proof of Theorem \ref{th-6.3} and is summarized in Remark \ref{rem-6.4}.

\begin{definition} \label{def-6.1}
We denote by $\rec$ the set of points $x \in \Lambda$ such that for some two-dimensional  plane $P_x$
we have  that $x+ P_x \subset \Lambda$.
\end{definition}
Note that if $K$ is convex  $\rec=\Lambda$.
\begin{lemma}\label{lemm-6.2}
For every $A\in \p2$ and every unit vector, $\hv$, in $\ere^3,$ we have that
\beq
\nabla \int_{-\infty}^\infty \hv\cdot A(x+\tau \hv) \, d\tau= \int_{-\infty}^{\infty} \hv\times B(x+\tau\hv)
\, d\tau,
\label{6.1}
\ene
in distribution sense in $\Lambda_{\hv}$ .
\end{lemma} \noindent {\it Proof:}

The following identity holds in distribution sense in $\Lambda_{\hv}$ ( this is just the
 triple vector product formula),
\beq
\hv \times (\nabla \times A)= \nabla(\hv \cdot A)- (\hv\cdot\nabla)A.
\label{6.2}
\ene
Then, for every $ \phi \in C^\infty_0(\Lambda_{\hv})$
$$
\begin{array}{l}
\int_{-\infty}^{\infty} \hv\times B(x+\tau\hv)
\, d\tau \, [\phi]= \int_{\ere^3}\, dx \int_{-\infty}^{\infty} \hv\times B(x+\tau\hv)
\, d\tau \, \phi(x) =\int_{\ere^3}\, dx   \lim_{r\rightarrow \infty}\int_{-r}^r \,d\tau\,
 \hv\times B(x)
\, \phi(x- \tau\hv)=\\\\
\int_{\ere^3} dx \lim_{r\rightarrow \infty}\int_{-r}^r \left( -\hv \cdot A(x) (\nabla \phi)(x-\tau \hv)
+ A(x) \left(\hv\cdot \nabla \phi\right)(x-\tau \hv)\right)=\\\\
\left( \nabla \int_{-\infty}^\infty \hv\cdot A(x+\tau \hv)\, d\tau \right)\left[ \phi\right]
+\lim_{r\rightarrow \infty} \int_{\ere^3} A(x)( \phi(x-r\hv)- \phi(x+r\hv))=
\left( \nabla \int_{-\infty}^\infty \hv\cdot A(x+\tau \hv)\, d\tau \right)\left[ \phi\right],
\end{array}
$$
where in the last equality we used the decay of $A$ and the fact that $\phi$ has compact support.
\begin{theorem} \label{th-6.3}(Reconstruction of the Magnetic Field and the Electric Potential)
Suppose that $B,V$ satisfy Assumption \ref{ass-5.1}. Then, for any flux, $\Phi$, and all $ A\in \p2$, the
high-velocity limits of  $S(A,V$)  in (\ref{5.35}) known  for all $\Lambda_0$, all unit vectors $\hv$
and all $\phi_0\in \mathcal H_2(\ere^3)$ with support
$\phi_0 \subset \Lambda_0$, uniquely determine $B(x)$ for almost every $ x \in \rec$.
Furthermore,  for any flux, $\Phi$, and all $ A\in \sp2$, the high-velocity limits of $S(A,V)$ in (\ref{5.38})
known for all $\Lambda_0$, all unit vectors $\hv$
and all $\phi_0, \psi_0\in \mathcal H_6(\ere^3)$ with support
$\phi_0$, support $ \psi_0 \subset \Lambda_0$, uniquely determine $V(x)$ for almost every $ x \in \rec$.
\end{theorem}

\noindent{\it Proof:} We proceed as in the proof of Theorem 1.1 of \cite{ew} (see also the proof of
Theorem 1.4 \cite{we1}) with the modifications
that are necessary to take the
obstacle into account and to reconstruct the magnetic field.

Let us fix a $x_0 \in \rec$. For each $j=1,2,3$ we take vectors unit vectors $\hu_j,\hv_j$ and $\varepsilon >0$
such that the following conditions are satisfied.
\begin{enumerate}
\item
$$\nonumber
\hu_j\cdot\hv_i=0, \, i,j \in \{ 1,2,3 \}.
$$
\item
The unit vectors
$$\nonumber
\hn_j:= \hu_j\times \hv_j, j=1,2,3,
$$
 are linearly independent.
\item
$$\nonumber
B_\varepsilon^{\ere^3}(x_0)+ p(\hu_j, \hv_j) \subset \Lambda,j=1,2,3,
$$
where $p(\hu_j,\hv_j)$ is the two-dimensional plane generated by $\hu_j,\hv_j$.
\end{enumerate}
For any $z=(z_1,z_2)\in \ere^2$ we define,
\beq
\phi_j(z):= e^{-i(z_1\hu_j+ z_2\hv_j) \cdot \mo} \phi_0, \, \psi_j(z):= e^{-i(z_1\hu_j+ z_2\hv_j)\cdot \mo} \psi_0, j=1,2,3, \phi_0
,\psi_0 \in C^\infty_0\left(B_\varepsilon^{\ere^3}(x_0)\right).
\label{6.3}
\ene
From the limit (\ref{5.35}) we uniquely reconstruct
$$
e^{i \int_{-\infty}^\infty \hv\cdot A(x+\tau\hv)\, d\tau}
$$
for all $x \in \Lambda_{\hv}$ and then, we reconstruct $ \int_{-\infty}^\infty \hv\cdot A(x+\tau\hv)\, d\tau+
2\pi n(x,\hv)$ with $n(x,\hv)$ an integer that is locally constant. By Lemma \ref{6.2} we also
reconstruct uniquely
\beq
\int_{-\infty}^{\infty} \hv\times B(x+\tau\hv)
\, d\tau
\label{6.4}
\ene
for a.e. $x \in \Lambda_{\hv}$.

Take now $\hv \in p(\hu_j,\hv_j)$. Hence, we uniquely reconstruct

\beq
 \int_{-\infty}^{\infty}  \hn_j\cdot B(x+\tau\hv)
\, d\tau=- \hn_j\cdot \left(\hv \times \int_{-\infty}^{\infty} \hv\times B(x+\tau\hv)
\, d\tau\right),
\label{6.5}
\ene
for a.e. $x \in\Lambda_{\hv}$.
We used the triple vector product formula, $ a\times (b\times c)= (a\cdot c) b- (a\cdot b)c$.
We now define $F_j: \ere^2 \rightarrow \CE$,
$$
F_j(z):= \left(    \hn_j\cdot B(x) \phi_j(z), \psi_j(z)\right).
$$
$F_j$ is continuous and
$$
\left| F_j(z)\right| \leq C(1+|z|)^{-\mu}, j=1,2,3.
$$
Moreover, we uniquely reconstruct from (\ref{5.35}) the Radon transforms,
$$
\tilde{F_j}(\hw;z):= \int_{-\infty}^\infty F_j(z+\tau \hw)d\tau =
 \left( \int_{-\infty}^{\infty}\hn_j\cdot B(x+\tau ( \hw_{1}\hu_{j} + \hw_{2}\hv_{j}   ))
\, d\tau \phi_j(z), \psi_j(z)\right),
$$
where $ z \, \in \mathbb{R}^{2}$ and $ \hw := (\hw_1, \hw_2) \,   \in \mathbb{R}^{2} $ has modulus one.

Inverting this Radon transform (see Theorem 2.17 of \cite{hel}, \cite{hel2}, \cite{na}) we uniquely
reconstruct $F_j(z)$
and in particular $F_j(0)= \left(  \hn_j\cdot B\phi_0, \psi_0\right)$ and hence, we uniquely reconstruct
 $\hn_j\cdot B(x),\, j=1,2,3$ for a.e. $x \in B^{\ere^3}_\varepsilon(x_0)$ and as the $\hn_j$ are linearly
 independent we uniquely reconstruct
 $B(x)$ for a.e. $x \in B^{\ere^3}_\varepsilon(x_0)$. Since $x_0\in \rec$ is arbitrary we uniquely reconstruct
 $B(x)$ for a.e. $x \in \rec$.

We now uniquely reconstruct $V$. Take any $x_0 \in \rec$. Let $\hu, \hat{\mathbf w}$ be orthonormal
vectors such that
$B_\varepsilon^{\ere^3}(x_0)+ p(\hu,\hat{\mathbf w})\subset  \Lambda_{\hv}$.
We define,
$$
\phi(z):= e^{-i(z_1\hu+ z_2\hat{\mathbf w}) \cdot \mo } \phi_0,\, \psi(z):= e^{-i(z_1\hu+ z_2\hat{\mathbf w}) \cdot \mo} \psi_0, \, \, \phi_0
, \psi_0 \in C^\infty_0\left(B_\varepsilon^{\ere^3}(x_0)\right),
$$

and the function $F: \ere^2 \rightarrow
\CE$,

$$
F(z):= \left(V(x) \phi(z), \psi(z)\right).
$$
$F$ is continuous and
$$
\left| F(z)\right| \leq C(1+|z|)^{-\alpha}.
$$
Moreover, since $B$ is already known in $ \rec $, we uniquely reconstruct
from (\ref{5.38}) the Radon transforms,
$$
\tilde{F}(\hat{\mathbf y};z):= \int_{-\infty}^\infty F(z+\tau \hat{\mathbf y})d\tau =
 \left( \int_{-\infty}^{\infty}V(x+\tau ( \hat{\mathbf y}_{1}\hu + \hat{\mathbf y}_{2}\hw   ))
\, d\tau \phi(z), \psi(z) \right),
$$
where $ z \, \in \mathbb{R}^{2}$ and $ \hat{\mathbf y} := (\hat{\mathbf y}_1, \hat{\mathbf y}_2) \,
\in \mathbb{R}^{2} $ has modulus one.

As above inverting these Radon transforms we uniquely reconstruct $F(z)$, and in particular
$F(0)= ( V \phi_0, \psi_0)$ what uniquely determines $V(x)$ for a.e. $x \in \ B_\varepsilon^{\ere^3}(x_0)$.
Since $x_0 \in \rec$ is arbitrary, $V(x)$ is uniquely reconstructed for a.e. $x \in \rec$.

\begin{remark} \label{rem-6.4} {\rm
Let us summarize the reconstruction  method given by Theorem {\ref{th-6.3}}.
From the high-velocity limit (\ref{5.35}) we uniquely reconstruct,

\beq
e^{i \int_{-\infty}^\infty \hv\cdot A(x+\tau\hv)\, d\tau},
\label{6.6}
\ene
and from this we uniquely reconstruct,

\beq
\int_{-\infty}^{\infty} \hv\times B(x+\tau\hv)
\, d\tau, x \in \Lambda_{\hv},
\label{6.7}
\ene
what gives us the Radon transform

\beq
\tilde{F_j}(\hw;z):= \int_{-\infty}^\infty F_j(z+\tau \hw)d\tau =
 \left( \int_{-\infty}^{\infty}\hn_j\cdot B(x+\tau ( \hw_{1}\hu_{j} + \hw_{2}\hv_{j}   ))
\, d\tau \phi_j(z), \psi_j(z)\right),
\ene \label{6.8}
where $ z \, \in \mathbb{R}^{2}$ and $ \hw := (\hw_1, \hw_2) \,   \in \mathbb{R}^{2} $ has modulus one.

Inverting this Radon transform  we uniquely reconstruct $F_j(z)$
and in particular $F_j(0)= \left(  \hn_j\cdot B\phi_0, \psi_0\right)$ and hence, we uniquely reconstruct
 $\hn_j\cdot B(x), j=1,2,3$ for a.e. $x \in B^{\ere^3}_\varepsilon(x_0)$ and as the $\hn_j$ are linearly
 independent we uniquely reconstruct
 $B(x)$ for a.e. $x \in B^{\ere^3}_\varepsilon(x_0)$. Since $x_0\in \rec$ is arbitrary we uniquely reconstruct
 $B(x)$ for a.e. $x \in \rec$. Note that to reconstruct $B $ almost everywhere in a neighborhood of a point $ x_{0} $  we only need the high-velocity
 limit of the scattering operator a neighborhood of three two-dimensional planes. For the inversion of the
 Radon transform see Theorem 2.17 of \cite{hel} and \cite{hel2}, \cite{na}.

 Remember  that given any $A \in \p2$ we can always find an $A \in \pot$
with the same scattering operator. We can take, for example,  $A_\Phi$. See equation (\ref{5.12}). Then, there
is no loss of generality taking $A \in \pot$. Note that (\ref{6.6}) is not a gauge invariant quantity.
If $\tilde{A}, A \in \pot$ and $\tilde{A}=A+d\lambda$, then,
$$
\int_{-\infty}^\infty \hv\cdot \tilde{A}(x+\tau \hv)d\tau = \int_{-\infty}^\infty \hv\cdot A(x+\tau \hv)d\tau
+\lambda_\infty(\hv)- \lambda_\infty(-\hv).
$$
We can, however, reconstruct (\ref{6.7}) from the gauge invariant quantity,

$$
\mathcal R(x,y):= e^{i \int_{-\infty}^\infty \hv\cdot [A(x+\tau\hv)- A(y+\tau\hv)]\, d\tau},
x,y \in \Lambda_{\hv}.
$$
We have that,
$$
\frac{1}{i}\,\, \overline{\mathcal R(x,y)} \,\,\nabla_x \mathcal R(x,y)
=\nabla_x \int_{-\infty}^\infty \hv\cdot A(x+\tau\hv)\, d\tau =\int_{-\infty}^{\infty}
\hv\times B(x+\tau\hv)\, d\tau, x \in \Lambda_{\hv}.
$$

We now  uniquely reconstruct $V$. Since $B$ is already known in $\Lambda_{\hbox{\rm rec}}$, for
any $\hv \in p(\hu,\hat{\mathbf w}) $
we uniquely reconstruct from (\ref{5.38}) the Radon transforms,
$$
\tilde{F}(\hat{\mathbf y};z):= \int_{-\infty}^\infty F(z+\tau \hat{\mathbf y})d\tau =
 \left( \int_{-\infty}^{\infty}V(x+\tau ( \hat{\mathbf y}_{1}\hu + \hat{\mathbf y}_{2}\hw   ))
\, d\tau \phi(z), \psi(z) \right),
$$
where $ z \, \in \mathbb{R}^{2}$ and $ \hat{\mathbf y} := (\hat{\mathbf y}_1, \hat{\mathbf y}_2) \,
\in \mathbb{R}^{2} $ has modulus one.

As above inverting these Radon transforms we uniquely reconstruct $F(z)$, and in particular
$F(0)= ( V \phi_0, \psi_0)$ what uniquely determines $V(x)$ for a.e. $x \in \ B_\varepsilon^{\ere^3}(x_0)$.
Since $x_0 \in \rec$ is arbitrary, $V(x)$ is uniquely reconstructed for a.e. $x \in \rec$.
}
\end{remark}

\section{The Aharonov-Bohm Effect} \sss
In this section we assume that $B\equiv 0$, i.e., that there is no magnetic field in $\Lambda$.
On the contrary, the electric potential, $V$, is not assumed to be zero. In other words, we will analyze the
Aharonov-Bohm effect in the presence of an electric potential. As we will show, for high-velocities the
electric potential gives  a lower-order contribution that plays no role in the Aharonov-Bohm effect. However,
it could  be of interest  to allow for a non-trivial
electric potential  from the experimental point of view.

For any $x \in \ere^3$ and any unit vector $\hv \in \ese^2$ we denote
$$
L(x,\hv):= x+\ere \hv,
$$
and we give to $L(x,\hv)$ the orientation of $\hv$. Suppose that $x,y \in \ere^3$, $\hv, \hw \in \ese^2$
satisfy $ \hv\cdot\hw \geq 0$ and that
$$
L(x,\hv) \cup L(y,\hw) \subset \Lambda.
$$
Take $ \rho >0$ so large that
$$
\hbox{\rm convex}\, \left( (x+ (-\infty, -\rho] \hv ) \cup  (y+ (-\infty, -\rho] \hw )\right)
\cup \,\hbox{\rm convex}\, \left( (x+ [\rho, \infty) \hv ) \cup  (y+ [\rho, \infty, ) \hw )\right)
\subset {B^{\ere^3}_r(0)}^c,
$$
where $ K \subset B^{\ere^3}_r(0)$, ${B^{\ere^3}_r(0)}^c$ is the complement of  $B^{\ere^3}_r(0)$ and
 the symbol convex$(\cdot)$ denotes the convex hull of the indicated set.

We denote by $\gamma(x,y,\hv,\hw)$ the continuous, simple, oriented  and closed curve
with sides, $ x+[-\rho, \rho]\hv$, oriented in the direction of $\hv$,
 $ y+[-\rho, \rho]\hw$, oriented in the direction of $-\hw$ and the oriented straight lines that join the
 points $x+\rho \hv$ with $y+ \rho\hw$ and $y-\rho \hw$ and $x-\rho \hv$.

 Suppose that $A$ is short-range (see Definition  \ref{def-5.3}). For example, we can take $A=\ac$.
 We denote  $x_{\perp,\hv}:= x- (x,\hv) \hv$.
 It
 follows from Stoke's theorem that   if $|x_{\perp,\hv}| \geq r$,
\beq
\int_{-\infty}^\infty\hv\cdot A(x+\tau \hv)\, d\tau= \int_{-\infty}^\infty\hv\cdot A(x_\perp+\tau
 \hv)\, d\tau= \lim_{s \rightarrow \infty}  \int_{-\infty}^\infty\hv\cdot A(sx_\perp+\tau
 \hv)\, d\tau=0.
\label{7.3b}
\ene
By Stoke's theorem and arguing as in the proof of (\ref{7.3b}) we prove that  for short-range $A$

\beq
\int_{\gamma (x,y,\hv,\hw)}A =
\int_{L(x,\hv)}A- \int_{L(y,\hw)}A.
\label{7.1}
\ene

Take any $ z\in \ere^3$ such that, $|(x+z)_{\perp,\hv}|\geq r, |(y+z)_{\perp,\hw}| \geq r$. By Stoke's theorem
and (\ref{7.3b}),
$$
\int_{L(x+z,\hv)} A= \int_{L(y+z,\hw)}A=0.
$$

Then, adding zero we write (\ref{7.1}) as

\beq
\int_{\gamma (x,y,\hv,\hw)}A =
\left(\int_{L(x,\hv)}A- \int_{L(x+z,\hv)}A \right)-\left( \int_{L(y,\hw)}A-\int_{L(y+z,\hw)}A\right).
\label{7.2}
\ene
The point is that the left- and right- hand sides of (\ref{7.2}) are gauge invariant, and in consequence
(\ref{7.2}) holds for any $A \in \0p2$.

It follows that from the high-velocity limit (\ref{5.35}) we can reconstruct $\int_{\gamma (x,y,\hv,\hw)}A$,
modulo $2 \pi$. We have proven the following theorem.

\begin{theorem} \label{th-7.1}
Suppose that $B\equiv 0$ and that $V$ satisfies Assumption \ref{ass-5.1}. Then, for any flux, $\Phi$,
and all $ A\in \0p2$, the high-velocity limits of  $S(A,V$)  in (\ref{5.35}) known  for $\hv$ and $\hw$
determines the fluxes
\beq
 \int_{\gamma (x,y,\hv,\hw)}A
\label{7.3}
\ene
modulo $2\pi$, for all curves $\gamma (x,y,\hv,\hw)$.
\end{theorem}

\begin{remark} \label{rem-7.2}{\rm
Theorem \ref{th-7.1} implies that from the high-velocity limit (\ref{5.35}) for $\hv$ and $\hw$ we can
reconstruct the fluxes
$$
\int_{\alpha}A
$$
for any closed curve $\alpha$ such that there is a surface (or chain) $\mathcal S$  in $\Lambda$ with
$\partial \mathcal S= \alpha -\gamma (x,y,\hv,\hw)$, because by Stoke's theorem,

$$
 \int_{\alpha}A =\int_{\gamma (x,y,\hv,\hw)}A+ \int_{\mathcal S}B= \int_{\gamma (x,y,\hv,\hw)}A.
 $$
 Remember also that given any $A \in \p2$ we can always find an $A \in \pot$
with the same scattering operator. We can take, for example,  $A_\Phi$. See equation (\ref{5.12}). Then, there
is no loss of generality taking $A \in \nb$.
Furthermore, notice that we can at most reconstruct the fluxes modulo $2 \pi$ because by (\ref{5.12})
$S(A_\Phi,V)= S(A,V)$ and the fluxes of $A_\Phi$ and $A$ in differ by integer multiples of $2 \pi$.
For general $A \in \0p2$ we recuperate the fluxes from equation (\ref{7.2}). However if $A$ is short-range
we can use the simpler formula (\ref{7.1}).
}
\end{remark}

\begin{remark} \label{rm-7.3}{\rm
As $\gamma(x,y,\hv,\hw)$ is a cycle, the homology class $[\gamma(x,y,\hv,\hw)]_{H_1(\Lambda;\ere)}$ is
well defined.

We denote,
\beq
\hr:= \left\langle\left\{ [\gamma(x,y,\hv,\hw)]_{H_1(\Lambda;\ere)}: L(x, \hv)\cup L(x,\hw)\subset \Lambda
\right \}\right\rangle.
\label{7.3d}
\ene
$\hr$ is a vector subspace of $H_1(\Lambda;\ere)$. Let us denote by $H^1_{ \hbox{\rm de  R, rec}}(\Lambda)$
the  vector subspace
of $H^1_{\hbox{\rm de  R}}(\Lambda)$ that is the dual to $\hr$, given by de Rham's Theorem.
Then, for all $\Phi$ and all $A \in \0p2$, from the high-velocity limit (\ref{5.35}) known for all $ \hv, \hw$
we reconstruct the projection of $A$ into $H^1_{\hbox{\rm de R, rec}}(\Lambda)$ modulo $2\pi$, as we now show.
Let
$$
\left\{ [\sigma_j]_{\hr} \right\}_{j=1}^m,
$$
be a basis of $\hr$, and let
$$
\left\{ [A_j]_{H^1_{\hbox{\rm de  R, rec}}(\Lambda)} \right\}_{j=1}^m,
$$
be the dual basis, i.e.,

$$
\int_{\sigma_j} A_k = \delta_{j,k}, j,k= 1,2, \cdots, m.
$$
Let us denote by $P_{\hbox{\rm rec}}$ the projector onto $H^1_{\hbox{\rm de  R, rec}}(\Lambda)$.
Hence, for any $A\in \p2$
$$
 P_{\hbox{\rm rec}}\left[ A\right]_{H^1_{\hbox{\rm de  R}}(\Lambda)}=
 \sum_{j=1}^m \lambda_j  [A_j]_{H^1_{\hbox{\rm de  R, rec}}(\Lambda)},
$$
and, furthermore, as

$$
\lambda_j= \int_{\sigma_j} A,
$$
we reconstruct $\lambda_j, j=1,2,\cdots,m$ (modulo $ 2 \pi $ ) from the high-velocity limit (\ref{5.35}) known for all $ \hv, \hw$.
}\end{remark}

\bull

We now give a precise definition of when a line $L(x,\hv)$ goes through a hole of $K$. Take $r >0$ such that
$K \subset  B^{\ere^3}_r(0)$. Suppose that $  L(x,\hv) \subset \Lambda$, and
$L(x,\hv) \cap B^{\ere^3}_r(0) \neq \emptyset$. we denote by  $c(x,\hv)$ the curve consisting of the segment
$L(x,\hv) \cap \overline{B^{\ere^3}_r(0)}$ and  an arc on $ \partial\overline{ B^{\ere^3}_r(0)}$ that connects the points
$L(x,\hv) \cap \partial  \overline{ B^{\ere^3}_r(0)}$. We orient $c(x,\hv)$ in such a way that the segment of straight
line has the orientation of $\hv$.

\begin{definition}\label{def-7.4}
A line  $L(x,\hv) \subset \Lambda$ goes through a hole of $K$ if $L(x,\hv) \cap B^{\ere^3}_r(0) \neq \emptyset$
and $[c(x,\hv)]_{H_1(\Lambda;\ere)}\neq 0$. Otherwise we say that $L(x,\hv)$ does not go through a hole of $K$.
\end{definition}
Note that this characterization of lines that go or do not go through a hole of $K$ is independent of the
$r$ that was used in the definition. This follows from the homotopic invariance of homology. See Theorem 11.2,
page 59 of \cite{gh}.

In an intuitive sense $[c(x,\hv)]_{H_1(\Lambda;\ere)}= 0$ means that
$c(x,\hv)$ is the boundary  of a surface (actually of a chain) that is contained in $\Lambda$ and then
it can not go through a hole of $K$. Obviously, as $K \subset  B^{\ere^3}_r(0)$,  if
$L(x,\hv) \cap B^{\ere^3}_r(0) = \emptyset$ the line $L(x,\hv)$ can not go through a hole of $K$.

\begin{definition} \label{def-7.5}
Two lines $L(x,\hv), L(y,\hw) \subset \Lambda$ that go through a hole of $ K $  go through the same hole if $[c(x,\hv)]_{H_1(\Lambda;\ere)}=
\pm [c(y,\hw)]_{H_1(\Lambda;\ere)}$. Furthermore, we say that the lines go through the hole in the same direction
if $[c(x,\hv)]_{H_1(\Lambda;\ere)}=
 [c(y,\hw)]_{H_1(\Lambda;\ere)}$.
 \end{definition}

\begin{lemma} \label{lemm-7.6}
Let $A,A_0 \in \nb$ with $A_0$  short-range and let $ \lambda$ be such that $A_0= A+d\lambda$.
Assume that $L(x,\hv)$ and $L(y,\hw)$ go through the same hole of $K$. Then,
\beq
\int_{-\infty}^\infty\hv\cdot A(x+\tau \hv)\, d\tau+ \lambda_\infty(\hv)-
\lambda_\infty(-\hv )=\pm \left(   \int_{-\infty}^\infty\hw\cdot A(y+\tau \hw)\, d\tau+ \lambda_\infty(\hw)-
\lambda_\infty(-\hw)\right),
\label{7.4}
\ene
 if $[c(x,\hv)]_{H_1(\Lambda;\ere)}=
\pm [c(y,\hw)]_{H_1(\Lambda;\ere)}$.

Moreover,
\beq
\int_{-\infty}^\infty\hv\cdot A(x+\tau \hv)\, d\tau+ \lambda_\infty(\hv)-
\lambda_\infty(-\hv )= \int_{-\infty}^\infty\hv\cdot A_0(x+\tau \hv)\, d\tau= \int_{c(x,\hv)} A_0=
 \int_{c(x,\hv)} A.
\label{7.5}
\ene
\end{lemma}

\noindent{\it Proof:}
 By  (\ref{7.3b}) and Stoke's theorem,

$$
\begin{array}{c}
\int_{-\infty}^\infty\hv\cdot A(x+\tau \hv)\, d\tau+ \lambda_\infty(\hv)-
\lambda_\infty(-\hv )= \int_{-\infty}^\infty\hv\cdot A_0(x+\tau \hv)\, d\tau=
\int_{c(x,\hv)} A_0=
\\\\
\pm \int_{c(y,\hw)}A_0 =\pm \left(
 \int_{-\infty}^\infty\hw\cdot A(y+\tau \hw)\, d\tau+ \lambda_\infty(\hw)-
\lambda_\infty(-\hw) \right).
\end{array}
$$

\begin{lemma} \label{lemm-7.7}
Let $A,A_0 \in \nb$ with $A_0$  short-range and let $ \lambda$ be such that $A_0= A+d\lambda$.
Assume that $L(x,\hv)$ does not go through  a hole of $K$. Then,
\beq
\int_{-\infty}^\infty\hv\cdot A(x+\tau \hv)\, d\tau+ \lambda_\infty(\hv)-
\lambda_\infty(-\hv )=0.
\label{7.7}
\ene
\end{lemma}
\noindent{\it Proof:} If $L(x,\hv) \cap B^{\ere^3}_r(0)= \emptyset$ it follows from (\ref{7.3b}) and Stoke's theorem that (\ref{7.7}) holds. Otherwise, $[c(x,\hv)]_{H_1(\Lambda, \ere)}=0$, and then,
by Stoke's theorem,
$$
\int_{c(x,\hv)} A=0.
$$
 Take $ z \in \partial \overline{B^{\ere^3}_r(0)}\cap c(x,\hv)$ such that $L(z,\hv)$ is tangent to
 $\partial \overline{B^{\ere^3}_r(0)}$. By the argument above,
$$
\int_{-\infty}^\infty\hv\cdot A(z+\tau \hv)\, d\tau+ \lambda_\infty(\hv)-
\lambda_\infty(-\hv )=0.
$$
Finally, using  once more Stoke's theorem we obtain that,
$$
0= \int_{c(x,\hv)} A= \int_{-\infty}^\infty\hv\cdot A(x+\tau \hv)\, d\tau - \int_{-\infty}^\infty\hv\cdot
A(z+\tau \hv)\, d\tau,
$$
and then, (\ref{7.7}) is proven.

\begin{remark}\label{remm-7.9}
{\rm
If $(x,\hv)\in \Lambda \times \ese^2$, there are neighborhoods $ B_x \subset \ere^3, B_{\hv}\subset \ese^2$
such that $(x,\hv) \in B_x\times B_{\hv}$ and if $(y,\hw) \in B_x\times B_{\hv}$ then, the following is true:
if $L(x,\hv )$ does not go true a hole of $K$, then, also $L(y,\hw )$ does not go through a hole of $K$.
If $L(x,\hv)$ goes through a hole of $K$, then, $L(y,\hw)$ goes through the same hole and in
the same direction. This follows from the homotopic invariance of homology, Theorem 11.2, page 59 of
\cite{gh}.}
\end{remark}

\begin{definition} \label{def-7.10}
For any $\hv \in \ese^2$ we denote by $ \Lambda_{\hv, \hbox{\rm out}}$ the set of points
$x \in \Lambda_{\hv}$ such that $L(x,\hv)$ does not go through a hole of $K$. We call this set the region
without holes of $\Lambda_{\hv}$.  The holes of $\Lambda_{\hv}$ is the set $\Lambda_{\hv, \hbox{\rm in}}:=
\Lambda_{\hv} \setminus  \Lambda_{\hv, \hbox{\rm out}} $.
\end{definition}

We define the following equivalence relation on $\Lambda_{\hv, \hbox{\rm in}}$. We say that
$ x R_{\hv} y$ if and only if

$L(x,\hv)$ and $L(y,\hv)$ go through the same hole and in the same direction. By $[x]$ we designate the
classes of equivalence under $R_{\hv}$.

We denote by $\left\{ \Lambda_{\hv, h} \right\}_{h \in \mathcal I}$ the partition of
$\Lambda_{\hv,\hbox{\rm in}}$ given by this equivalence relation. It is defined as follows.
$$
\mathcal I := \{ [x]  \}_{x\in \Lambda_{\hv, \hbox{\rm in}}}.
$$

Given $ h \in \mathcal I$ there is $x \in \Lambda_{\hv, \hbox{\rm in}}$ such that $h=[x]$. We denote,
$$
\Lambda_{\hv,h}:= \{y \in \Lambda_{\hv,\hbox{\rm in}}: y R_{\hv} x  \}.
$$
Then,
$$
\ds
\Lambda_{\hv, \hbox{\rm in}}= \cup_{h \in \mathcal I} \Lambda_{\hv,h},\,\,\,\,  \Lambda_{\hv,h_1}\cap
\Lambda_{\hv,h_2}= \emptyset, \,h_1 \neq h_2.
$$
We call $\Lambda_{\hv,h}$   the hole $h$ of $K$  in the direction of $\hv$.
Note that
\beq
\{\Lambda_{\hv,h}\}_{h \in \mathcal I} \cup \{\Lambda_{\v, \hbox{\rm out}}\}
\label{7.8}
\ene
is an open disjoint cover of $\Lambda_{\hv}$.

\begin{definition} \label{def-7.11}
For any $\Phi$, $ A \in \p2$, $\hv \in \ese^2$,   and    $ h \in \mathcal I$ we define,
$$
F_h:= \int_{c(x,\hv)} A,
$$
 where $x$ is any point in $\Lambda_{\hv,h}$. Note that $F_h$ is independent  the    $x \in \Lambda_{\hv,h}$
that we choose. $F_h$ is the flux of the magnetic field over any surface (or chain) in $\ere^3$ whose boundary
is $c(x,\hv)$. We call $F_h$ the magnetic flux on the hole $h$ of $K$.
\end{definition}

Let us take $\phi_0 \in \mathcal H_2(\ere^3)$ with compact support in $\Lambda_{\hv}$. Then, since (\ref{7.8})
is a disjoint open cover of $\Lambda_{\hv}$,

\beq
\phi_0= \sum_{h\in \mathcal I} \varphi_h+ \varphi_{\hbox{\rm out}},
\label{7.9}
\ene
with $\varphi_h, \varphi_{\hbox{\rm out}} \in \mathcal H_2(\ere^3), \varphi_h$ has compact support
in $\Lambda_{\hv,h}, h\in \mathcal H$, and $\varphi_{\hbox{\rm out}}$ has compact support in
$\Lambda_{\hv,\hbox{\rm out}}$. The sum is finite because $\phi$ has compact support.
We denote,
$$
\phi_{\v}:= e^{im\v\cdot x} \phi_0, \, \varphi_{h,\v}:= e^{im\v\cdot x}\varphi_h, \,
\varphi_{\hbox{\rm out}, \v}:= e^{im\v\cdot x} \varphi_{\hbox{\rm out}}.
$$
\begin{theorem}\label{th-7.12}
Suppose that $B\equiv 0$ and that $V$ satisfies Assumption \ref{ass-5.1}. Then, for any $\Phi$  and
any $A \in \nb$,
\beq
S(A,V)\, \phi_{\v} = e^{-i(\lambda_\infty(\hv )-\lambda_\infty(-\hv))} \,\left( \sum_{h\in \mathcal I}\,
e^{i F_h}\varphi_{\v,h}+ \varphi_{\hbox{\rm out}, \v}\right)+ O\left( \frac{1}{v}\right).
\label{7.10}
\ene
\end{theorem}

\noindent {\it Proof:} The theorem follows from Theorem \ref{th-5.6} and Lemmas \ref{lemm-7.6}, \ref{lemm-7.7}.

\begin{corollary} \label{cor-7.13}
Under the conditions of Theorem \ref{th-7.12}
\beq
\left(S(A,V)\, \phi_{\v},\varphi_{\v,h}\right)= e^{-i(\lambda_\infty(\hv )-\lambda_\infty(-\hv))} e^{i F_h}
 + O\left( \frac{1}{v}\right),
   \, h \in \mathcal I,
\label{7.11}
\ene

\beq
\left(S(A,V)\, \phi_{\v},\varphi_{\v,\hbox{\it out}}\right)= e^{-i(\lambda_\infty(\hv )-\lambda_\infty(-\hv))}
+ O\left( \frac{1}{v}\right).
\label{7.12}
\ene
Moreover, the high-velocity limit of $S(A,V)$ in the direction $\hv$  determines
$ \lambda_\infty(\hv )-\lambda_\infty(-\hv)$ and the fluxes $ F_h, h\in \mathcal I$, modulo $2\pi$.
\end{corollary}

\noindent {\it Proof:} The corollary follows immediately from Theorem \ref{7.10}.

\begin{remark} \label{rem-7.13}{\rm
Equations (\ref{7.11}, \ref{7.12}) are reconstruction formulae that allows us to reconstruct
$ \lambda_\infty(\hv )-\lambda_\infty(-\hv)$ and the fluxes $ F_h, h\in \mathcal I$, modulo $2\pi$,
 from the high-
velocity limit of the scattering operator in the direction $\hv$. Recall that
$ \lambda_\infty(\hv )-\lambda_\infty(-\hv)$ is independent of the particular short-range potential
that we use to define $\lambda$. Remember also that given any $A \in \p2$ we can always find an $A \in \pot$
with the same scattering operator. We can take, for example,  $A_\Phi$. See equation (\ref{5.12}). Then, there
is no loss of generality taking $A \in \nb$.

Note that it is quite remarkable that we can determine  $ \lambda_\infty(\hv )-\lambda_\infty(-\hv)$
since it is not a gauge invariant quantity. According to the standard interpretation of quantum mechanics only
gauge invariant quantities are physically relevant. Note that if $A$ is short-range $\lambda_\infty$ is
constant. In this case  $\lambda_\infty(\hv )-\lambda_\infty(-\hv)\equiv 0$  and  it drops out from all
our formulae. We see that one possibility   is to consider that only short-range potentials are physically
admissible. This is consistent with the usual interpretation of quantum mechanics in three dimensions.
However, we can also go beyond the standard interpretation of quantum mechanics and consider the class of
long-range potentials $\pot$ as physically admissible. This raises the interesting question of what is the
physical significance of the $ \lambda_\infty(\hv )-\lambda_\infty(-\hv)$.
}\end{remark}

\begin{example} \label{ex-7.14}{\rm
Here we consider a simple example where we give an explicit description of the holes. Furthermore, the fluxes
of the holes are the fluxes of the magnetic field over cross  sections of the tori. We reconstruct
all the fluxes modulo $2\pi$  and also we determine the cohomology class of the magnetic potential modulo
$2\pi$, from the high-velocity limit of the scattering operator in only one direction.

Given a vector $z\in \ere^3$  and  $ a > b >0$    we denote by $ T(z,a,b)$ the following set
$$
T(z,a,b):= \left\{ z+ a (\cos \theta, \sin\theta,0)+ b \left(x (\cos \theta, \sin \theta,0)+y (0,0,1)
\right):\, \theta \in [0,2\pi], (x,y)\in \overline{B^{\ere^2}_1(0)}\right\}.
$$
The map $ F_{z,a,b}: T \rightarrow T(z,a,b)$ given by
$$
F_{z,a,b}((\cos\theta,\sin \theta), (x,y))\rightarrow z+ a (\cos \theta, \sin\theta,0)+
b \left(x (\cos \theta, \sin \theta,0)+y (0,0,1)
\right)
$$
is a diffeomorphism.

\noindent {\bf The Obstacle}

\noindent We now define the obstacle $K$. We assume that $\v= (0,0,1)$.

 As before the connected  components of $K$ are $K_j, j=1,2,\cdots,L$.
Let us denote $J=\{1,2,\cdots,m\}$ and  $I=\{m+1,\cdots,L\}$. If $m=L$, then, $I=\emptyset$.
We assume that $K$ satisfy the following assumptions.
\begin{enumerate}
\item
There are vectors $z_j \in \ere^3$ and numbers $a_j > b_j, j=1,2,\cdots,m$ such that,
$$
K_j= T(z_j,a_j,b_j), \forall j \in J, \,\, \,
K_j \cong \overline{B^{\ere^3}_1(0)}, j \in I.
$$
\item
$$
\left( \hbox{\rm convex}\,(K_j)+ \ere \v\right) \cap \left(\hbox{\rm convex} \,(K_l)+ \ere \v\right)=
\emptyset, j,l \in J,\,\,\,
\left( \hbox{\rm convex}\,(K_j) +\ere \v\right) \cap \left( K_l+ \ere \v\right)=
\emptyset, j\in J ,l \in I.
$$
\end{enumerate}
We denote as before by $\hbox{\rm convex}\, (\cdot)$ the convex hull of the indicated set.

\noindent {\bf The Curves $\mathbf \gamma_j, \tilde{\gamma_j}, \hat{\gamma_j}$}

Let $ \theta_j  $ be such that $  z_j = r_j (\cos(\theta_j), \sin(\theta_j) , 0) + (0, 0, (z_j)_3)$.

The curves $\gamma_j, j \in J$ are given by
$$
\gamma_j(t):= z_j+ a_j (\cos t, \sin t,0),
$$
and the curves $\tilde{\gamma_j}, j \in J$, are
$$
\tilde{\gamma_j}:= z_j + a_j (\cos\theta_j, \sin\theta_j,0)+ b_j \left( \cos t (\cos\theta_j, \sin\theta_j, 0)
+\sin t (0,0,1)\right).
$$
Furthermore, the curves $ \hat{\gamma_j}, j \in J$, are
$$
\hat{\gamma_j}:=  z_j + a_j (\cos\theta_j, \sin\theta_j,0)+ (b_j+ \delta/2)
\left( \cos t (\cos\theta_j, \sin\theta_j, 0)
+\sin t (0,0,1)\right),
$$
where $\delta >0$ so small that, $ \delta < a_j-b_j,$ and

$$
\left( \hbox{\rm convex}\,(K_{j,\delta})+ \ere \v\right) \cap \left(\hbox{\rm convex} \,(K_{l,\delta})+
 \ere \v\right)=
\emptyset, j,l \in J, \,\,
\left( \hbox{\rm (convex}\,K_{j,\delta}) +\ere \v\right) \cap \left( K_{l,\delta}+ \ere \v\right)=
\emptyset, j\in J ,l \in I.
$$
The subindex $\delta$ denotes the set of points that are at distance up to $\delta$ of the indicated set.

\noindent {\bf The Flux $\Phi$}

\noindent We define the following sets
$$
h_j:= \left\{ z_j + t(\cos\theta,\sin\theta): \theta \in [0,2\pi], t \in [0, a_j-b_j) \right\} +
\ere \hv,
j \in J.
$$
We have that,
\beq
 [c(x,\hv)]_{H_1(\Lambda;\ere)}= [c(y,\hv)]_{H_1(\Lambda;\ere)} , \forall x, y \in h_j, j \in J.
 \label{7.12b}
 \ene
Since $c(x,\hv)$ and $c(y,\hv)$ are homotopic in $\Lambda$,  this follows from the homotopic invariance of
homology. See Theorem 11.2, page 59 of \cite{gh}. Then, we can associate a flux $\Phi_j$ to each $h_j, j \in J$
as follows,
$$
\Phi_j = \int_{c(x,\hv)} A, \, \hbox{\rm for some}\, x \in h_j, j \in J.
$$

We have that,
\beq
[c(y,\hv)]_{H_1(\Lambda;\ere)}=0, \forall y \in \left( \Lambda_{\hv} \setminus \left(\cup_{j\in J} h_j \right)
\right)
 \cap \left( B^{\ere^3}_r(0) + \ere\hv  \right).
\label{7.13}
\ene
Let us prove this. As the segment of straight line in $c(y,\hv)$ does not belong to any of the sets
$\hbox{\rm convex}\,(K_j)+\ere \hv, j \in J$, we have that, for any $ j \in J$ there is a surface (or a chain)
$\sigma_j$ contained in the complement of
$K_j$ such that $ \partial \sigma_j= c(y,\hv)$.
Let $\left\{ \left[G^{(j)}\right]_{H^1_{\hbox{\rm de  R}}}(\Lambda)\right\}_{j=1}^m$ be the basis of
$H^1_{\hbox{\rm de  R}}$ constructed in Proposition \ref{prop-2.2}. Then, as $ d G^{(j)}=0$ it follows
from Stoke's theorem that
$$
\int_{c(y,\hv)} G^{(j)}=0, \forall j\in J.
$$
Hence, (\ref{7.13}) follows from  de Rham's Theorem, Theorem 4.17, page 154 of \cite{w}.

Let us prove now that,
\beq
\Phi_j = \Phi(\hat{\gamma_j}), j \in J.
\label{7.14}
\ene
For any $j \in J$ we define,
$$
 x_j:= z_j + a_j (\cos\theta_j, \sin\theta_j,0)- (b_j+ \delta/2)
 (\cos\theta_j, \sin\theta_j, 0), \,\,
 $$
 $$
y_j:= z_j + a_j (\cos\theta_j, \sin\theta_j,0)+ (b_j+ \delta/2)
 (\cos\theta_j, \sin\theta_j, 0).
$$
We choose the curves $c(x_j, \hv), c(y_j,\hv)$ in such a way  that the   arc in $c(y_j,\hv)$ is contained in the
arc in $c(x_j,\hv)$. Let $c_j$ be the curve obtained by taking the segments of straight line in $c(x_j,\hv)$
and in $c(y_j,\hv)$ and the two arcs that are obtained by cutting from the arc in $c(x_j,\hv)$ the arc in
$c(y_j,\hv)$. We orient $c_j$ in such a way that the segment of straight line in $c(x_j,\hv)$ has the
orientation of $\hv$. Then, in homology,
\beq
[c_j]_{H_1(\Lambda;\ere)}= [c(x_j,\hv)]_{H_1(\Lambda;\ere)}-[c(y_j,\hv)]_{H_1(\Lambda;\ere)}.
\label{7.15}
\ene
This follows from de Rham's Theorem -Theorem 4.17, page 154 of \cite{w}- since for any  closed 1-form $D$,
$$
\int_{c_j} D = \int_{c(x_j,\hv)} D- \int_{c(y_j,\hv)} D.
$$
The curves $\hat{\gamma}_j$ and $ c_j$ are homotopically equivalent in $\Lambda$. Hence, by the homotopical
invariance of homology, Theorem 11.2, page 59 of \cite{gh},
\beq
[c_j]_{H_1(\Lambda;\ere)}= [\hat{\gamma_j}]_{H_1(\Lambda;\ere)}.
\label{7.15b}
\ene
Then, by   (\ref{7.13}, \ref{7.15}),

\beq \label{7.15b0}
[\hat{\gamma_j}]_{H_1(\Lambda;\ere)}= [c(x_j,\hv)]_{H_1(\Lambda;\ere)},
\ene
and hence,
$$
\int_{c(x_j,\hv)} A = \int_{\hat{\gamma_j}} A, j \in J,
$$
what proves (\ref{7.14}).

\noindent {\bf The Holes of $K$}

\noindent Recall that    $\Lambda_{\hv, \hbox{\rm out}}$ and $ \Lambda_{\hv,\hbox{\rm in}}$ were defined
in Definition \ref{def-7.10}, that the holes of $K$ are the sets $\Lambda_{\v,h}, h \in \mathcal I$, that $F_h$ is the flux
over the hole $\Lambda_{\v,h}, h \in \mathcal I$, that
$\Lambda_{\hv, \hbox{\rm in }}= \cup_{h \in \mathcal I } \Lambda_{\hv,h}$.

Then, we have that,
\begin{enumerate}

\item
 The index set $\mathcal I$  can be taken as $ \mathcal I= \{h_j\}_{j\in J}\sim J$. Moreover,
  denoting $ \Lambda_{\hv, j}= \Lambda_{\hv,h_j}$, we have that $ \Lambda_{\hv,j}= h_j$ and
   $\Lambda_{\hv, \hbox{\rm in}}= \cup_{j\in J} h_j$.
\item
We designate, $F_j := F_{h_j}$. Then,

$$
F_j= \Phi(\hat{\gamma_j}), \, j \in J.
$$
\item
$$
\Lambda_{\hv, \hbox{\rm out}}= \ere^3 \setminus \left(\Lambda_{\hv, \hbox{\rm in}} \cup_{j=1 }^L
(K_j+\hv \ere)\right).
$$
\end{enumerate}

Let us prove this. By (\ref{7.13}) $[c(y,\hv)]_{H_1(\Lambda;\ere)}=0, \forall y
\in \left( \Lambda_{\hv} \setminus \cup_{j\in J} h_j \right)
$ such that $ B^{\ere^3}_r(0) \cap L(y,\hv) \neq \emptyset$. Then,
$$
\left( \Lambda_{\hv} \setminus \cup_{j\in J} h_j
\right)
 \cap \left( B^{\ere^3}_r(0) + \ere\hv  \right) \subset \Lambda_{\hv,\hbox{\rm out} }.
$$
But by our definition the complement in $\Lambda_{\hv}$ of $ B^{\ere^3}_r(0)+\ere \hv$ is contained in
$ \Lambda_{\hv,\hbox{\rm out} }$. It follows that,
$$
\left( \Lambda_{\hv} \setminus \cup_{j\in J} h_j
\right) \subset \Lambda_{\hv,\hbox{\rm out} }.
$$
Moreover, if $ x\in h_j$ for some $j \in J$,  since $[\hat{\gamma_j}]_{H_1(\Lambda;\ere)}\neq 0$,
it follows from (\ref{7.12b}) and (\ref{7.15b0}) that $[ c(x_{j},\hv)]_{H_1(\Lambda;\ere)} = [ c(x,\hv)]_{H_1(\Lambda;\ere)} \neq 0$, and then,
$  x \notin \Lambda_{\hv,\hbox{\rm out} }$. Then, we have proven that,

\beq
\left( \Lambda_{\hv} \setminus \cup_{j\in J} h_j
\right) =\Lambda_{\hv,\hbox{\rm out} },
\label{7.16}
\ene
and hence,
$$
\Lambda_{\hv,\hbox{\rm in}}= \cup_{j\in J} h_j.
$$
Item 3 is now obvious. By (\ref{7.12b}) if $x,y \in h_j,$ then, $ [c(x,\hv)]_{H_1(\Lambda;\ere)}=
[c(y,\hv)]_{H_1(\Lambda;\ere)}$. Hence, $x R_{\hv} y$ what implies that  $h_j$ is contained in some hole of
$K$. But by (\ref{7.12b}) and (\ref{7.15b0}) if $x \in h_j, y \in h_l, j \neq l$, then, $ [c(x,\hv)]_{H_1(\Lambda;\ere)} \neq
[c(y,\hv)]_{H_1(\Lambda;\ere)}$ because as the $ [\hat{\gamma_j}]_{H_1(\Lambda;\ere)}, j \in J$ are a basis of
 $H_1(\Lambda;\ere)$  they are different. In consequence, $x$ and $y$ belong to different holes of $K$.
Then, since (\ref{7.16}) holds, we have proven  item 1. Item 2 follows from (\ref{7.14}).

By Corollary \ref{cor-7.13} and Remark \ref{rem-7.13}, this proves that from the high-velocity limit of
$S(A,V)$ in the direction of $\hv$ we reconstruct all the fluxes
$\Phi(\hat{\gamma_j}), j \in J,$ modulo $2\pi$.

Let us now prove that  from the high-velocity limit of
$S(A,V)$ in the direction of $\hv$ we also reconstruct the cohomology class
$[A]_{H^1_{\hbox{\rm de  R}}(\Lambda)}$
modulo $2\pi$, in the sense that we reconstruct  modulo $2\pi$ the coefficients of
 $[A]_{H^1_{\hbox{\rm de  R}}(\Lambda)}$
in any basis of $H^1_{\hbox{\rm de  R}}(\Lambda)$.

Let $\left\{ [M_j]_{H^1_{\hbox{\rm de  R}}(\Lambda)}\right\}_{j=1}^m$ be any basis of
$H^1_{\hbox{\rm de  R}}(\Lambda)$ and let $\left\{ [\Gamma_j]_{H_1(\Lambda; \ere)}\right\}_{j=1}^m$
be the dual  basis of $H_1(\Lambda; \ere)$  given by de Rham's Theorem,

$$
\int_{\Gamma_j} M_l = \delta_{j,l}, j,l \in J.
$$
Let $\{\alpha_j\}_{j \in J}$ be the expansion coefficients of $A$,

$$
[A]_{H^1_{\hbox{\rm de R}}(\Lambda)}= \sum_{j \in J} \alpha_j [M_j]_{H^1_{\hbox{\rm de R}}(\Lambda)},
$$

$$
\alpha_j = \int_{\Gamma_j} A.
$$

By Proposition \ref{prop-b.1}   $\left\{ [\hat{\gamma_j}]_{H_1(\Lambda; \ZETA)}\right\}_{j=1}^m$
is a basis of $H_1(\Lambda; \ZETA)$. Then,
$$
 [\Gamma_j]_{H_1(\Lambda; \ZETA)}= \sum_{l \in  J} n(j,l) [\hat{\gamma_l}]_{
 H_1(\Lambda; \ZETA)},
$$
where the coefficients $ n(j,l)$ are integers. Finally,

$$
\alpha_j = \int_{\Gamma_j} A= \sum_{l \in L} n(j,l) \int_{\hat{\gamma_l}}A=
 \sum_{l \in L} n(j,l) \Phi(\hat{\gamma_l}), j \in J,
$$
and since we have already determined the $\Phi(\hat{\gamma_l})$ modulo $2\pi$,  the coefficients
$\alpha_j, j \in J$ are determined modulo $2 \pi$.
}
\end{example}

\section{The Tonomura et al. Experiments}
\sss
 The fundamental experiments of Tonomura et al. \cite{to1,to2}, gave a conclusive evidence of the existence of
 the Aharonov-Bohm effect. For a detailed account see \cite{pt}.

 Tonomura et al. \cite{to1,to2}  did their experiments in the case of  toroidal magnets. This corresponds
 to our Example \ref{ex-7.14} with only one torus, i.e., $L=1, J=\{1\}$. In  very careful and precise
 experiments they managed to superimpose behind the toroidal magnet two electron beams. One of them traveled
 inside the hole of the toroidal magnet and the other -the reference beam- outside it. They measured the interference fringes
 between the two beams produced by the magnetic flux inside the torus.

We show now that our results give a rigorous  mathematical proof that quantum mechanics predicts the
interference fringes observed by Tonomura et al. \cite{to1,to2} in their remarkable experiment.

 An equivalently description of these experiments is to consider that both electron beams traveled inside the
 hole of the torus, one of them with a nonzero  magnetic flux inside the torus, and the other -the
 reference beam-
 with the magnetic flux inside the torus set to zero. Since long-range magnetic potentials add a global
 constant phase that does not affect the interference pattern, we take, for simplicity, a short-range
 magnetic potential.  According to Theorem \ref{th-7.12}, for the particle
 that goes inside the hole with the magnetic flux present, up to an error of order $1/v$, we have that,

 \beq
S(A,V)\phi_{\hv}= e^{i\frac{q}{\hbar c} {\mathbf \Phi}}\, \phi_{\hv},
\label{8.1}
\ene
where we have taken physical units, with $\mathbf \Phi$ the flux of the physical magnetic field $\mathbf
B$ and $\phi_{\hv}= e^{i \frac{M}{\hbar}\hv\cdot x }\phi_0$. See Section 4.
For the particle that goes outside the hole of the magnet, or equivalently inside the hole with the magnetic
field set to zero,

\beq
S(A,V)\phi_{\hv}=  \phi_{\hv}.
\label{8.2}
\ene
If we superimpose both asymptotic states we obtain the wave function,

\beq
\left( 1+ e^{i\frac{q}{\hbar c}\mathbf \Phi} \right) \phi_{\hv},
\label{8.3}
\ene
up to an error of order $1/v$. This shows the interference patterns that were observed experimentally by
Tonomura et al. \cite{to1,to2}.
For example,
if $\frac{q}{\hbar c} \mathbf \Phi$ is an odd multiple of $\pi$ there is destructive interference and there is
a dark zone behind the hole of the magnet, as observed experimentally.

Tonomura et al. \cite{to1,to2} also considered the case when the reference beam is slightly tilted. In this
case the reference beam is given by
$$
\phi_{\hv +{\mathbf v}_0}= e^{i \frac{M}{\hbar} {\mathbf v}_0\cdot x}\phi_{\hv},
$$
and (\ref{8.2}) is  replaced by,
$$
S(A,V)\phi_{\hv+{\mathbf v}_0}=  \phi_{\hv+{\mathbf v}_0}= e^{i{\frac{M}{\hbar} \mathbf v}_0
\cdot x}\,\phi_{\hv}.
$$
In this case we obtain the wave function
$$
e^{i\frac{M}{\hbar} {\mathbf v}_0\cdot x}  \left(1+ e^{- i\frac{M}{\hbar} {\mathbf v}_0\cdot x}
e^{i\frac{q}{\hbar c}}
  \right) \phi_{\hv},
$$
up to an error of order $1/v$.
We see that the factor,
$$
 \left(1+ e^{- i\frac{M}{\hbar} {\mathbf v}_0\cdot x}
e^{i\frac{q}{\hbar c}}
  \right)
$$
produces the parallel fringes that were observed experimentally by Tonomura et al. \cite{to1,to2}.

\section{Appendix A}\sss
In this appendix we prove, for the reader's convenience, that $H_s(\natural\, k\,T;\erc )=0, s \geq  2$,
 that $H_1(\natural k T; \erc)\cong \oplus_{i=1}^k \erc $, and that $\left\{ [Z_j]_{H_1(\natural k T;
\erc)}\right\}_{j=1}^k$
is a basis of $H_1(\natural k T; \erc)$.
$\erc$ is  $\ZETA$ or $\ere$.

Recall that we defined,  $\gamma_\pm : [0,1] \rightarrow T: \gamma_\pm(t)=(e^{\pm 2\pi it},0,0)$.
\begin{prop}\label{prop-a.1} $H_s(T; \erc)=0, s \geq 2$ and
$H_1(T;\ZETA) \cong \ZETA$ and $\left\{ [\gamma_\pm]_{H_1( T;
\ZETA)}\right\}$ are  basis of $H_1(T;\ZETA)$.
\end{prop}
\noindent {\it Proof:} We define $\tilde{\gamma}_\pm
:[0,1]\rightarrow \ese^1: \tilde{\gamma}_\pm(t) := e^{\pm 2\pi it}$
and let $I_{\ese^1}: \ese^1 \rightarrow T$ be the inclusion given by
$ I_{\ese^1}(s):=(s,0,0)$. Clearly, $ I_{\ese^1}\circ
\tilde{\gamma}_\pm = \gamma_\pm$. It is easy to see that $\ese^1$ is
homotopically equivalent to $T$ and that the inclusion $I_{\ese^1}:
\ese^1 \rightarrow T$ is a homotopic equivalence. It follows that
$I_{\ese^1}$ induces an isomorphism in holomogy given by
$H_s(I_{\ese^1})$ (see theorem 11.3, page 59 \cite{gh}). Then,
$H_s(T;\erc)\cong H_s(\ese^1;\erc)$ and hence, we have that
$H_s(T;\erc)=0, s \geq 2$ by Corollary 15.5, page 84 of \cite{gh}.
For $s=1$ and $\erc=\ZETA$, the isomorphism is given in the
following way (see page 49 \cite{gh}). Let $\sigma_i:
[0,1]\rightarrow T$ be continuous functions and let $n_i \in \ZETA$.
let us assume that $\sum n_i \sigma_i$ is a cycle (its boundary is
zero). Then, $H_1(I_{\ese^1})[\sum n_i \sigma_i]_{H_1(\ese^1;
\ZETA)}:= [\, \sum n_i I_{\ese^1}\circ \sigma_i]_{H_1(T;\ZETA)}$.

As $I_{\ese^1}\circ \tilde{\gamma}_\pm= \gamma_\pm$, it follows that
$H_1(I_{\ese^1})[\tilde{\gamma}_\pm]_{H_1(\ese^1; \ZETA)}=
[\gamma_\pm]_{H_1(T;\ZETA)}$. Then, to prove the Proposition it is
enough to prove that $H_1(\ese^1;\ZETA )\cong \ZETA$ and that
$\left\{ [\tilde{\gamma}_\pm]_{H_1(\ese^1; \ZETA)}\right\}$ are
basis of $H_1(\ese^1;\ZETA)$.

By Theorem 12.1, page 63 of \cite{gh}, there is a homomorphism $\Xi:
\Pi_1(\ese^1;1)\rightarrow H_1(\ese^1;\ZETA)$ that sends a homotopy
class to its homology class. In our case $\Xi$ is an isomorphism
since $\Pi_1(\ese^1;1)$ is abelian. Actually, $\Pi_1(\ese^1;1)\cong
\ZETA$. See Theorem 4.4, page 17 of \cite{gh}. Then, $\ZETA \cong
\Pi_1(\ese^1;1)\cong H_1(\ese^1;\ZETA)$. To prove that $\left\{
[\tilde{\gamma}_\pm]_{H_1(\ese^1; \ZETA)}\right\}$ are  basis of
$H_1(\ese^1;\ZETA)$ it is enough to prove that $\left\{
[\tilde{\gamma}_\pm]_{\Pi_1(\ese^1; 1)}\right\}$ are  basis of
$\Pi_1(\ese^1;1)$. The isomorphism $\Lambda: \Pi_1(\ese^1; \ZETA)
\rightarrow \ZETA$ is given (see Theorem 4.4, page 17 of \cite{gh})
as follows. Given a path $\sigma$ with $
[\sigma]_{\Pi_1(\ese^1;1)}\in \Pi_1(\ese^1;1)$ let $\sigma':
[0,1]\rightarrow \ere$ satisfy $ \sigma'(0)=0$ and $ e^{2\pi
i\sigma'(t)}= \sigma(t)$. Then,
$\Lambda[\sigma]_{\Pi_{1}(\ese^1;1)}= \sigma'(1)$. In our case, if
we take $\tilde{\gamma}_\pm'(t):=\pm t, \tilde{\gamma}_\pm'(0)=0$
and $\tilde{\gamma}_\pm'(1)=\pm 1$. It follows that, $ \Lambda
[\tilde{\gamma}_\pm]_{\Pi_1(\ese^1;1)}= \pm 1$. As $\pm1$ are basis
of $\ZETA$ it follows that $[\tilde{\gamma}_\pm]_{\Pi_1(\ese^1;1)}$
are basis of $\Pi_{1}(\ese^1; 1)$ and this concludes the proof that
$[\gamma_\pm]_{H_1(T;\ZETA)}$ are basis of $H_1(T;\ZETA)$.

\begin{prop}\label{prop-a.2}
For $s \geq 2, H_s(\natural k T;\erc)=0$. Furthermore,
$H_1(\natural k T;\ZETA)\cong \oplus_{i=1}^k \ZETA$, and
 $\left\{ [Z_j]_{H_1(\natural k  T;
\ZETA)}\right\}_{j=1}^k$ is a  basis of $H_1(\natural k T;\ZETA)$.
\end{prop}
\noindent{\it Proof:} We prove the Proposition by induction in $k$. For $k=1, Z_1= \gamma_+$ and the result
follows from Proposition \ref{prop-a.1}. Let us assume that
$H_s(\natural\, (k-1) \,T;\erc) \cong \oplus_{i=1}^{k-1} \erc$ and that
$\left\{ [Z_j]_{H_1(\natural\, (k-1) \, T;
\ZETA)}\right\}_{j=1}^{k-1}$ is a  basis of $H_1(\natural \, (k-1) \, T;\ZETA)$.
Let $X_1$ and $X_2$ be open
subsets of $\natural\, k\, T$ such that
$$
\cup_{j\leq k-1} \, l_j(T) \,  \subseteq X_1, \,\,l_k(T) \subseteq X_2,\, X_1 \simeq\, \cup_{j\leq k-1}
 \,l_j(T) \approx \natural\, (k-1) T,\,
\, X_2\simeq l_k(T)\approx T,
$$
and $X_1\cap X_2$ is contractible, i.e. $X_1 \cap X_2 \simeq$ to a single point. The symbol $\simeq$ means
homotopic equivalence and  $\approx$ means homeomorphism.

By  Example 17.1, page 98 of \cite{gh} $(\natural \, k \, T, X_1,X_2)$ is an exact triad and we can apply the
sequence of Mayer-Vietoris (17.7 page 99 and 17.9 page 100 of \cite{gh}).

$$
H_s(X_1\cap X_2;\erc) \rightarrow H_s(X_1;\erc)\oplus H_s(X_2;\erc)\rightarrow H_s(\natural\, k\, T;\ZETA)
\rightarrow H_{s-1}^\sharp (X_1\cap X_2;\erc).
$$
As $X_1\cap X_2$ is homotopically equivalent to a point -that we denote by $\{*\}$- we have that
$H_s(X_1\cap X_2;\erc)\cong H_s(\{*\};\erc)=0, H_{s-1}^\sharp(X_1\cap X_2;\erc)\cong H_{s-1}^\sharp(\{*\};\erc)
=0$
(see Theorem 11.3, page
59, Example 9.4, page 47 and Example 9.7, page 48 of \cite{gh}). Hence, we obtain the isomorphism,

\beq
H_s(X_1;\erc)\oplus H_s(X_2;\erc)\rightarrow H_s(\natural\, k\, T;\ere).
\label{a.0}
\ene
This isomorphism is given by (see 17.4, page 99 of \cite{gh})
\beq
\left( [c_1]_{H_s(X_1;\erc)}, [c_2]_{H_s(X_2;\erc)}\right) \rightarrow - [c_1]_{H_s(\natural\, k\, T;\erc)}+
[c_2]_{H_s(\natural\, k\, T;\erc)}.
\label{a.1}
\ene
As $ \cup_{j\leq k-1} l_j(T) \simeq X_1, \, l_k(T) \simeq X_2$, the inclusions $ \cup_{j\leq k-1} l_j(T) \hookrightarrow X_1, \,
l_k(T) \hookrightarrow X_2$ induce isomorphisms in homology (see Theorem 11.3, page 59 of \cite{gh}).
We have, then, the following isomorphisms.
\beq
H_s(\natural \, (k-1)\,T;\erc) \stackrel{\cong} \rightarrow  H_s(\cup_{j\leq k-1} l_j(T);\erc)
\stackrel{\cong}\rightarrow H_s(X_1; \erc),
\label{a.2}
\ene
\beq
H_s(T;\erc)\stackrel{\cong}\rightarrow  H_s(l_k(T); \erc) \stackrel{\cong}\rightarrow H_s(X_2;\erc).
\label{a.3}
\ene
By our induction hypothesis and (\ref{a.0}, \ref{a.1}, \ref{a.2}, \ref{a.3})
$H_s(\natural\, k \,T;\erc) \cong \oplus_{i=1}^{k} \erc$. Hence, by Proposition \ref{prop-a.1}
$H_s(\natural \,
k \, T; \erc )=0, s \geq 2$.
Moreover, by the induction  hypothesis and (\ref{a.2}),
it also follows that
$\left\{ [Z_j]_{H_1(X_1;
\ZETA)}\right\}_{j=1}^{k-1}$ is a basis of $ H_1(X_1;\ZETA)$.
By Proposition \ref{prop-a.1} and (\ref{a.3}) $H_1(X_1;\ZETA) \cong \ZETA$; furthermore,
by the definition of $ Z_k$ (see {\ref{1.2})) and as the homeomorphism $l_k: T \rightarrow l_k(T)$ induces an
isomorphism in homology it follows from  Proposition \ref{prop-a.1} that $[Z_k]_{H_1(l_k(T);\ZETA)}$
is a basis of $H_1(l_k(T);\ZETA)$ and then, by (\ref{a.3}) it is also a basis of $H_1(X_2;\ZETA)$.
Finally, it follows from (\ref{a.1}) that $H_1(\natural k T;\ZETA) \cong \oplus_{i=1}^k \ZETA$ and $\left\{ [Z_j]_{H_1(\natural k  T;
\ZETA)}\right\}_{j=1}^k$ is a  basis of $H_1(\natural k T;\ZETA)$.

\begin{prop}\label{prop-a.3}
$H_1(\natural k T;\ere) \cong \oplus_{i=1}^k \ere$ and $\left\{ [Z_j]_{H_1(\natural k  T;
\ere)}\right\}_{j=1}^k$ is a  basis of $H_1(\natural k T;\ere)$.
\end{prop}

\noindent{\it Proof:} The homology group $H_1(\natural\, k\, T;\ere)$ is a module over the ring $\ere$,
i.e. it is a vector space (page 47 of \cite{gh}). If $G$ is an abelian group we can also define the
homology groups as in page 153 of \cite{h}. In this case $H_1(\natural\, k\, T;G)$ is a group. As $\ere$ is a
group and a ring we can define the homology groups as modules and as groups. To differentiate them we will
denote by $H_1(\natural \, k\, T;\ere)$ the homology module considering $\ere$ as a ring, and by
$\tilde{H}_1(\natural\, k\, T;\ere)$ considering $\ere$ as a group. Actually, $H_1(\natural \, k\, T;\ere)$
and $\tilde{H}_1(\natural \, k\, T;\ere)$ are equal as sets and as groups. By the theorem of universal
coefficients -Corollary 3 A.4, page 264 of \cite{h}- there is the exact sequence,
$$
0\rightarrow \tilde{H}_s(\natural \,k \, T;\ZETA)\otimes \ere \rightarrow \tilde{H}_s(\natural \, k\,; \ere)
\rightarrow \,\hbox{\rm Tor}\,( \tilde{H}_{s-1}(\natural \, k\, T;\ZETA), \ere)\rightarrow 0.
$$
As $\ere$ is torsion free, $\hbox{\rm Tor}\,( \tilde{H}_{s-1}(\natural \, k\, T;\ZETA), \ere)=0$. See
Proposition
3A.5, page 265 of \cite{h}. In consequence,
\beq
\tilde{H}_s(\natural\, k\, T;\ZETA)\otimes \ere \stackrel{\cong}\rightarrow \tilde{H}_s(\natural\, k\, T;\ere).
\label{a.4}
\ene
The isomorphism, $I$, is given as follows. Let $\sigma$ be a singular simplex and take $r\in \ere$.
Then,
$$
I([\sigma]\otimes r)=[r\sigma].
$$
See equation $(iv)$ and Lemma 3.A1, pages 261, 262 of \cite{h}. By Proposition \ref{prop-a.2}
$H_1(\natural k T;\ZETA) \cong \oplus_{i=1}^k \ZETA$ and $\left\{ [Z_j]_{H_1(\natural k  T;
\ZETA)}\right\}_{j=1}^k$ is a  basis of $H_1(\natural k T;\ZETA)$. Then,

$$
\oplus_{i=1}^k \ere \cong \tilde{H}_1(\X; \ZETA)\otimes \ere.
$$
The isomorphism is given by,
$$
\begin{array}{rrrr}\oplus_{j=1}^k \ere \longrightarrow &\oplus_{j=1}^k (\ZETA\otimes\ere)
\longrightarrow & (\oplus_{j=1}^k
\ZETA) \otimes \ere \longrightarrow & \tilde{H}_1(\X;\ZETA)\otimes \ere\\\\
(r_1,\cdots,r_k) \rightarrow & (1\otimes r_1\cdots,1\otimes r_k)\rightarrow & (1,0,\cdots,0)
\otimes r_1+\cdots +(0,0,\cdots1)\otimes r_k \rightarrow &\sum_{j=1}^k [Z_j]_{\tilde{H}_1(\X;\ZETA)}\otimes r_j.
\end{array}
$$
It follows that the morphism
$$
I':\oplus_{j=1}^k \ere \rightarrow \tilde{H}_1(\X;\ere): I'((r_1,\cdots,r_k)):= \sum_{j=1}^k [r_j Z_j
]_{\tilde{H}_1(\X; \ere)},
$$
is an isomorphism of groups.

We now prove that this implies that $\{ [Z_j]_{H_1(\X;\ere)} \}_{j=1}^k$ is a basis of
$H_1(\X; \ere)$ as a
vector space. As $\tilde{H}_1(\X;\ere)$ and $H_1(\X;\ere)$ are equal as sets and as groups the morphism

$$
I':\oplus_{j=1}^k \ere \rightarrow H_1(\X;\ere): I'((r_1,\cdots,r_k)):= \sum_{j=1}^k [r_j Z_j
]_{H_1(\X; \ere)},
$$
is an isomorphism of groups. By the structure of vector space of $H_1(\X;\ere)$ we have that,
$\sum_{j=1}^k [r_j Z_j]_{H_1(\X;\ere)}= \sum_{j=1}^k  r_j[ Z_j]_{H_1(\X;\ere)}$. As $I'$ is an
isomorphism of groups we have that $\forall \sigma  \in H_1(\X;\ere)$ there are real numbers
$\{r_j\}_{j=1}^k$ such that $\sigma =\sum_{j=1}^k  r_j[ Z_j]_{H_1(\X;\ere)}$. This means that
$\{ [Z_j]_{H_1(\X;\ere)} \}_{j=1}^k$ generates $H_1(\X;\ere)$. Moreover, if $0=\sum_{j=1}^k r_j[ Z_j
]_{H_1(\X;\ere)}= \sum_{j=1}^k  [r_j Z_j]_{H_1(\X;\ere)}$ we have that, $(r_1,r_2,\cdots,r_k)=0$
and we conclude that $\{ [Z_j]_{H_1(\X;\ere)} \}_{j=1}^k$ is a linearly independent set and since it also
generates $H_1(\X;\ere)$ it is a basis.

\section{Appendix B}\sss
In this appendix we prove, for completeness, the following proposition.

\begin{prop} \label{prop-b.1}
$\left\{ [\hat{\gamma_j}]_{H_1(\Lambda; \ZETA)}\right\}_{j=1}^m$
is a basis of $H_1(\Lambda; \ZETA)$.
\end{prop}

\noindent{\it Proof:} For simplicity we will omit $\ZETA$ in the homology groups in this proof.

\noindent Step 1.

\noindent As in the proof of (\ref{1.9}) we prove that $ H_2(\ere^3, \ere^3 \setminus K)\cong H_1(\ere^3 \setminus K)$.
Moreover the isomorphism is given by (page 75 of \cite{gh})

\beq
[\sigma]_{H_1(\ere^3, \ere^3\setminus K)}\rightarrow [\partial \sigma]_{H_1(\ere^3 \setminus K)}.
\label{b.1}
\ene

\noindent Step 2.

\noindent Define $K_\varepsilon:=\{x \in \ere^3 : \,{\mathrm dist}(x, K)\leq \varepsilon\}$.
   Since $ \overline{\ere^3 \setminus K_{\varepsilon}} \subset ({\ere^3\setminus K}\stackrel{\circ})$ it follows from the
   excision theorem (page 82 of \cite{gh}) that the inclusion $(K_\varepsilon ,K_\varepsilon\setminus K)
   \hookrightarrow (\ere^3, \ere^3\setminus K)$ induces an isomorphism in  homology.

\noindent Step 3.

\noindent Let $K_{\varepsilon,j}, j=1,2, \cdots,L $ be the connected components of $K_\varepsilon$
 for $\varepsilon$ small
enough. Then,  $K_{\epsilon,j}=\{x \in \ere^3: {\mathrm dist}(x, K_{j})\leq \varepsilon\}.$
By Proposition 13.9, page 72 of \cite{gh}
$$
H_2(K_\varepsilon, K_\varepsilon\setminus K)\cong \oplus_{j=1}^L H_2(K_{\varepsilon,j}, K_{\varepsilon,j}
\setminus K_j).
$$

\noindent Step 4.

\noindent We have the following homotopic equivalence $ K_{\varepsilon,j} \setminus K_j \simeq \partial
K_{\varepsilon,j}$, that induces the isomorphism in homology
$$
H_k(K_{\varepsilon,j}\setminus K_j)\cong H_k(\partial K_{\varepsilon,j}).
$$
Let us consider the exact  homology sequences of the pairs
$(K_{\varepsilon,j},K_{\varepsilon,j}\setminus K_j)$
and $(K_
{\varepsilon,j}, \partial K_{\varepsilon,j})$. The first  starts at
$H_k(K_{\varepsilon,j}\setminus K_j)$ and ends at $H_{k-1}(K_{\varepsilon,j})$ and the second  starts at
$H_k(\partial K_{\varepsilon,j})$ and ends at $H_{k-1}(K_{\varepsilon,j})$. By the five lemma
(page 77 of \cite{gh})
the inclusion $(K_{\varepsilon,j},  \partial K_{\varepsilon, j})
\hookrightarrow (K_{\varepsilon,j}, K_{\varepsilon,j}\setminus
K_j)$ induces the isomorphism in homology,

$$
H_k(K_{\varepsilon,j}, \partial K_{\varepsilon,j}) \cong H_k(K_{\varepsilon,j},K_{\varepsilon,j}
\setminus K_j ).
$$
\noindent Step 5.

\noindent  By the exact homology sequence for the pair $(K_{\varepsilon,j}, \partial K_{\varepsilon,j} )$
we obtain the sequence
$$
\rightarrow H_2(K_{\varepsilon,j}) \rightarrow H_2(K_{\varepsilon,j},\partial K_{\varepsilon,j})
\stackrel{\Delta_2}
{\rightarrow} H_1(\partial K_{\varepsilon,j})\stackrel{I}{\rightarrow} H_1(K_{\varepsilon,j})\rightarrow,
$$
where $\Delta_2$ is taking boundary and $I$ is the inclusion.
By Proposition \ref{prop-a.2} $H_2(K_{\varepsilon,j})=0$. Hence we obtain the exact sequence

\beq
0 \rightarrow H_2(K_{\varepsilon,j},\partial K_{\varepsilon,j})
\stackrel{\Delta_2}
{\rightarrow} H_1(\partial K_{\varepsilon,j})\stackrel{I}{\rightarrow} H_1(K_{\varepsilon,j})\rightarrow.
\label{b.2}
\ene
Let $\Gamma_{j} \subset \{1,2,\cdots\,m\}$ be such that $\{[\gamma_i]_{H_1(K_{\epsilon,j})}\}_{i \in \Gamma_j}$
is a basis of $H_1(K_{\epsilon,j})$ (see Subsection 2.4).

 Let
$\{\alpha_i\}_{i \in \Gamma_j}, \{\beta_i\}_{i \in \Gamma_j}$ be the curves defined in Example 2A.2, page 168
of  \cite{h}. Note that we can choose $\alpha_i= \hat{\gamma_i}$ (see (\ref{1.6}), just take
$ \frac{\varepsilon}{2} $ instead of $ \varepsilon $ in $ K_{\varepsilon} $ ).
Moreover as $\gamma_i \simeq \beta_i$ we have that  (see Theorem 11.2, page 59 of \cite{gh})
$[\beta_i]_{H_1(K_{\varepsilon,j})}= [\gamma_i]_{H_1(K_{\varepsilon,j})}$. Then, by Example 2A.2, page 168
of  \cite{h},
$$
\left\{ [\hat{\gamma}_i]_{H_1(\partial K_{\varepsilon,j})},[\beta_i]_{H_1(\partial K_{\varepsilon,j})} \right
\}_{i \in \Gamma_j}
$$
is a basis of $H_1(\partial K_{\varepsilon,j})$.

It is clear that $I ([\hat{\gamma_i}]_{H_1(\partial K_{\varepsilon,j})})=0, i \in \Gamma_j$.
Moreover, $I([\beta_i]_{H_1(\partial K_{\varepsilon,j})})= [\beta_i]_{H_1( K_{\varepsilon,j})}
= [\gamma_i]_{H_1( K_{\varepsilon,j})}$. Hence, $\hbox{\rm  Kern}\, I= \big\langle \left\{
[\hat{\gamma_i}]_{H_1(\partial K_{\varepsilon,j})}\right\}_{i \in \Gamma_j} \big\rangle$, the free
$\ZETA-$ module} or the free group  generated by
$\{ [\hat{\gamma_i}]_{H_1( \partial K_{\varepsilon,j})}\}_{i \in \Gamma_j}$.
We obtain then that,
$$
H_2(K_{\varepsilon,j}, \partial K_{\varepsilon,j}) \stackrel{\Delta_2}{\rightarrow} \,\hbox{\rm Kern}(I)=
\big\langle \left\{
[\hat{\gamma_i}]_{H_1(\partial K_{\varepsilon,j})}\right\}_{i \in \Gamma_j} \big\rangle.
$$

It follows that to construct a basis of $H_2(K_{\varepsilon,j}, \partial K_{\varepsilon,j})$ it is enough to
compute the inverse image under $ \Delta_2$ of the $\left\{
[\hat{\gamma_i}]_{H_1(\partial K_{\varepsilon,j})}\right\}_{i \in \Gamma_j}$. Let us take then,
$[\sigma_i]_{H_1(K_{\varepsilon,j},\partial K_{\varepsilon,j})}$ such that, $\partial \sigma_i= \hat{\gamma_i}$.
Hence, $\left\{[\sigma_i]_{H_1(K_{\varepsilon,j}, \partial K_{\varepsilon,j})}\right\}_{i \in \Gamma_j}$ is a
basis of $H_1(K_{\varepsilon,j}, \partial K_{\varepsilon,j})$.

Finally, by steps 4 and 5 $ \left\{[\sigma_i]_{H_2(K_{\varepsilon,j},  K_{\varepsilon,j}\setminus
 K_j)}\right\}_{i \in \Gamma_j}$ is a basis of $H_2(K_{\varepsilon,j},  K_{\varepsilon,j}\setminus
 K_j)$. By step 3  $\left\{[\sigma_i]_{H_2(K_{\varepsilon},  K_{\varepsilon}\setminus
 K)}\right\}_{i=1}^m$ is a basis of $H_2(K_{\varepsilon},  K_{\varepsilon}\setminus
 K)$. By step 2 $\left\{[\sigma_i]_{H_2(\ere^3,  \ere^3\setminus
 K)}\right\}_{i=1}^m$ is a basis on $H_2(\ere^3,  \ere^3\setminus
 K)$. By step 1 $\left\{[\hat{\gamma_i}]_{H_1(\ere^3\setminus K)}\right\}_{i=1}^m$ is a basis of
 $H_1(\ere^3\setminus K)$.

 \noindent{\bf Acknowledgement}

 \noindent This work was partially done while we were visiting the Department of Mathematics and Statistics of  the
 University of Helsinki. We thank Prof. Lassi P\"aiv\"arinta for his kind hospitality.

\end{document}